\documentclass{aa}  

\usepackage{graphicx}
\usepackage{txfonts}
\usepackage{booktabs}
 \usepackage{lscape}
 
\usepackage{tabularx,colortbl}

\begin{document}

   \title{The DUST-GAS AGN TORUS AS CONSTRAINED FROM X-RAY AND MID-INFRARED OBSERVATIONS}

   \author{D. Esparza-Arredondo\inst{1}
          \and
          O. Gonzalez-Mart\'in \inst{1}
          \and D. Dultzin \inst{2}
          \and J. Masegosa \inst{3}
          \and C. Ramos-Almeida \inst{4,5}
          \and Ismael Garc\'ia-Bernete \inst{6}
          \and Jacopo Fritz \inst{1}
          \and Natalia Osorio-Clavijo \inst{1}
          }

   \institute{Instituto de Radioastronom\'ia y Astrof\'isica (IRyA-UNAM), 3-72 (Xangari), 8701, Morelia, Mexico\\
              \email{d.esparza@irya.unam.mx}
         \and
             Instituto de Astronom\'ia (IA-UNAM), Mexico city, Mexico
          \and
             Instituto de Astrof\'isica de Andaluc\'ia (CSIC), Glorieta de la Astronom\'ia s/n 18008, Granda, Spain
          \and
            Instituto de Astrof\'isica de Canarias (IAC), C/Vía Láctea, s/n, E-38205, La Laguna, Spain
          \and 
            Departamento de Astrof\'isica, Universidad de La Laguna (ULL), E-38205 La Laguna, Spain
          \and
            Department of Physics, University of Oxford, Oxford OX1 3RH, UK
             }

   \date{Received ?? ; accepted ??}
\titlerunning{A two-phase agn torus}
 
  \abstract
   {In the last decades, several multiwavelength studies have been dedicated to exploring the properties of the obscuring material in active galactic nuclei (AGN). Various models have been developed to describe the structure and distribution of this material and constrain its physical and geometrical parameters through spectral fitting techniques. However, questions, including how the torus mid-infrared (mid-IR) and X-ray emission are related remain unanswered.}
   {In this work, we study whether the dust continuum at mid-IR and gas reflection at X-rays have the same distribution in a sample of AGN.}
   {We carefully selected a sample of 36 nearby AGN with \emph{NuSTAR} and \emph{Spitzer} spectra available that satisfy the following two criteria: 1) the AGN component dominates the mid-IR spectra (i.e. the stellar and interstellar medium components contribute less than 50\% to the spectrum) and 2) the reflection component has a significant contribution to the X-ray spectrum. We discarded the sources whose reflection component could be produced by ionized material in the disk. We derived the properties of the nuclear dust and gas through a spectral fitting, using models developed for mid-IR and X-ray wavelengths assuming smooth and clumpy distributions for this structure.
   }
   {We found that a combination of smooth and clumpy distributions of gas and dust, respectively, is preferred for $\sim 80$\% of sources with good spectral fits according to the Akaike criterion. However, considering extra information about each individual source, such as the absorption variability, we found that $\sim 50$\% of our sources are best described by a clumpy distribution of both dust and gas. The remaining $\sim$50\% of our sources can still be explained with a smooth distribution of gas and a clumpy distribution of dust. Furthermore, we explored the torus dust-to-gas ratio, finding that it is [0.01-1] times that of the interstellar medium.}
   {The results presented in this paper suggest that the distribution of the gas and dust in AGN is complex. We find at least six scenarios to explain the observed properties of our sample. In these scenarios, three gas-dust distribution combinations are possible: clumpy-clumpy, smooth-smooth, and smooth-clumpy. Most of them are in agreement with the notion that gas could also be located in the dust-free region, which is consistent with the dust-to-gas ratio found.}
   


   \maketitle
%

\section{Introduction}
\label{sec:intro}

The term active galactic nucleus (AGN) is used to define the center of galaxies that contain a super-massive black hole ($\rm{M_{BH} \sim 10^{6}}$-$\rm{10^{9} \, M_{\odot}}$, SMBH) which is efficiently fed by material through an accretion disc \citep[e.g.][]{Kormendy95,Kormendy13}. The accretion disc, in turn, is embedded in a region of gas clouds with high-velocity dispersion which is known as the broad-line region \citep[BLR, see e.g.][for a review]{Sulentic00,Marziani10}. These components are surrounded by an obscuring region composed of gas and dust which classically is referred to as the ``torus'' \citep{Antonucci85, Urry95}. The classical unification model proposes that all classes of AGN observed are intrinsically the same but appear different merely due to the orientation effect \citep[see][for a review]{Netzer15}. However, the individual properties of the torus might also explain the AGN types and they might be linked to changes in the accretion state ruled by the bolometric luminosity ($\rm{L_{bol}}$) and/or the SMBH mass \citep[][]{Khim17}.

Several techniques have been used to study the properties of the torus. They can be roughly divided into three categories: (1) interferometry \citep{Tristram09,Burtscher13,Alonso-Herrero18,Garcia-Burillo19,Impellizzeri19,Combes19}; (2) reverberation \citep[e.g.][]{Suganuma06,Koshida14,Almeyda17}; and (3) spectral energy distribution (SED) fitting \citep[e.g.][]{Ramos-Almeida09, Murphy09,Ramos-Almeida11, Garcia-Bernete19, Brightman11,Liu14, Furui16,Yaqoob12, Esparza-Arredondo19}. Both interferometry and infrared (IR) reverberation are restricted to the brightest AGN due to the high signal-to-noise required and spatial scales at play. This work is based on the third technique, which allows us to target larger AGN samples. Most of the AGN energy is produced in the accretion disc mainly by photons at ultraviolet (UV) and optical wavelengths. Dust absorbs part of this disc emission, which is then heated and re-emitted as IR radiation \citep{Rieke78}. A portion of this UV/optical disc emission is also reprocessed by an optically thin corona of hot electrons plasma above the accretion disc that scatters the energy in the X-ray bands due to inverse Compton \citep[][and references therein]{Netzer15,RamosAlmeida17}. This comptonization produces one of the three main components seen in AGN at X-rays: the intrinsic continuum. The second, and third components are the reflection of the intrinsic continuum, and the iron emission line at $\rm{6.4\, keV}$ (FeK$\rm{\alpha}$). These two components are produced due to the scattering of X-ray emission reflected by the inner walls of the torus or the BLR. While the FeK$\rm{{\alpha}}$ line can be produced by material with column densities ($\rm{N_H}$) as low as $\rm{N_{H} = 10^{21-23} \, cm^{-2}}$, the Compton hump can only be seen by the reprocessing of X-ray photons in a Compton thick material $\rm{N_{H} > 10^{24} \, cm^{-2}}$. Note that the FeK$\rm{\alpha}$ line and the compton hump components might also be associated with reprocessing of the intrinsic emission at the accretion disc \citep{Fabian98, Laor91}.

Both the infrared continuum and the reflection components at X-rays depend on the shape (geometry and density) of the reprocessing material \citep[e.g.][]{Ghisellini94,Nenkova08a}. Thus, comparing spectra and models helps to infer the properties of the AGN torus, otherwise unreachable for the vast majority of the AGN. Different sets of IR and X-ray torus models have been developed in attempts to reproduce the observed spectra (pioneer works by \citealt{Krolik88,Granato94}; and see \citealt{RamosAlmeida17} for a review).

At mid-infrared (mid-IR) wavelengths, the stationary models that assume a smooth distribution of the dust were the first explored due to computational reasons \citep[known as smooth models, e.g][]{Pier92,Efstathiou95,Manske98, Fritz06}. {These smooth models were followed by models that adopt a clumpy distribution for the dust} \citep[e.g.][]{Rowan-Robinson95,Nenkova08a,Nenkova08b, Hoenig06,Schartmann08,Hoenig10}, a clumpy two-phase medium \citep{Stalevski12, Siebenmorgen15}, and more recently, a clumpy dusty disc with a polar outflow \citep{Hoenig17}. The smooth and clumpy models have been tested in large samples of AGN and have achieved to find differences in the torus parameters between type-1 and type-2 AGN \citep[e.g][]{Ramos-Almeida09,Ramos-Almeida11,Hoenig10,Lira13,Garcia-Bernete19} and even a dependency with the AGN luminosity \citep{Gonzalez-Martin17,Gonzalez-Martin19b}. However, obtaining the full set of parameters for the models has been difficult so far \citep{Ramos-Almeida14, Gonzalez-Martin19b}.

At X-ray wavelengths, several smooth and clumpy torus models have been developed to explore the reflection component produced by gas around the accretion disc \citep[e.g.][]{Murphy09,Ikeda09,Brightman11}. Although with slight differences in morphology, these models have allowed to constrain several properties of the material that originates the reprocessed emission. Only a few works have compared these models in an effort to constrain the model parameters \citep[e.g.][]{Liu14,Furui16,Balokovic18}.

Although somehow accepted, the link between the dust-producing mid-IR continuum and the gas-producing X-ray reflection is not clear yet. Recently, few works combine mid-IR and X-ray observations and models either by full spectral energy distribution (SED) fitting for a large collection of objects \citep[e.g.][]{Yang20,Ogawa20} or through simultaneous fitting of both ranges for individual objects \citep[see][]{Esparza-Arredondo19}. The advantage of the simultaneously fitting technique is its capability to better constrain all the torus parameters. However, prior to perform the simultaneous fitting, the emission at both wavelengths needs to be confronted.

We present the results obtained from fitting a sample of AGN using the clumpy and smooth models to reproduce their mid-IR and X-ray emission. We discuss our results for each wavelength and how compatible they are with each other. Our goal is to understand if the same structure of material can produce both spectral components in AGN. A brief description of the models used in this work is shown in Section \ref{sec:ModelsDesc}. The sample selection and spectral fitting methodology are described in Section \ref{Sec:sample}.

The results for X-ray and mid-IR wavelengths are described in Sections \ref{sec:Xrayresults}  and \ref{sec:midIRresults}, respectively. The confidence range of error calculated here is 1 sigma. In Section \ref{sec:Disc} we discuss our results within the framework of our goals. Our conclusions are given in Section \ref{sec:Conclusion}. Throughout this work, we assume a cosmology with $\rm{H_{0} = 70\, kms^{-1}\,Mpc^{-1}}$, $\Omega_{M} = 0.27$, and $\Omega_{\lambda} = 0.73$.

\begin{table*}
\begin{center}
\begin{scriptsize}
\resizebox{19cm}{!}{
\begin{tabular}{llll}
\hline \hline
Model & Parameter & Range & Description \\
\hline \hline
Smooth model & i & [$0.01^{\circ}$, $90^{\circ}$] & Inclination angle respect to the equatorial plane \\
by \citet{Fritz06} & $\sigma$ & [$20^{\circ}$, $60^{\circ}$] & Half-opening angle of the torus \\
   & $Y$ & [10, 150] & Ratio between the inner and outer radius \\ 
   & $\gamma$ & [0.01, 6] & Index of the logarithmic elevation density distribution \\
    & $\beta$ & [-1, -0.01] & Index of the radial profile of the density distribution \\ 
   & $\tau_{9.7 \mu m}$ & [0.1, 10] &  Equatorial optical depth at 9.7$\mu$m \\  
\hline
Clumpy model & i & [$0.01^{\circ}$, $90^{\circ}$] & Inclination angle respect to the polar plane \\
by \citet{Nenkova08a,Nenkova08b} & $\sigma$ & [$15^{\circ}$, $70^{\circ}$] & Half-opening angle of the torus \\
   & $Y$ & [5, 100]  & Ratio between the inner and outer radius \\ 
  & N0 & [1, 15]& Number of clouds in the equatorial plane of the torus \\
   & $q$ & [0.01, 2.5] & Slope of the radial distribution of clouds \\ 
   & $\tau_{\nu}$ & [10, 300] & Optical depth of each cloud \\ 
\hline \hline
Smooth model (borus02) & $\theta_{inc}$ & [$19^{\circ}$, $87^{\circ}$] & Inclination angle respect to the polar plane \\
by \citet{Balokovic18} & $\rm{\theta_{tor}}$ & [$0^{\circ}$, $84^{\circ}$] & Half-opening angle of the torus respect to the polar plane \\
   & $\rm{N_H}$ & [22.0, 25.5] & The average column density of the torus [cm$^{-2}$] \\
   & $\Gamma$ & [1.4, 2.6] & Photon index of power law \\
   & E$\rm{_{cut}}$ & [20,2000] & High-energy cutoff [KeV] \\
   & A$\rm{_{Fe}/A_{Fe,\odot}}$ & [0.01,10.0] & Relative abundance of iron \\
\hline
Clumpy model (UXClumpy) & $\theta_{inc}$ & [$0^{\circ}$,$90^{\circ}$] & Inclination angle, relative to the inner (flat) disc portion. \\
by \citet{Buchner19} & $\rm{\theta_{tor}}$ & [$6^{\circ}$, $90^{\circ}$] & Cloud dispersion (half-opening angle of the torus) \\
    & $\rm{N_H}$ & [20.0, 26.0] & The average column density of the line-of-sight [cm$^{-2}$] \\
   & $\Gamma$ & [1.0, 3.0] & Photon index of power law \\
  & E$\rm{_{cut}}$ & [60,400] & High-energy cutoff [KeV] \\
  & $CTkcover$ & [0, 0.6] & Covering fraction of inner ring \\
\hline \hline
\end{tabular}
}
\caption{Summary of mid-IR and X-ray models used in this paper, including parameter range and description.}
\label{tab:ModelParam}
\end{scriptsize}
\end{center}
\end{table*}
\section{Description of the models}
\label{sec:ModelsDesc}

The models presented in this section have been created through different radiative transfer codes which include the physics required to account for mid-IR and X-ray main continuum features. All of them assume a torus-like morphology that obscures the accretion disc for certain viewing angles. Mid-IR models reproduce the continuum including re-emission due to dust outside the dust sublimation radius. X-ray models reproduce the reflection component and the FeK$\rm{\alpha}$ emission line, assuming that both components come from a distant reflecting material composed by neutral gas. The X-ray models chosen here are the more recent, flexible (more free parameters), and solve some problems that affected their predecessors \citep[see][]{Liu14}. The mid-IR models are chosen to match in geometry those available at X-rays. Table \ref{tab:ModelParam} shows the parameters involved including coverage for each of them. Below we give a short description highlighting the main reasons why we chose them. We refer the reader to the primary papers for a complete description of them. 

\subsection{Mid-IR models}

$\bullet$ \underline{Smooth model} \citep[by][]{Fritz06}: This model considers a flared disc which is created as two concentric spheres with the polar cones removed and delimited by inner and outer radius. The model assumes a continuous/smooth distribution of dust with almost equal silicate and graphite proportions. The smooth model is a simple representation of the torus but it is capable of giving a good approximation of the infrared SED of AGN \citep[e.g.][]{Esparza-Arredondo19}. This is the only mid-IR smooth torus model with a complete SED library in the literature to compare with the available X-ray models.

$\bullet$ \underline{Clumpy model} \citep[by][]{Nenkova08a,Nenkova08b}: This model considers a toroidal distribution of dusty clumps with standard Galactic composition (of $53\%$ silicates and $47\%$ graphite) of dust. We favor this model against the other available clumpy torus models \citep[in particular][]{Hoenig10} because it provides the largest SED library and it has been extensively used due to its probed ability to explain the mid-IR emission of AGN at large range of luminosities \citep[e.g.][]{Ramos-Almeida09,Gonzalez-Martin17,Martinez-Paredes17,Garcia-Bernete19}.

A comparison between these two mid-IR torus models is presented in \citet{Feltre12}. Recently, \citet{Hoenig17} presented a new radiative transfer model consisting on an inflowing disc dominating the NIR emission and an outflowing wind emitting in the MIR \citep{Hoenig19}. This model is motivated by the detection of a significant fraction of polar emission in the MIR in several nearby AGN \citep[e.g.][]{Lopez-Gonzaga16, Leftley18}. Although some effort has already been made to look for the X-ray signatures of this polar component \citep[e.g.][]{Buchner18,Liu19}, there are not available X-ray SED libraries for this type of model. For this reason we do not explore this disc+wind torus models here, in spite of their adequacy to reproduce IR spectra \citep[e.g.][]{Gonzalez-Martin19b}.


\subsection{X-ray models}
\label{sec:XrayModelsDesc}

$\bullet$ \underline{Smooth model} \citep[by][]{Balokovic18}: This smooth torus model is called borus02. The geometry of this model is a uniform-density sphere with two conical polar cutouts, where the opening angle is a free parameter. This geometry is fully compatible with the mid-IR smooth torus model by \citet{Fritz06}. We chose this model due to its capability to separate the column density of the line-of-sight (LOS) and the reprocess component compared to similar models \citep[e.g.][]{Brightman11}.


$\bullet$ \underline{Clumpy model} \citep[by][]{Buchner19}: The so-called UXClumpy model assumes toroidal distribution with the column density being a monotonic function decreasing towards the poles. The geometry and parameters of this model were chosen to match the clumpy mid-IR model proposed by \citet{Nenkova08a,Nenkova08b} \citep[see Table 2 in][]{Buchner19}, making this combination ideal for our work.

\citet{Balokovic18} and \citet{Buchner19} compared the borus02 and UXClumpy models, respectively, with previous models (e.g. BNtorus by \citealt{Brightman11} and ctorus by  \citealt{Liu14}). They are the most recent among the available models with smooth and clumpy torus geometries. The viewing angle, $\rm{\theta_{inc}}$, photon index, $\rm{\Gamma}$, and half opening angle of the torus, $\rm{\theta_{tor}}$, cover similar ranges (see Table\,\ref{tab:ModelParam}). In both models, the $\rm{N_H}$ parameter ranges are similar. However, the smooth model considers that the $\rm{N_H}$ from the torus and that of the LOS could be different (we test this possibility in Section\,\ref{Sec:sample}).

\section{Sample and Spectral fitting}
\label{Sec:sample}

\subsection{Mid-IR selection and spectral fitting}
\label{Subsec:midIRselection}

The initial sample included 169 AGN with available mid-IR IRS/\emph{Spitzer} spectra within the CASSIS\footnote{https://cassis.sirtf.com} archive \citep{Lebouteiller11} and X-ray \emph{NuSTAR} \citep{Harrison13} observations. We only considered sources with more than 10 ksec of \emph{NuSTAR}, in order to have spectra with good signal-to-noise, and full $\sim [5 \mu m - 35 \mu m]$ wavelength coverage for the IRS/\emph{Spitzer} spectra. The final sample also included a minimum number of bins in the resulting X-ray spectrum (see below).

First, we converted IRS/\emph{Spitzer} spectra into X-ray spectral fitting package XSPEC \citep{Arnaud96} format using {\sc flx2xsp} task within HEASOFT\footnote{https://heasarc.gsfc.nasa.gov/docs/software/heasoft/}. Following the technique developed by \citet{Gonzalez-Martin19b}, we fitted each spectrum using four baseline models:
\begin{eqnarray}
M1(IR) & = & zdust \times {dust \, model} \label{eqs:1} \\
M2(IR) & = & zdust \times {dust \, model} + Stellar \label{eqs:2} \\
M3(IR) & = & zdust \times  {dust\, model} + ISM \label{eqs:3} \\
M4(IR) & = & zdust \times {dust\, model} + ISM + Stellar \label{eqs:4}
\end{eqnarray}

\noindent where $zdust$ component is the foreground extinction by dust grains \citep{Pei92}. We found that the inclusion of zdust in stellar or ISM components or both is neglectable since returns similar results to those obtained when it is not considered. The $ISM$ and $stellar$ components are the interstellar medium and stellar population of $\rm{10^{10}}$ years and solar metallicity. Note that a different choice of stellar population age does not affect the results. These components were taken from \citet{Smith07} and \citet{Bruzual03}, respectively. Finally, the $dust \, model$ correspond to the smooth \citep{Fritz06} or clumpy \citep{Nenkova08a,Nenkova08b} torus models designed to describe the IR AGN emission. The initial parameters were set to the mean value over the parameter range. We computed the $\chi^2$ statistic for the best fit through finding the absolute minimum within each parameter range. We then used f-statistics to test whether the inclusion of the stellar (eq.\,\ref{eqs:2}), ISM (eq.\,\ref{eqs:3}), or the stellar+ISM (eq.\,\ref{eqs:4}) components significantly improves the simpler model when f-test probability is below $10^{-4}$.

We selected objects showing less than 50\% of stellar component compared to that of the torus component at $\rm{5 \mu m}$ and less than 50\% of ISM component compared to that of the torus component at $\rm{30 \mu m}$. \citet{Gonzalez-Martin19a} demonstrated that these restrictions are a good compromise to recover the model parameters using \emph{Spitzer}/IRS spectra. We imposed that this condition is fulfilled when using both (clumpy and smooth) AGN dust models. In this way, we excluded 71 sources that are not AGN dominated at mid-IR wavelengths from our initial sample. This yielded to a sample of 98 sources where the AGN dust dominates at mid-IR wavelengths that are sampled by \emph{Spitzer}/IRS spectra. 

\begin{figure}
    \centering
    \includegraphics[width=1.0\columnwidth]{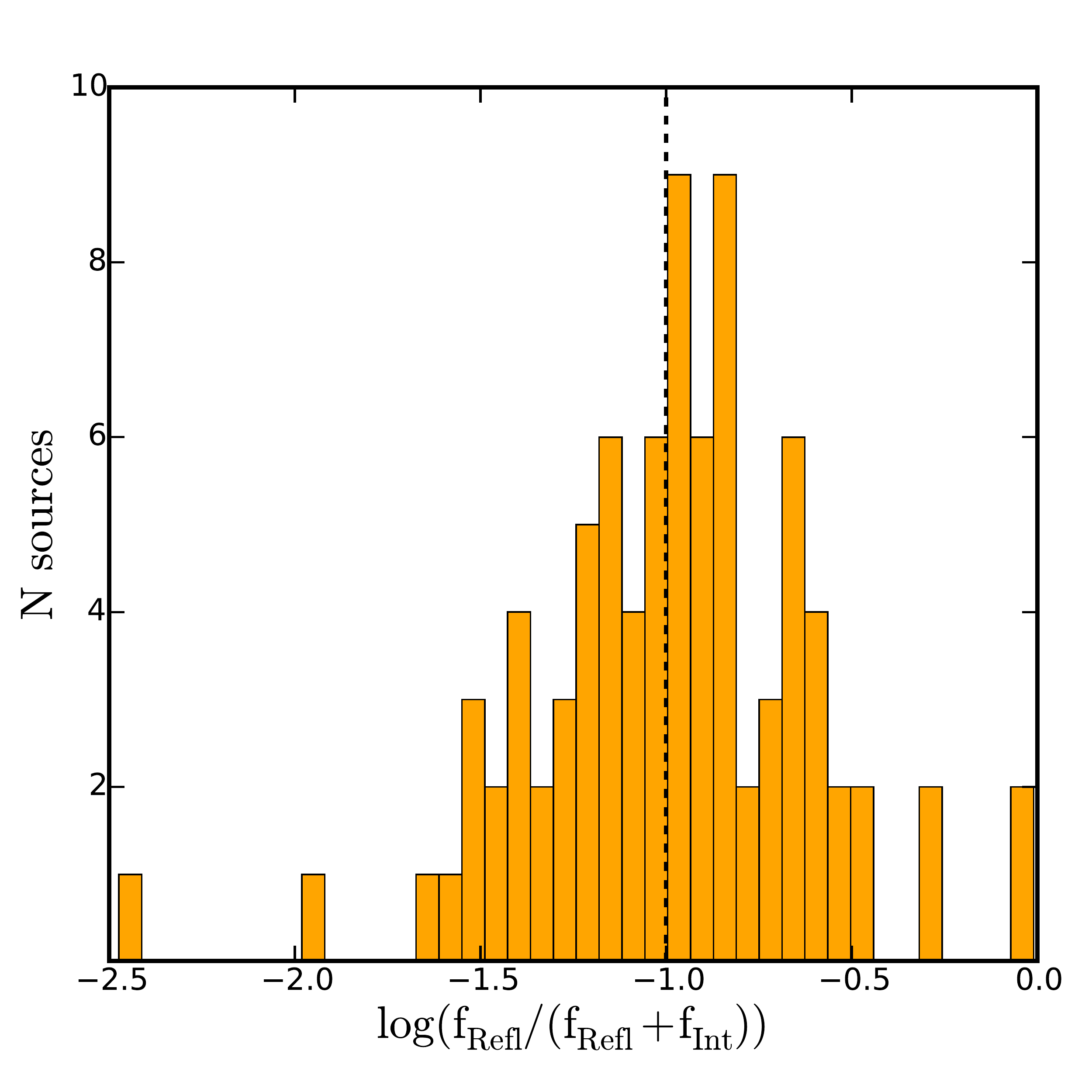}
    \caption{Histogram of the reflection fraction in the 3-70 keV band for the sample of mid-IR AGN dominated spectra (86 sources, see text).}
    \label{fig:histSelectXray}
\end{figure}

\begin{table*}
\begin{small}
   \caption{\label{tab:observations} General properties of sample}
    \centering
    \begin{tabular}{lccccccccc}
    \hline \hline
Object  &  \multicolumn{2}{c}{Coordinates} & Type$^{*}$& z & $L_X$ & M$\rm{_{BH}}^{a}$ & Obsid & Exptime & NH  \\
   &  RA & DEC &  & & $log(L_{2-10KeV})$ & log($M/M_{\odot}$) & & (k sec) & Variability \\
(1) & (2) & (3) & (4) & (5) &  (6)  & (7) &  (8) &  (9) & (10) \\ 
\hline
Mrk1018 &   02 06 15.98 &   -00 17 29.22 &   Sy1.8 &   0.0424 & 42.89$\pm_{0.03}^{0.03}$  &  8.03$^{1}$ &   60301022003 &  43.3 & N$^{1}$\\ 
Mrk590 &   02 14 33.56 &   -00 46 00.18 &   Sy1.2 &   0.0213 & 42.57$\pm_{0.02}^{0.02}$ & 7.20$^{2}$  &   90201043002 & 51.0 & N$^{2}$\\
PG0804+761 &   08 10 58.66 &   +76 02 42.45 &   Sy1 &   0.1000 &  42.55$\pm_{0.01}^{0.01}$ & 8.73$^{1}$  &   60160322002 & 17.3 & N$^{3}$ \\
RBS0770 &   09 23 43.00 &   +22 54 32.57 &   Sy1.2 &   0.0323 &  43.80$\pm_{0.01}^{0.01}$ & 7.34$^{1}$ &   60061092002 & 18.8 & - \\
I11119+3257 &   11 14 38.89 & +32 41 33.48 &   Sy1 &   0.1890 & 44.12$\pm_{0.08}^{0.06}$ & 8.4$^{3}$   &   60101045002 & 10.4 & Y$^{4}$\\ 
PG1211+143 &   12 14 17.67 &   +14 03 13.18 &   Sy1 &   0.0904 & 43.85$\pm_{0.01}^{0.01}$ & 7.49$^{2}$ &   60001100007 & 74.8 & N$^{5}$ \\ 
RBS1125 &   12 32 03.62 &   +20 09 29.49 &   Sy1 &  0.0630 & 43.82$\pm_{0.03}^{0.03}$ &  7.76$^{1}$ &   60061229002 & 19.9 & - \\ 
Mrk231 &   12 56 14.23 &   +56 52 25.24 &   Sy1 &  0.0422 & 42.78$\pm_{0.04}^{0.04}$ & 8.39$^{3}$  &   80302608002 &  82.0 & N$^{6}$ \\
Mrk1383 &   14 29 06.57 &   +01 17 06.15 &   Sy1 &   0.0866 & 44.30$\pm_{0.01}^{0.01}$ & 8.65$^{3}$ &   60061254002 & 32.4 & N$^{7}$ \\
Mrk1392 &   15 05 56.55 &   +03 42 26.34 &   Sy1.8 &   0.0361 & 43.18$\pm_{0.02}^{0.02}$ &  8.17$^{13}$  &   60160605002 &   21.0 & - \\
Mrk1393 &   15 08 53.95 &   -00 11 48.99 &   Sy1.5 &   0.0543 & 43.88$\pm_{0.01}^{0.01}$ & 7.87$^{1}$   &   60376005002 &  30.8 & Y$^{8}$\\
PG1535+547 &   15 36 38.40 &   +54 33 33.21 &   Sy1 &   0.0451 & 42.80$\pm_{0.06}^{0.05}$ & 7.19$^{4}$   &   60201023004 &  82.1 & N$^{9}$ \\
ESO141-G055 &   19 21 14.15 &   -58 40 12.98 &   Sy1.2 &   0.0371 & 43.94$\pm_{0.01}^{0.01}$ & $-$ &   60201042002 & 93.0 & - \\
NGC7213 &   22 09 16.21 &   -47 10 00.08 &   Sy1.5 &   0.0051 & 41.98$\pm_{0.01}^{0.01}$ &  7.99$^{2}$  &   60001031002 & 10.1 & N$^{10}$ \\ 
MCG+01-57-016 &   22 40 17.06 &   +08 03 13.52 &   Sy1.8 &   0.0249 & 42.79$\pm_{0.02}^{0.02}$ &  7.20$^{13}$  &   60061343002 & 21.3 & - \\ 
\hline 
UM146 &   01 55 22.02 &   +06 36 42.42 &   Sy2 & 0.0144 & 41.84$\pm_{0.07}^{0.06}$  & 6.23$^{13}$  &   60465002002 & 28.4 & - \\ 
NGC788 &   02 01 06.46 &   -06 48 57.15 &   Sy2 &   0.0136 & 42.43$\pm_{0.07}^{0.06}$ & 7.51$^{2}$ &   60061018002 & 15.4 & N$^{12}$ \\
NGC1052 &   02 41 04.79 &   -08 15 20.75 &   Sy2 &   0.0048 & 41.74$\pm_{0.01}^{0.01}$ & 8.19$^{2}$  &   60201056002 &  59.6 & Y$^{13}$ \\
NGC1358 &   03 33 39.68 &   -05 05 22.23 &   Sy2 &   0.0134 & 42.38$\pm_{0.03}^{0.02}$ & 7.88$^{2}$ &   60301026002 &  49.8 & - \\
J05081967+1721483 &   05 08 19.71 &   +17 21 48.09 &   Sy2 &   0.0175 & 42.97$\pm_{0.02}^{0.02}$ & $-$ &   60006011002 & 15.5 & - \\
Mrk3 &   06 15 36.45 &   +71 02 15.24  &   Sy2 &   0.0143 & 43.53$\pm_{0.01}^{0.01}$ & 8.65$^{2}$ &   60002048004 &  33.4 & N$^{14}$ \\ 
ESO428-G014 &   07 16 31.21 &   -29 19 28.89 &   Sy2 &   0.0054 & 39.83$\pm_{0.45}^{0.22}$ & 7.3$^{5}$ &   60001152002 & 40.2 & - \\ 
Mrk78 &   07 42 41.70 &   +65 10 37.43 &   Sy2 &   0.0371 & 43.11$\pm_{0.05}^{0.05}$ &  7.87$^{2}$  &   60061336002 &  24.1 & -\\
Mrk1210 &   08 04 05.86 &   +05 06 49.81 &   Sy2 &   0.0135 & 44.63$\pm_{0.02}^{0.02}$  & 7.1$^{6}$  &   60061078002 &  15.4 & Y$^{14}$ \\ 
J10594361+6504063 &   10 59 43.62 &   +65 04 06.37  &   Sy2 &   0.0836 & 43.66$\pm_{0.02}^{0.02}$  & $-$  &   60061207002 &  25.9 & - \\ 
NGC4388 &   12 25 46.82 &   +12 39 43.45    &   Sy2 &   0.0045 & 41.87$\pm_{0.02}^{0.02}$ & 8.54$^{7}$  &   60061228002 &  21.3 & N$^{11}$ \\ 
NGC4507 &   12 35 36.63 &   -39 54 33.66 &   Sy2 &   0.0118 &  43.17$\pm_{0.01}^{0.01}$ & 8.25$^{7}$  &   60102051004 &  34.4 & Y$^{11}$ \\ 
NGC4939 &   13 04 14.33 &   -10 20 22.32 &   Sy2 &   0.0085 & 42.0$\pm_{0.04}^{0.04}$ & $-$  &   60002036002 &   22.0 & Y$^{16}$\\ 
ESO097-G013 &   14 13 09.91 &   -65 20 20.47   &   Sy2 &   0.0009 & 40.78$\pm_{0.02}^{0.02}$ & 7.95$^{8}$  &   60002039002 &  53.8 & N$^{14}$\\ 
IC4518W &   14 57 40.50 &   -43 07 54.00 &   Sy2 &   0.0162 & 42.68$\pm_{0.06}^{0.05}$  & $-$ &   60061260002 &   7.78 & - \\ 
ESO138-G1 &   16 51 20.27 &   -59 14 04.28 &   Sy2 & 0.0091 & 42.43$\pm_{0.04}^{0.04}$ & 6.7$^{9}$ &   60201040002 &  45.6 & N$^{14}$\\
NGC6300 &   17 16 59.47 &   -62 49 13.98 &   Sy2 &   0.0029 & 41.86$\pm_{0.01}^{0.01}$  &  5.45$^{10}$  &   60261001004 &  23.5 & N$^{14}$\\ 
ESO103-G35 &   18 38 20.32 &   -65 25 39.14 &   Sy2 &   0.0133 &  43.23$\pm_{0.01}^{0.01}$ & 7.0$^{11}$  &   60301004002 & 43.7 & -\\ 
MCG+07-41-03 &   19 59 28.36 &   +40 44 02.09 &   Sy2 & 0.0561 & 44.61$\pm_{0.01}^{0.01}$ & $-$  &   60001083002 & 43.6 & Y$^{15}$ \\
IC5063 &   20 52 02.33 &   -57 04 07.60  &   Sy2 &   0.0088 & 42.55$\pm_{0.02}^{0.01}$ &  7.74$^{2}$  &   60061302002 & 18.4 & Y$^{10}$\\ 
PKS2356-61 &   23 59 04.36 &   -60 54 59.41 &   Sy2 &   0.0963 & 43.96$\pm_{0.04}^{0.04}$ & 8.96$^{12}$   &   60061330002 &  23.0 & N$^{17}$\\
\hline 
\end{tabular}
\tablefoot{(*) Classification obtained from NASA/IPAC Extragalactic Database (NED)\footnote{https://ned.ipac.caltech.edu/}. Note that UM\,146 is classified as Sy2. However, its classification is controversial; It is also classified as Sy1.9 according to NED. The instrinsic (not absorbed) X-ray luminosity obtained through \emph{clumin} tool from Xspec. References $\rm{M_{BH}}$: (1) \citet{Kamraj18}; (2) \citet{Woo02}; (3) \citet{Sani10}; (4) \citet{Vestergaard06}; (5) \citet{Fabbiano19};(6) \citet{Clouse11}; (7) \citet{Nicastro03}; (8) \citet{WaltonJ13}; (9) \citet{Piconcelli11}; (10) \citet{Awaki05}; (11) \citet{Wilkes01}; (12) \citet{Ursini18}; and (13) mass calculated using the $\rm{M_{BH} \propto \sigma}$ relationship, using the $\sigma$ values reported in the hyperleda database \footnote{$http://leda.univ-lyon1.fr$ \citep{Makarov14}.}. References for the $\rm{N_H}$ variability: (1) \citet{Krumpe17}; (2) \citet{Denney14}; (3) \citet{Papadakis03}; (4) \citet{Tombesi17}; (5) \citet{Reeves18}; (6) \citet{Teng14}; (7) \citet{Scott04}; (8) \citet{Wang09}; (9) \citet{Ballo08}; (10)\citet{Burtscher16}; (11)\citet{Risaliti02}; (12) \citet{Hernandez-Garcia15}; (13)\citet{Osorio-Clavijo20}; (14) \citet{Hernandez-Garcia16}; (15)\citet{Laha20}; (16)\citet{Guainazzi05}; and (17) \citet{Ursini18}.}
\end{small}
\end{table*}

\subsection{X-ray selection}

We used the High Energy Astrophysics Archive Research Center, HEASARC\footnote{https://heasarc.gsfc.nasa.gov/docs/archive.html} to download the data, and processed the \emph{NuSTAR} spectra of the 98 AGN with mid-IR \emph{Spitzer} spectra dominated by the AGN dust. Note that, when multiple observations were available for a single object, we took the longest exposure time to guarantee the best signal-to-noise ratio available. 

\emph{NuSTAR} data processing was done using the data analysis software \emph{NuSTARDAS} (v.1.4.4) distributed by HEASARC. The calibrated, cleaned, and screened event files were generated using the {\sc nupipeline} task (CALDB 20160502). We left the default pipeline option of not using any Atlantic Anomaly (SAA) filtering. SAA high background periods do not significantly affect our observations, according to the \emph{NuSTAR} SAA filtering reports. A circular region of $1'$ $- 2'$ radius (depending on the brightness of the source) was taken to extract the source and background spectrum on the same detector and to compute the response files (RMF and ARF files) using the {\sc nuproducts} package within \emph{NuSTARDAS}. Finally, we used the {\sc grppha} task within the FTOOLS to group the spectra with at least 60 counts per bin. After reducing the data, we discarded 12 sources due to the low-quality of the \emph{NuSTAR} observations with less than 70 bins in the 3-70 keV range.

We fitted the remaining 86 \emph{NuSTAR} spectra with a power-law model including neutral partial covering (modeled with $zpcfabs$ within the XSPEC software) plus a reflection model. We are aware that attenuation from Compton-scattering might have an impact at energies above 20-30\,keV and is significant particularly when the column density of the gas in the LOS is larger than $\rm{\sim 10^{24} \, cm^{-2}}$ \citep{Maiolino07}. The only available model within XSPEC is CABS. Unfortunately, this model uses an inadequate approximation that results in a wrong spectral hardening, as reported by \citet[][]{Murphy09,Yaqoob12}. 
In fact, we made some tests, finding that the inclusion of CABS results in unrealistically high luminosity (with intrinsic luminosity above 10,000 times larger than the observed luminosity). Due to the lack of Compton-scattering models properly accounting for Compton-scattering, we neglected this contribution. Note that many other works use the same approximation including only photoelectric absorption \citep[e.g.][]{Xu17,Ricci17, Ronchini19,Falocco20,Wang20}. Therefore, the baseline model use in this work has the form:
\begin{equation}
    M(X-ray) = zpcfabs \times cutoffpl + {reflection}
    \label{Eq:BaselineXray}
\end{equation}

\noindent Note that the existence of the 6.7 and 6.96\,keV emission lines was also tested by including Gaussian profiles at a fixed energy. For sample selection purposes, we chose the smooth torus model ($borus02$) for the reflection component \citep{Balokovic18}. We fixed the Ecut parameter from $borus02$ to 400\,keV. We linked the photon index, high-energy cutoff, and column density parameters of the intrinsic and reflection components to the same value. 
In Section\,\ref{Sec:sample}, we also tested if allowing different column densities along the LOS to the intrinsic continuum compared to that of the reflection component produced significantly improves results. The viewing angle cannot be constrained using this baseline model and data (see Section\,\ref{Sec:sample}). Therefore, we fitted each object to three fixed viewing angles (19$^{\circ}$,45$^{\circ}$, and 87$^{\circ}$, which are the minimum, medium, and maximum values of this model, respectively). To choose the best fit for each source, we compared the $\chi^2/d.o.f.$ value obtained for each viewing angle. We followed this fit methodology for the clumpy torus model (UXClumpy) also considering three different viewing angles (1$^{\circ}$,45$^{\circ}$, and 90$^{\circ}$, which are the minimum, medium, and maximum values of this model, respectively). These are slightly different due to the differences in the parameter space for these models (see Table\,\ref{tab:ModelParam}).

We then computed the observed fluxes (i.e. absorbed) for intrinsic continuum ($\rm{f_{intr}}$) and reflection component ($\rm{f_{refl}}$) using $\rm{cflux}$ in XSPEC in an energy band between 3-70 KeV to both components. Additionally, we calculated the reflection fraction relative to the total flux as $\rm{f_{refl}/(f_{refl}+f_{intr})}$.
Figure\,\ref{fig:histSelectXray} shows the histogram of the reflection fraction. We chose the 47 sources where the flux of the reflection component contributes at least $10\%$ to the total flux (i.e., $\rm{f_{refl}/(f_{refl}+f_{intr}) > 0.1}$, dashed vertical line in Figure\,\ref{fig:histSelectXray}). This ensured that the reflection component has a significant contribution to the X-ray spectrum to make the analysis of its shape.

The purpose of this paper is to study the X-ray reflection produced by distant and neutral material to compare it with the results obtained from the dust emission at mid-IR. However, the X-ray reflection can also be produced by ionized material in the disk. If the latest is the case, the comparison aim of this paper is not possible. Therefore, we tested the need for this disk reflection component by adding the \textit{relxill} component to the best fit obtained above. In this way, we identified 11 sources where disk reflection might be dominating the X-ray emission. Many of them are already reported by their disk reflection signatures and all of them  are type 1 as expected (see Appendix \ref{app:diskreflection} for more details). We discarded these 11 sources from the analysis, so the final sample contains 36 AGN.
Table \ref{tab:observations} shows the main observational details of the sample. Twenty-one objects are type 2 Seyfert (Sy2), and fifteen are type 1 Seyfert (Sy1). Our sample covers four orders of magnitude in X-ray luminosities ($\rm{Log(L_{2-10 keV}) \simeq 39.8-44.6}$).

\begin{table*}
\begin{small}
   \caption{\label{tab:AkaikeCriterio} Best model results according with Akaike criterion}
    \centering
    \begin{tabular}{lcccc|lcccc}
    \hline \hline
    \multicolumn{5}{c}{Seyfert 1} & \multicolumn{5}{c}{Seyfert 2} \\
    \cmidrule(lr){1-5} \cmidrule(r){6-10}
    Object  & \multicolumn{2}{c}{X-ray} & \multicolumn{2}{c}{mid-IR} & Object & \multicolumn{2}{c}{X-ray} & \multicolumn{2}{c}{mid-IR} \\
    \cmidrule(lr){2-3} \cmidrule(r){4-5} \cmidrule(lr){7-8} \cmidrule(r){9-10}
     &  $\epsilon$ & Model &  $\epsilon$ & Model &  &  $\epsilon$ & Model & $\epsilon$ & Model \\
     \hline
     Mrk1018 & 5.50E+00$^{*}$ &  S/C  & 6.45E+05 & Clumpy  &  UM146 & 6.71E-01 & S/C  &  8.90E+08 & Clumpy \\
    Mrk590 & 2.55E-01 & S/C  & 1.03E+27 & Clumpy  &   NGC788 & 2.88E+04 & Smooth  &  1.02E+45 & Clumpy \\
    PG0804+761 & 6.97E+02 & Smooth  & 8.53E-24 & Smooth  &   NGC1052 & 6.48E+00 & S/C  &  3.49E+42 & Clumpy \\
    RBS0770 & 1.73E+00 & S/C  & 1.24E-01 & S/C  & NGC1358 & 2.94E+00 & S/C  &  8.06E-01 & S/C \\
    I11119+3257 & 7.65E-04 & Clumpy  & 2.05E-38$^{*}$ &  Smooth  & J05081967+1721483 & 1.62E-03 & Clumpy  &  1.68E+17 & Clumpy \\
    PG1211+143 & 2.25E+07 & Smooth  & 3.31E-07 & Smooth  & Mrk3 & 1.54E+101 & Smooth  &  6.44E+42$^{*}$ &  Clumpy \\
    RBS1125 & 5.59E+01 & S/C  & 2.70E+29 & Clumpy  & ESO428-G014 & 4.90E+02$^{*}$ &  Smooth  &  1.26E+11 & Clumpy \\
    Mrk231 & 2.07E+00 & S/C  & 1.20E+03$^{*}$ &  Clumpy  & Mrk78 & 6.77E+01 & S/C  &  1.76E+22 & Clumpy \\
    Mrk1383 & 2.55E+05 & Smooth  & 4.93E+10 & Clumpy  & Mrk1210 & 1.26E-05 & Clumpy  &  1.35E+29$^{*}$ &  Clumpy \\
    Mrk1392 & 1.21E-04 & Clumpy  & 2.26E+26 & Clumpy  & J10594361+6504063 & 1.07E+02 & Smooth  &  3.43E+00 & S/C \\
    Mrk1393 & 4.81E+00 & S/C  & 4.84E+42 & Clumpy  & NGC4388 & 1.71E+21 & Smooth  &  2.92E+14$^{*}$ &  Clumpy \\
    PG1535+547 & 1.15E-01$^{*}$ &  S/C  & 1.06E-10 & Smooth  & NGC4507 & 1.51E+57 & Smooth  &  6.31E+36 & Clumpy \\
    ESO141-G055 & 2.13E+16 & Smooth  & 2.44E+00 & S/C  & NGC4939 & 3.06E+02 & Smooth  &  7.53E+29 & Clumpy \\
    NGC7213 & 1.68E+05 & Smooth  & 3.14E+177 & Clumpy  & ESO-097-G013 & 3.20E+129 & Smooth  &  4.99E+120$^{*}$ &  Clumpy \\
    MCG+01-57-016 & 1.57E-02 & S/C  & 3.61E+19 & Clumpy  & IC4518W & 1.02E-02 & S/C  &  2.15E+12 & Clumpy \\
               &          &      &          &         & ESO138-G1 & 7.01E-02 & S$^{\bullet}$/C  &  1.13E+24 & Clumpy \\
               &          &      &          &         & NGC6300 & 5.92E+01 & S$^{\bullet}$/C  &  1.13E+44$^{*}$ &  Clumpy \\
               &          &      &          &         & ESO103-G35 & 3.90E+09 & Smooth$^{\bullet}$  &  8.36E+73 & Clumpy \\
               &          &      &          &         & MCG+07-41-03 & 1.02E+36 & Smooth &  2.57E+93$^{*}$ &  Clumpy \\
               &          &      &          &         & IC5063 & 5.16E-07 & Clumpy  &  1.52E+34 & Clumpy \\
               &          &      &          &         & PKS2356-61 & 7.91E+03 & Smooth  &  4.52E+00 & S/C \\
    \hline 
\end{tabular}
\tablefoot{Evidence ratio for the Akaike method and resulting model best to each source. S/C is included when either Smooth or Clumpy models provide an equally good fits. Bullets indicate that the best fit is obtained using the baseline model with $\rm{N_{H_{tor}}}$ independent to the $\rm{N_{H_{los}}}$. Asterisks indicate poor fits with $\rm{\chi^2/d.o.f. > 1.2}$ (see text).}
\end{small}
\end{table*}

We fitted the 36 \emph{NuSTAR} spectra of our sample using the baseline model presented in equation \ref{Eq:BaselineXray}. For the reflection component, we chose the two models discussed in Section \ref{sec:ModelsDesc}: borus02 and UXClumpy, for the smooth and clumpy torus models, respectively. For each of them, we fitted the spectra three times testing different inclination angles (see previous paragraphs). The average difference on $\chi^2/d.o.f.$ obtained between angles is $\Delta \chi^2 (i_1 - i_2)/d.o.f.\sim 0.3$. Only five objects showed a significant improvement on the best fit ($\chi^2/d.o.f. < 1.2 $) by selecting a viewing angle $\rm{\theta_{inc} =90^\circ}$ for the smooth model and none when using the clumpy model. We explore how the other torus parameters are affected by the choice of fixing the viewing angle in Section \ref{sec:Disc}.

Initially, we assumed that the column density along the LOS toward the intrinsic continuum and that of the medium producing the reflection component are the same. Thus, we used as best fit the one producing the minimum $\chi^2/d.o.f.$ value for each source. We also fitted the data to two baseline models where the column density of the LOS and the reprocessed material have different values (i.e. are not linked; $\rm{N_{H_{los}} \neq N_{H_{tor}}}$). At first, we chose the values obtained from the previous analysis (i.e. with these two parameters linked; $\rm{N_{H_{los}} = N_{H_{tor}}}$). We discuss the best scenario based on f-test statistics in the next sections.

\begin{figure}
    \centering
    \includegraphics[width=1.0\columnwidth]{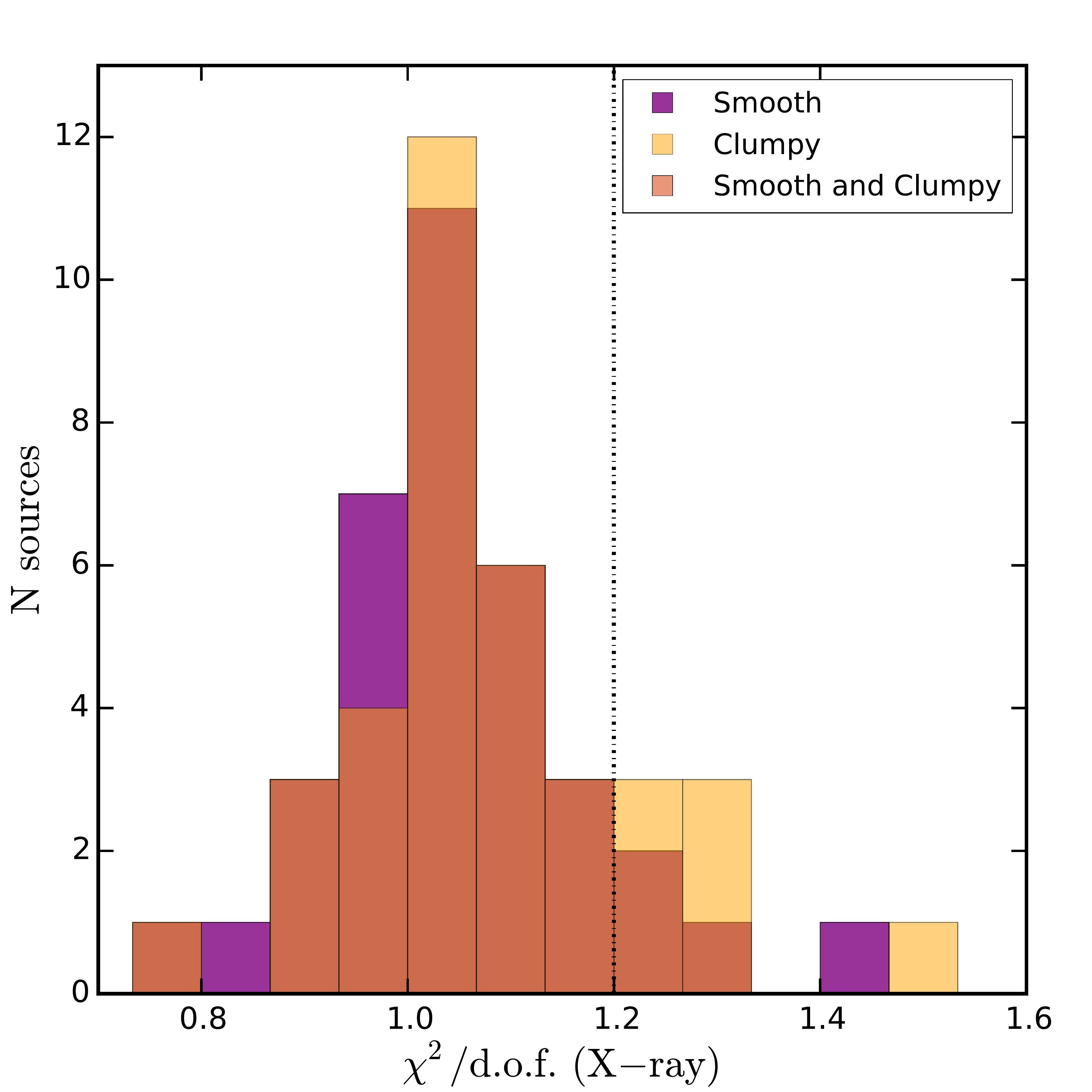}
   \caption{Distribution of the best-fit statistics for the smooth (purple), clumpy (yellow), and both (dark-salmon) baseline models in X-ray. The vertical dotted lines shows the threshold of $\rm{\chi^2/d.o.f.=1.2}$, where poorer fit are considered in this analysis (see text). }
    \label{fig:Hist_badChi}
\end{figure}

\begin{table}
    \caption{Mean and standard deviation values of X-ray parameters obtained for Sy1 and Sy2 when we used the smooth and clumpy baseline models to fit the \emph{NuSTAR} spectra.}
    \label{tab:MeanParam}
    \centering
    \begin{footnotesize}
    \begin{tabular}{cccccccc}
    \hline \hline
    Parameter & baseline & \multicolumn{3}{c}{Sy1}  & \multicolumn{3}{c}{Sy2} \\[3pt]
            &  model  & $\#$  & mean & std & $\#$ & mean & std \\[3pt]
    (1) & (2) & (3) & (4) & (5) & (6) & (7) & (8) \\[3pt]
    \hline
    $\rm{log(N_{H_{tor}})}$ & smooth &  14  &  23.9  &  0.6  &  21  &  23.8  &  0.6  \\[3pt]
        & clumpy &  15  &  23.4  &  0.6  & 21 &  23.4  &  0.7  \\[3pt] 
    $\rm{Cf}$ & smooth &  6  &  0.4  &  0.2  &  16  &  0.8  &  0.2  \\[3pt]
        & clumpy &  12  &  0.4 &  0.2  & 12 &  0.8  &  0.1  \\[3pt] 
    $\rm{\Gamma}$ & smooth &  15  &  1.9  &  0.2  &  18  &  1.8  &  0.3  \\[3pt]
        & clumpy &  15  &  2.1 &  0.2  & 21 &  1.8  &  0.2  \\[3pt]
    $\rm{\theta_{tor}}$ & smooth &  4  &  33.5  &  11.5  &  5  &  34.3  &  22.3  \\ 
        & clumpy &  8  &  10.6  &  10.3  & 11 &  16.0  &  11.1  \\ 
      {\tiny $\rm{log\left(\frac{f_{Refl}}{f_{Refl} + f_{Int}} \right)}$}  & smooth &  15  &  -0.8  &  0.2  &  21  &  -0.7  &  0.3  \\ 
        & clumpy &  15  &  -0.8  &  0.3  & 21 &  -0.6  &  0.3  \\
    \hline                                   
\end{tabular}
\end{footnotesize}
\tablefoot{Column(1): Parameter name. Column(2): X-ray baseline model. Column(3): number of sources considered to calculate the mean and standard values. We only considered AGNs for which the parameters are well constrained. Columns (4) and (7): Mean values to Sy1 and Sy2, respectively. Columns (7) and (8): standard deviation values to Sy1 and Sy2, respectively.}
\end{table}

\section{X-ray results}
\label{sec:Xrayresults}
\subsection{Smooth versus Clumpy baseline models}\label{sec:Xrayresults1}

In order to understand if there is a preference towards a smooth or a clumpy baseline model, we statistically compare the spectral fit obtained. Figure\,\ref{fig:Hist_badChi} shows the $\rm{\chi^2/d.o.f}$ distribution obtained for both the smooth and clumpy torus models. In general both models produce good results for above 86\% of the sources. We test if the iron lines at 6.70\,keV and/or 6.97\,keV could improve the spectral fitting of sources of our sample. Particularly, the use of the iron line at 6.70\,keV, as an additional component to the total model, improves the fit of three sources: NGC\,7213, ESO-097-G013, and MCG+07-41-03. Three AGN (Mrk\,1018, PG\,1535+547, and ESO\,428-G014; two Sy1s and one Sy2s) show a significantly lower quality spectral fit with $\rm{\chi^2/d.o.f > 1.2}$ irrespective of the model used. Additionally, two sources (Mrk3 and ESO\,097-G013; both Sy2s) have a $\rm{\chi^2/d.o.f > 1.2}$ when using the clumpy torus model (see bad spectral fits in Appendix \ref{sec:app2}).

We estimate the ``evidence ratio'' using the Akaike information criterion (AIC) for both models. This evidence ratio allows us to compare if one model is better than another one using $\epsilon = \rm{ W[AIC]_{clumpy}}$/ $\rm{W[AIC]_{smooth}}$, where $\rm{W[AIC]_{clumpy}}$ and $\rm{W[AIC]_{smooth}}$ are the ``Akaike weight'' \citep[see Eqs.\,5-7 by][]{Emmanoulopoulos16}. The clumpy baseline model is 100 times more likely than the smooth torus model when $\rm{\epsilon \leq 0.01}$. The smooth baseline model is 100 times more likely than the clumpy torus model when $\rm{\epsilon \geq 100}$.
The results are shown in Table\,\ref{tab:AkaikeCriterio}. For 16 sources ($\sim 44\%$ of the sample), the smooth baseline model is preferred while the clumpy baseline model is preferred for five sources ($\sim 14\%$ of the sample). Both models fit similarly well the data in $\sim 42\%$ of the sample (15 objects). There is no difference between Sy1 and Sy2; the smooth and clumpy torus models produce a better fit for 5 (11) and 2 (3) Sy1s (Sy2s), respectively. Similarly, we find no differences in the best model according to the X-ray luminosity. 

The fit does not significantly improve when the LOS column density is allowed to vary with respect to the column density of the reflecting material (i.e. $\rm{N_{H_{tor}} \neq N_{H_{los}}}$). Only the smooth baseline model allows us to test this option. We find that three sources are better fitted letting free the column densities independently. We identify these sources with a white dot next to the model name in Table\,\ref{Tab:Bestfit_XrayValues}. Indeed, for these three sources the smooth torus model with $\rm{N_{H_{tor}} \neq N_{H_{los}}}$ produces significantly better fits than those obtained with the clumpy. 

\subsection{Smooth versus Clumpy model parameters}

Table\,\ref{Tab:Bestfit_XrayValues} reports the results for the parameters obtained for the smooth and clumpy baseline models at X-rays.
These models allow us to infer two parameters (column density of the reflecting material, $\rm{N_{H_{tor}}}$, and half opening angle of the torus, $\rm{\theta_{tor}}$) that are directly related to the reflection component and, therefore, can give us information on the reflecting material. Additionally, we obtain the reflection fraction at the 3-70\,keV band, computed as the ratio between the flux of the reflection component ($\rm{f_{Refl}}$) over the sum of the fluxes of the reflection component and the intrinsic emission component ($\rm{f_{Int}}$), i.e. $\rm{f_{Refl}}$ / $\rm{(f_{Refl} + f_{Int})}$. The errors of this relationship are computed through the method of propagation of uncertainties. 
We also compute the covering factor of the patchy obscurer, $\rm{Cf}$, and the photon index, $\Gamma$, of the intrinsic continuum. Table\,\ref{tab:MeanParam} reports the mean and standard deviation for these parameters dividing the sample into Sy1 and Sy2 when we use the smooth and clumpy models.

\begin{figure*}[!t]
    \centering
    \includegraphics[width=1.0\columnwidth]{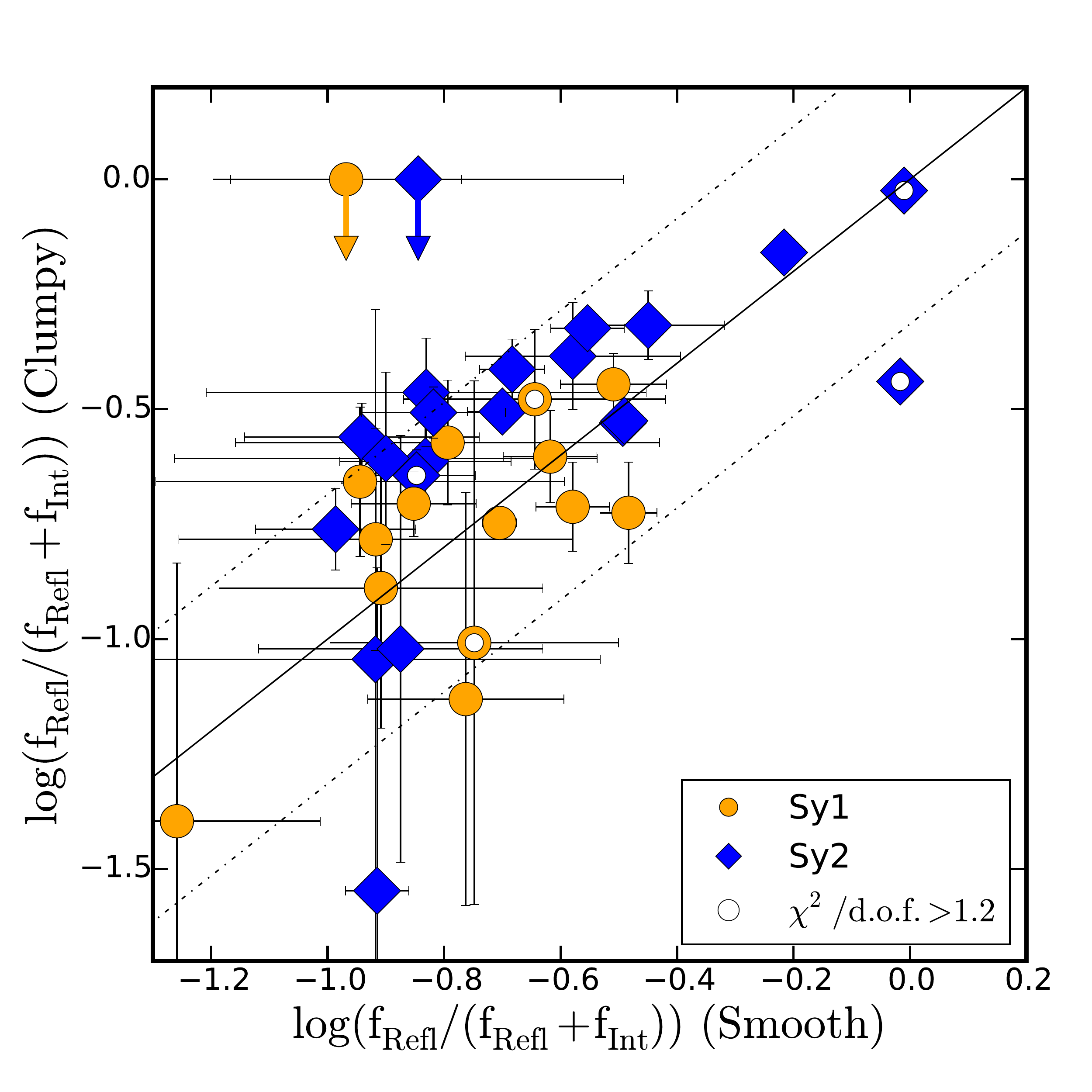}
    \includegraphics[width=1.0\columnwidth]{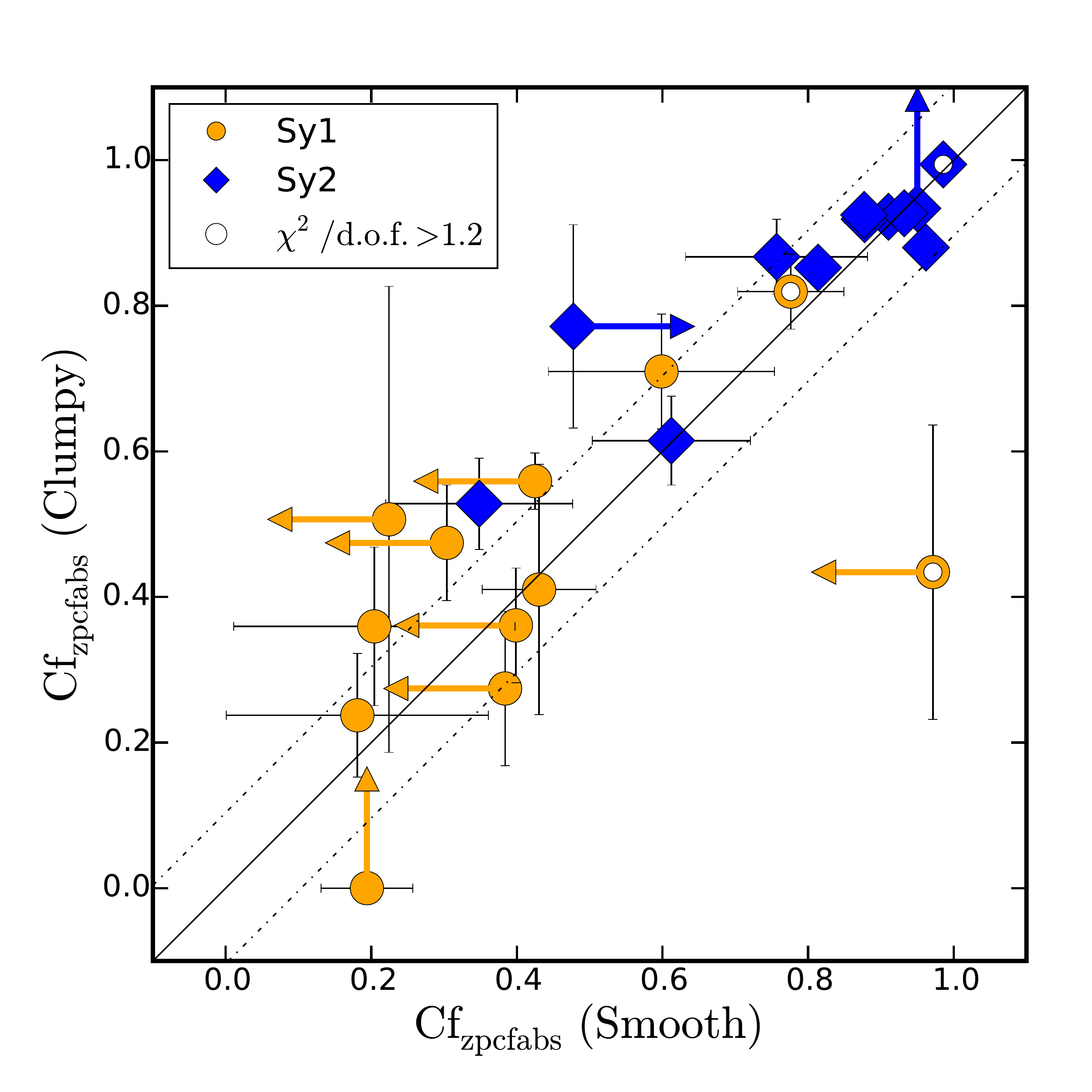}
    \includegraphics[width=1.0\columnwidth]{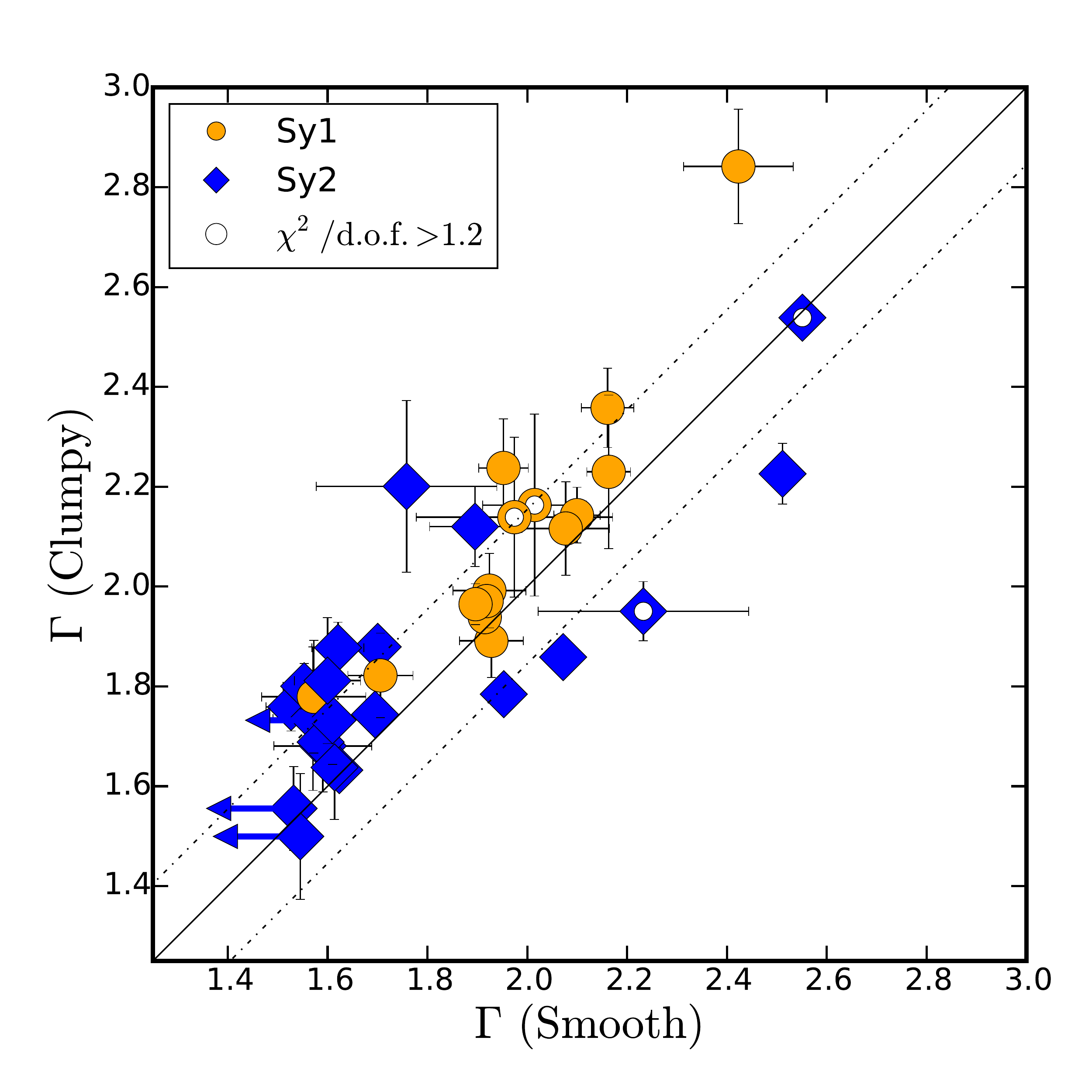}
    \includegraphics[width=1.0\columnwidth]{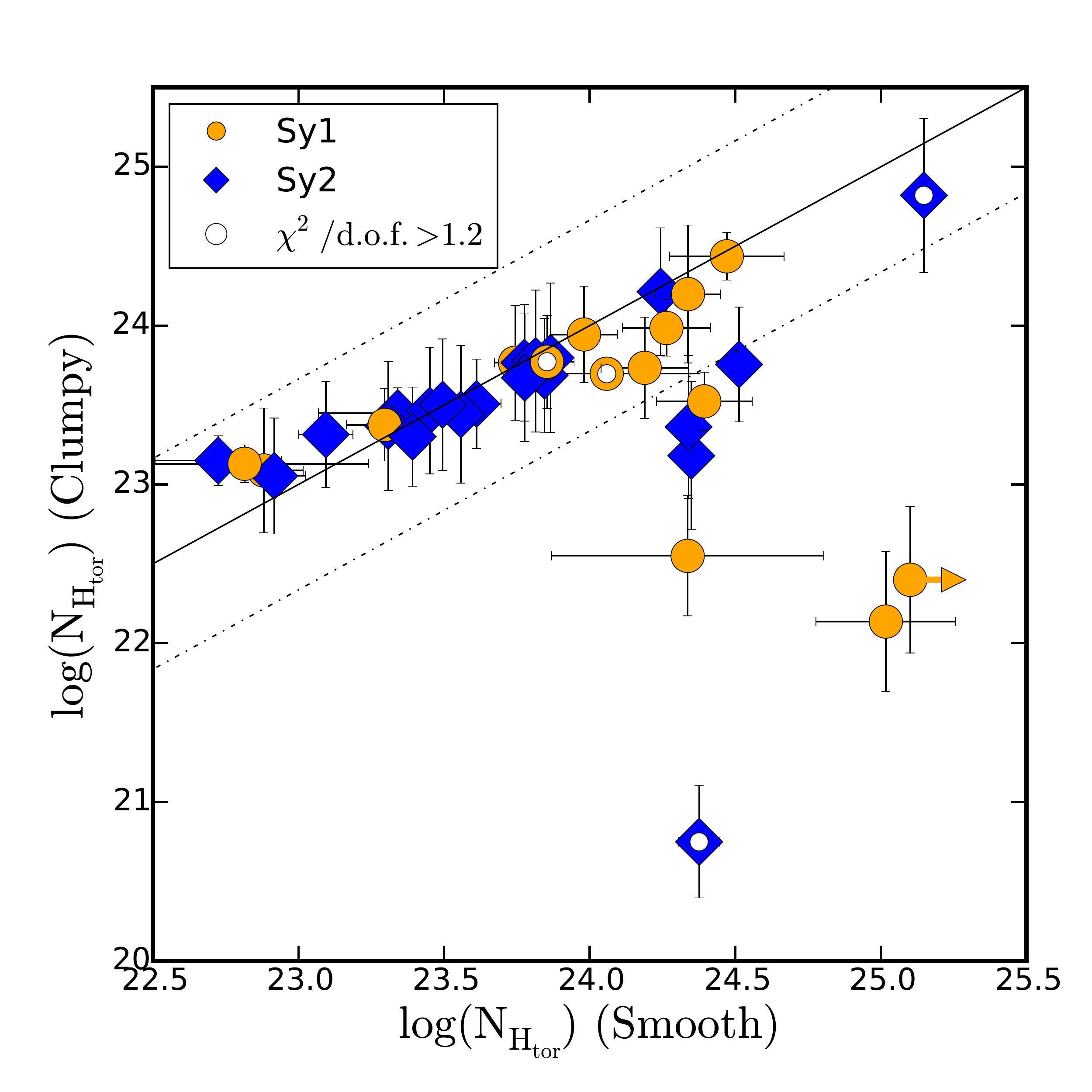}
\caption{(top-left): Fraction of the flux accounted for the reflection component relative to the total flux for the  clumpy versus smooth baseline models. (top-right): Covering factors (i.e. ratio between the scattered and intrinsic emission, $\rm{Cf}$) from the clumpy versus smooth baseline models. (bottom-left): Photon index ($\Gamma$) using the clumpy versus that using the smooth models. (bottom-right): Column densities ($\rm{N_{H}}$) using the clumpy versus that using the smooth models. The Sy1 and Sy2 are shown as orange dots and blue diamonds, respectively. Arrows indicate lower and upper limits. The solid and dotted lines show the one-to-one relationship and 2$\rm{\sigma}$ dispersion according to the error bars of the data, respectively. The error bars were calculated through the method of propagation of uncertainties (see text).}
    \label{fig:X-rayPar}
\end{figure*}

Figure\,\ref{fig:X-rayPar} show the comparison between the parameters obtained when using the smooth and the clumpy models at X-rays. In these figures, we include only source parameters restricted at least for one of the two models. Unfortunately, the half opening angle of the torus, $\rm{\theta_{tor}}$, is restricted in very few objects preventing a direct comparison between models (see Table\,\ref{tab:MeanParam}).

As explained in Section\,\ref{Sec:sample}, we use the fraction of the reflection component to select sources where it is significant (at least 10\% compared to the total flux of the source). Figure\,\ref{fig:X-rayPar} (top-left) allows us to investigate if the selection criterion depends on the baseline model. Only for ESO\,097-G013 we find that the reflection fraction computed with the smooth model is larger than that obtained for the clumpy model. However, the $\rm{\chi^2/d.o.f.}$ obtained when using the clumpy model for this source is above $\rm{\chi^2/d.o.f.> 1.2}$. This result confirms that, regardless of the baseline model used, the chosen criterion gives us the same result.

The covering factor, $\rm{Cf}$, measures the scattered flux compared to the intrinsic flux through a patchy absorber. Note that the $\rm{Cf}$ values come from $zpcfabs$ component. Figure\,\ref{fig:X-rayPar} (top-right) shows the comparison of the $\rm{Cf}$ obtained through the smooth and clumpy models. Sy2 galaxies have larger covering factors than Sy1 (see also Table\,\ref{tab:MeanParam}). This result holds irrespective of the model used. Indeed, \citet{Marchesi19} found that the $\rm{Cf}$ at X-rays is recovered independently of the model used, by comparing two smooth torus models (MyTorus and borus02). However, see Section\,\ref{sec:XrayvsmidIR} for a comparison between the $\rm{Cf}$ obtained through X-ray and mid-IR observations.

Figure\,\ref{fig:X-rayPar} (bottom-left) compares the photon indices of the intrinsic continuum, $\Gamma$, for both baseline models. Roughly $83 \%$ of the sample is consistent with the one-to-one relation (six sources are outside, i.e. above 2-$\sigma$). Note that most of the sources show systematically larger photon indices when using the clumpy baseline model compared to the smooth baseline model. This is also visible in the mean values reported in Table\,\ref{tab:MeanParam}.

Figure\,\ref{fig:X-rayPar} (bottom-right) shows the comparison of the column densities of the reflecting material obtained through the smooth and clumpy models. We find that $\sim 83\%$ of the sources follow a linear relationship, indicating that for a large fraction of the sources the column density is independent on the model used. Six sources are significantly outside the one-to-one relation (i.e. above 2$\rm{\sigma}$). Four and two of these sources are Sy1 (RBS0770, Mrk1392, MCG\,+01-57-016, and ESO\,141-G055) and Sy2 (ESO\,428-G014 and NGC\,6300), respectively. We find that the mean column density of the torus, $\rm{<N_{H_{tor}}>}$, is larger for Sy1 than for Sy2 using the smooth model while it is similar when using the clumpy model (see also Table \ref{tab:MeanParam}). We ruled out that the selection of the inclination angle affect these results (see Appendix\,\ref{sec:app3}). \citet{Furui16} compared the SED resulting from smooth and clumpy gas distribution in the torus models using the MONACO radiative transfer code. They found large differences between the SEDs produced by both models for large absorbing column densities ($\rm{N_{H} > 10^{24}cm^{-2}}$). These differences are not found when two smooth gas torus models \citep[MyTorus and borus02 models by][respectively]{Murphy09,Balokovic18} are compared \citep[][]{Marchesi19}. Our result is in good agreement with \citet{Furui16}, since both models show discrepant $\rm{N_H}$ values at the Compton-thick regime. 
In practice, while the smooth gas torus model infers $\rm{N_{H} > 10^{24}cm^{-2}}$, the column density obtained with the clumpy gas torus model tends to be only mild ($\rm{N_{H} = 10^{22}-10^{23} \, cm^{-2}}$).

We can use additional observational evidence to back up the smooth or clumpy distribution of the X-ray emitting material inferred from the modelling. Variability in the column density along the LOS has been reported for some objects \citep[e.g.][]{Risaliti02, Markowitz14,Laha20}.
In Table\,\ref{tab:observations}, we indicate whether absorption variability (or the lack of it) has been reported for the sources in our sample. We found $\rm{N_H}$ variability information for 23 objects (see Col. 10 in Table \ref{tab:observations}). We consider a source as $\rm{N_H}$ non-variable/variable if multiple observations of the object have been published that analyse the long term variability finding the lack/presence of changes on the $\rm{N_H}$ values. The sources (15 objects) without variable absorption are well fitted with a smooth gas torus model at X-rays. On the other hand, five out of the eight sources with absorption variability are best fitted with the clumpy gas torus model at X-ray (exceptions are NGC\,4507, NGC\,4939, and MCG+07-41-03). Thus, the detection or non-detection of variability in the column density along the LOS correlates quite well with the choice of a clumpy or smooth gas distribution at X-rays.

Therefore, we considered taking the non-variability and variability of LOS absorption as an indication for the smooth and clumpy distribution of the gas, respectively. Thus, this implies to considered different baseline models for three sources, irrespective of the model obtained through statistical analysis described in section \ref{sec:Xrayresults1}. We selected the clumpy distribution of the gas for NGC\,4507, NGC\,4939, and MCG+07-41-03, although they are best fitted with a smooth distribution (see section \ref{sec:Xrayresults1}). Furthermore, we selected the clumpy distribution at X-rays for Mrk\,1392 and NGC\,1052 and the smooth distribution for Mrk\,1018, Mrk\,590, Mrk\,231, PG\,1535+547, ESO\,138-G1, and NGC\,6300, although both models produce statistically similar results. This is highlighted in Table\,\ref{tab:midIRvsXrays_param} with an asterisk next to the preferred model at X-rays.


We constrained the angular width of the torus, $\rm{\theta_{tor}}$, for both models only in six sources ($\rm{\sim 17 \%}$). The clumpy model is able to restrict this parameter for a larger number of sources (19 sources, i.e. $\rm{\sim 53 \%}$) than the smooth model (8 sources, i.e. $\rm{\sim 22\%}$). This small number of constrained values prevent us from comparing this parameter. Although the X-ray viewing angle needs to be fixed and there are no statistically significant differences between the edge-on, face-on, or intermediate views \citep[in fact this parameter is fixed in most of the previous works done at X-rays, e.g.][]{Balokovic18}, we found that the optical classification of the sources into Sy1 and Sy2 is consistent with the inclination angles (with the lowest $\rm{\chi^2/dof}$) obtained for the $\rm{\sim 78\%}$ and $\rm{57 \%}$ of the sample for the smooth and clumpy baseline models, respectively (i.e. Sy1s have $\rm{\theta_{inc} \leq 45^{\circ}}$ ($\rm{\theta_{inc} \geq 45^{\circ}}$) and Sy2s have $\rm{\theta_{inc} \geq 45^{\circ}}$ ($\rm{\theta_{inc} \leq 45^{\circ}}$) with smooth (clumpy) baseline model).

\cite{Tanimoto18} found a correlation between the half-opening and inclination angles using the reflection model by \citet{Ikeda09}. According to \cite{Tanimoto18}, this means that along the LOS we are intercepting the outer edge of the torus, which seems unrealistic if claimed for all the objects. We compared the inclination and half-opening angles obtained from a smooth baseline model for 13 sources where the half-opening angle is completely or partially constrained. The difference between these two angles are large except for four objects ($\rm{5^{\circ}-6^{\circ}}$ for ESO-097-G013, Mrk1386, ESO138-G1, and ESO428-G014). We did not find this correlation for the clumpy models either. Interestingly, the viewing and half opening angles are consistent with the classification of each source under the unification scheme, except for NGC\,1358, when using the smooth model. The vast majority of the objects do not show consistent viewing and half opening angle values when using the clumpy model.


\begin{figure}
    \centering
    \includegraphics[width=0.98\columnwidth]{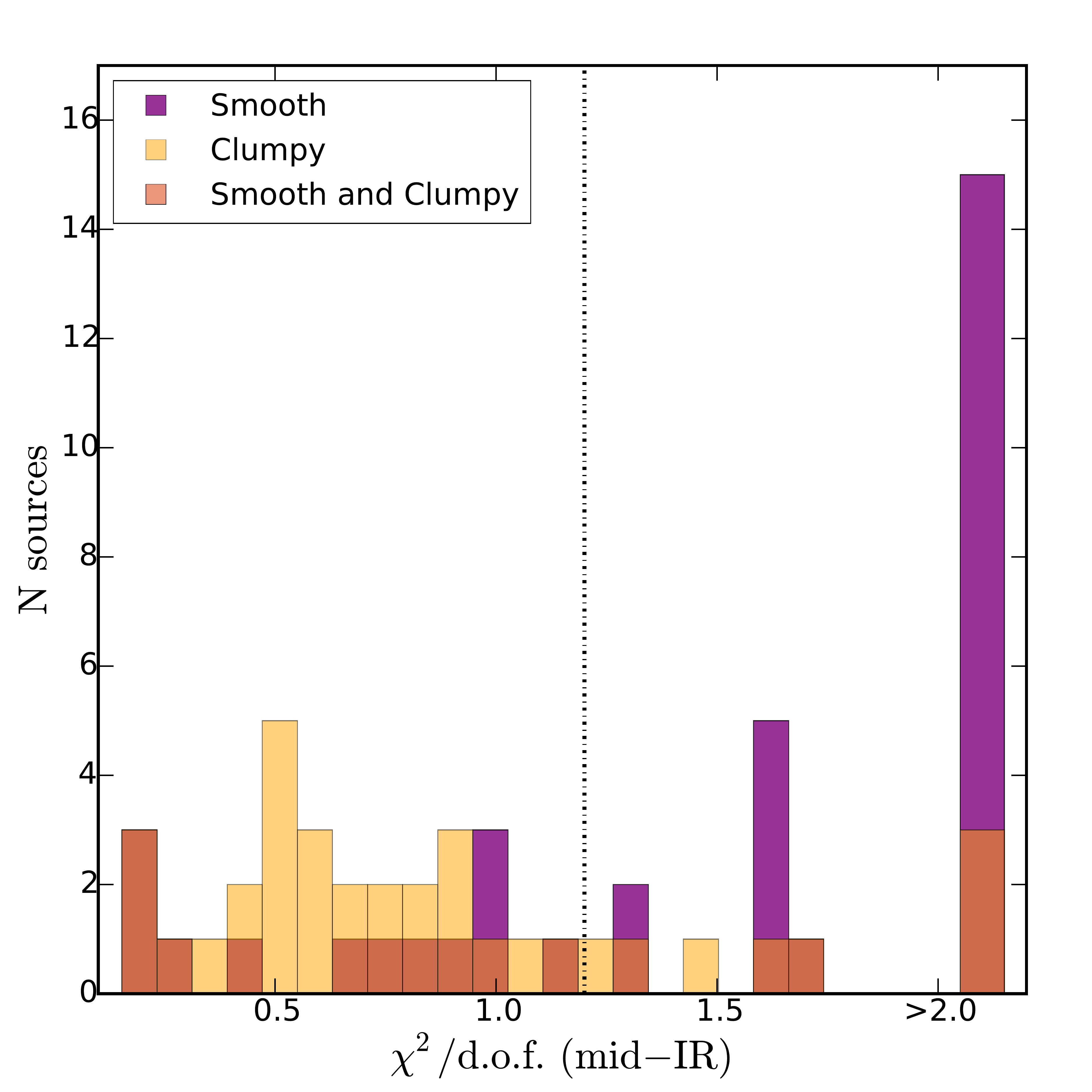}
    \caption{Distribution of the best-fit statistics for the smooth (purple), clumpy (yellow) and both (dark-salmon) baseline models in mid-IR. The vertical dotted lines shows the threshold of $\rm{\chi^2/d.o.f.=1.2}$, where poorer fit are considered in this analysis (see text). }
    \label{fig:Hist_badChi_midIR}
\end{figure}

\section{Mid-IR results}
\label{sec:midIRresults}

Table\,\ref{tab:midIRParam} shows the $\rm{\chi^2/d.o.f.}$, contribution of the three components used at mid-IR (i.e. AGN dust, stellar, and ISM), and parameter values that we obtain for the smooth and clumpy models. Figure\,\ref{fig:Hist_badChi_midIR} shows the distribution of $\rm{\chi^2/d.o.f.}$ The smooth and clumpy models produce good spectral fittings (i.e. $\rm{\chi^2/d.o.f. < 1.2}$) for 33\% (7 Sy1s and 5 Sy2s) and 72\% (11 Sy1s and 15 Sy2s) of the sample, respectively. The smooth and clumpy models show bad spectral fit with $\rm{\chi^2/d.o.f. > 1.2}$ for 67\% sources (8 Sy1s and 16 Sy2s) and 28\% sources (4 Sy1s and 6 Sy2s), respectively. Objects showing bad fits using the clumpy model also show bad fits with the smooth model (exceptions are PG\,0804+761 and PG\,1211+143).

Table\,\ref{tab:AkaikeCriterio} shows the ``evidence ratio'' and preferred model when using the smooth and clumpy mid-IR models. For 20 sources with good spectral fit ($\rm{\sim 55\%}$), we find that the clumpy model is preferred. Meanwhile, the smooth model is preferred only for three sources ($\rm{\sim 8 \%}$). For the remaining five sources both models produce statistically similar fits to the data. 
There are not significant differences between Sy1 and Sy2. Similarly, no differences for the best fits are found for different ranges of the X-ray luminosity.

\begin{table}
    \caption{Mean and standard deviation values of the mid-IR parameters obtained for Sy1 and Sy2 when using the smooth and clumpy baseline models.}
    \label{tab:SummaryParam_midIR}
    \centering
    \begin{footnotesize}
    \begin{tabular}{cccccccc}
    \hline \hline
    Parameter & baseline & \multicolumn{3}{c}{Sy1}  & \multicolumn{3}{c}{Sy2} \\
          &  model  & $\#$  & mean & std & $\#$ & mean & std  \\ 
(1) & (2) & (3) & (4) & (5) & (6) & (7) & (8)  \\
    \hline
    i  &  smooth  &  3  &  23.4  &  18.3  &  3  &  36.4  &  27.7   \\[3pt]
       &  clumpy  &  2  &  81.6  &  2.3  &  2  &  42.8  &  7.0  \\[3pt]
    $\sigma$  &  smooth  &  0  &  --  &  --  &  1  &  --  &  -- \\[3pt]
              &  clumpy  &  2  &  48.5  &  5.9  &  3  &  42.2  &  7.9 \\[3pt]
    Y  &  smooth  &  3  &  11.7  &  0.5  &  5  &  33.5  &  15.8 \\[3pt]
       &  clumpy  &  9  &  14.5  &  4.3  &  9  &  17.8  &  8.33 \\[3pt]
    $\tau_{9.7 \mu m}$  & smooth  &  5  &  4.9  &  2.3  &  1  &  --  & -- \\[3pt]
    $\tau_{\nu}$  &  clumpy  &  10  &  65.8  &  55.3  &  12  &  55.6  &  48.6 \\[3pt]
    \hline
    $\beta$  &  smooth  &  4  &  -0.51  &  0.19  &  1  & --  & -- \\[3pt]
    $\gamma$ &  smooth  &  3  &  4.5  &  0.85  &  2  &  2.84  &  2.81 \\[3pt]
    $N_0$  &  clumpy  &  8  &  4.35  &  1.45  &  10  &  6.19  &  2.34 \\[3pt]
    q  &  clumpy  &  5  &  1.41  &  0.48  &  9  &  1.37  &  0.47 \\
    \hline                                   
\end{tabular}
\end{footnotesize}
\tablefoot{Column(1): Parameter name. Column(2): mid-IR baseline model. Column(3): number of sources considered to calculate the mean and standard values. We only considered AGNs for which the parameters are well constrained. Columns (4) and (7): Mean values to Sy1 and Sy2, respectively. Columns (7) and (8): standard deviation values to Sy1 and Sy2, respectively.}
\end{table}

\subsection{Smooth versus Clumpy model parameters}

Table\,\ref{tab:midIRParam} reports the best-fitting parameters obtained when using the mid-IR models. We explore the possibility to create plots that compare parameters between models (as reported for the X-ray spectral fitting, see Section\,\ref{sec:Xrayresults}) but, unfortunately, many of them are not well restricted. Table\,\ref{tab:SummaryParam_midIR} shows a summary of the number of sources with good fit (12 and 26 sources using smooth and clumpy models, respectively) where we can completely (Cols.\,3 and 6) restrict each parameter. 

We also show in Table\,\ref{tab:SummaryParam_midIR} the mean and standard deviation per parameter, considering only restricted parameters (i.e. with an error bar well within the parameter space). We find that the parameters better restricted in both models are the ratio between the outer and the inner radius of the torus, $Y$, and the optical depth along the equator of the torus, ($\rm{\tau_{\nu}}$ or $\rm{\tau_{9.7\mu m}}$, depending on the model used). The mean $Y$ value for Sy2 is larger than that for Sy1 irrespective of the model used. Moreover, the mean $\rm{\tau_{\nu}}$ is similar for both Sy1 and Sy2. Meanwhile, the viewing angle ($i$) and the half angular width of the torus ($\rm{\sigma}$) are restricted only for a few objects. There are no differences between the number of parameters that are restricted in Sy1 and Sy2 for any of the two models. We find that the number of clouds in the equator of the torus ($\rm{N_0}$) and the radial distribution of the clouds ($q$) are restricted in 18 and 14 sources, respectively, for the clumpy model. The mean $\rm{N_0}$ for Sy2 is larger than that for Sy1 and the mean $q$ is similar for both Sy1 and Sy2. These results are in good agreement with those obtained in previous works \citep[e.g.][and references therein]{Garcia-Bernete19}.
The parameters associated to the vertical and radial distribution of dust, $\rm{\beta}$ and $\gamma$, are restricted only for a few objects in the smooth model. In a previous work \citep[][]{Esparza-Arredondo19}, we showed that the torus parameters are better restricted through mid-IR and X-ray simultaneous fits. This will be the subject of a forthcoming investigation (Esparza-Arredondo et al. in prep.).

\begin{figure}
    \centering
    \includegraphics[width=1.0\columnwidth]{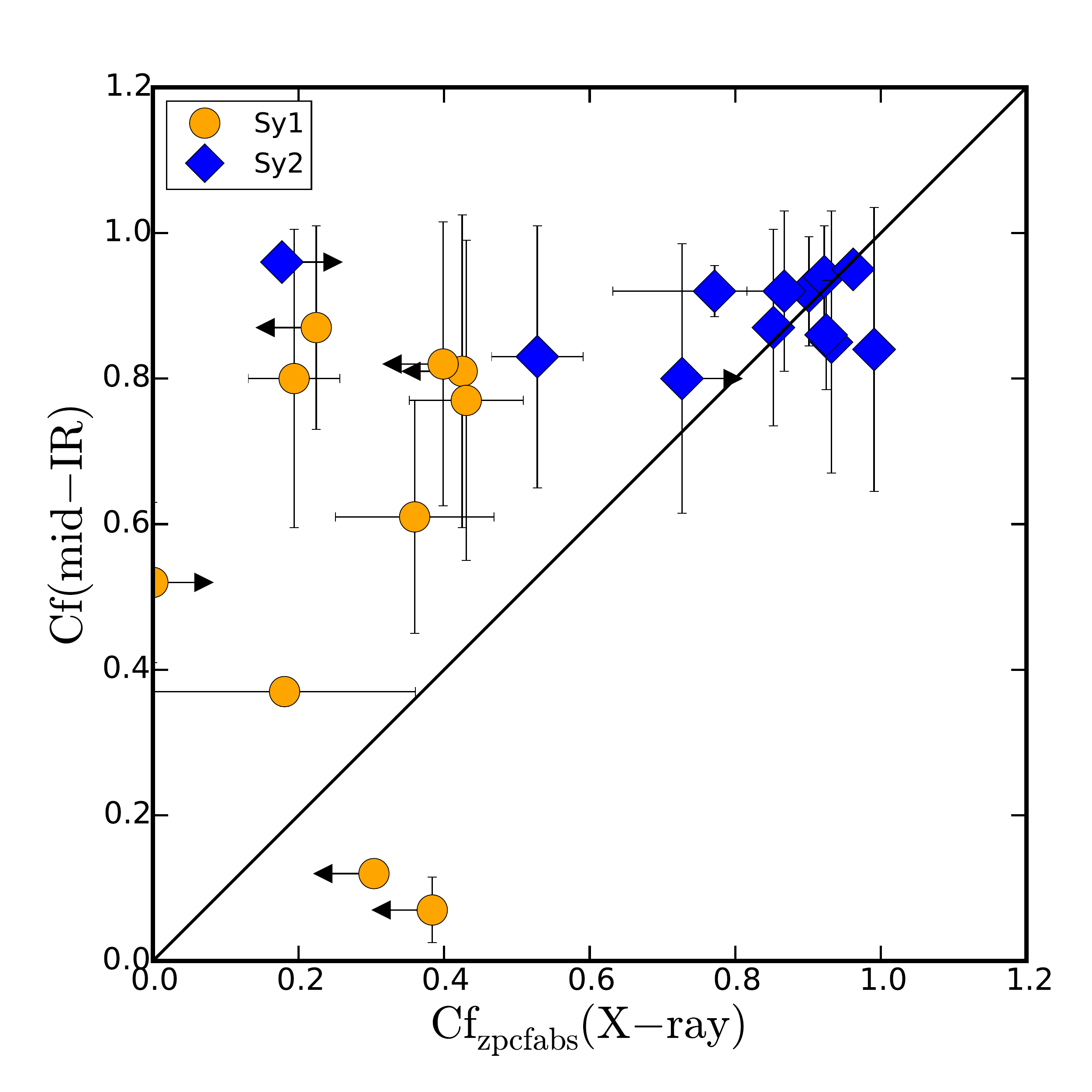}
    \caption{Relationship between covering factors. The Sy1 and Sy2 are shown as dots and diamonds, respectively. The solid line shows the one-to-one relationship. Note that the X-ray covering factor is obtained throughout the partial covering.}
    \label{fig:Cfs_XrayvsIR}
\end{figure}

\section{X-ray versus mid-IR results} \label{sec:XrayvsmidIR}

In this section, we present the results obtained from a direct comparison of the properties derived from mid-IR and X-ray parameters. First, we analyze whether the covering factor derived from X-rays through $zpcfabs$ component can be compared with the one derived from the mid-IR data. Then, we explore the dust-to-gas ratio of the AGN torus through column density and optical depth.

\subsection{Covering factor}

The covering factor of the AGN is one of the key parameters to describe the AGN population. It might be different among AGN types \citep{Ramos-Almeida09,Mateos16,Garcia-Bernete19} and it might also evolve along the AGN duty cycle of activity \citep{Lawrence91,Krongold03,Ricci17}. This idea is supported by the decrease of the mid-IR to optical luminosity ratio as the AGN bolometric luminosity increases \citep{Maiolino07}. While  \citet{Ramos-Almeida14} suggest that the covering factor is a robust estimate using mid-IR SED fitting \citep[although also model dependent, see][]{Gonzalez-Martin19b}, this might not be the case for the covering factors inferred from the partial covering applied to the intrinsic continuum at X-rays. To perform this comparison, we obtained the mid-IR $\rm{Cf}$ using the posterior distributions of the parameters. We computed the Cf for the clumpy and smooth models at mid-IR wavelengths following these equations:
\begin{eqnarray}
    \textit{\rm{Cf}} \, (Smooth) = \frac{ln(\tau_{9.7})}{\gamma} \\
    \textit{\rm{Cf}} \, (Clumpy) = 1 - \int_0^{\pi/2} cos \left( \beta e^{N_{0} e^{-\beta^2/\sigma^2}} \right) d\beta
    \label{eq:Cf}
\end{eqnarray}

Fig.\,\ref{fig:Cfs_XrayvsIR} shows the mid-IR versus X-ray $\rm{Cf}$ for all the objects with good spectral fits at both wavelengths. Note that we use for each source the best-fit model (either smooth or clumpy), taking into account if the source show absorption variations (see Section\,\ref{sec:Xrayresults}). As it can be seen, no correlation is found between mid-IR and X-ray $\rm{Cf}$. Most Sy2 show $\rm{Cf}$ $\rm{> 0.7}$ at X-ray and are also consistent with large $\rm{Cf}$ at mid-IR wavelengths. Meanwhile, Sy1 with $\rm{Cf}$ $\rm{< 0.7}$ at X-ray have larger $\rm{Cf}$ at mid-IR. If the covering factors at X-ray are dominated by the material at the host galaxy, then, we expect similar behaviors from both AGN types. However, we did find that type-2 AGN tend to have larger covering factors (irrespective of the baseline model used) than type-1 AGN. This discrepancy in the $\rm{Cf}$ of type-1 AGN could be explained if the soft emission is coming from nuclear X-ray radiation leaking through patchy obscuring gas in the LOS which may not be associated with torus-scale gas. Alternatively, this emission might be due to the scattering of nuclear radiation in distant gas. Indeed, the soft X-ray emission of Sy1 is complex and additional components are usually required to fit them. For instance, Mrk\,1018 is modelled by \citet{Noda18} including a soft Comptonization component to reproduce the soft emission. This soft Comptonization component is suggested to be originated in the accretion disk. Thus, the $\rm{Cf}$ estimated at X-rays might be wrongly estimated due to the complexity of the soft X-ray emission of Sy1. Spectra below 3\,keV (perhaps with \emph{XMM}-Newton) are required to properly account for the complexity of the soft emission, obtaining a better determination of the X-ray $\rm{Cf}$ for Sy1. Therefore, we opted to focus the subsequent analysis and discussion on the $\rm{Cf}$ obtained at mid-IR.



\begin{figure}
    \centering
    \includegraphics[width=1.0\columnwidth]{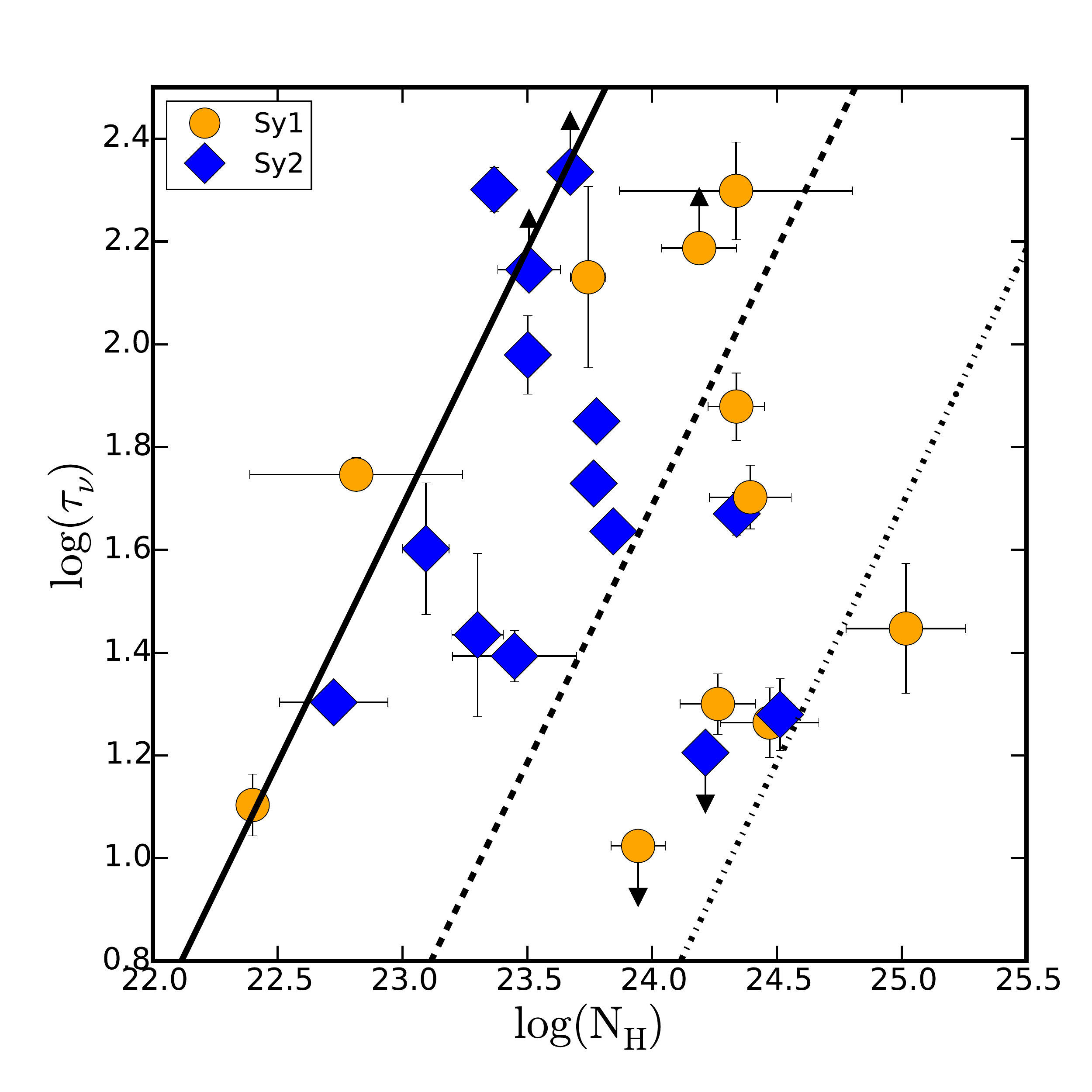}
    \caption{Relationship between column density and optical depth. The Sy1 and Sy2 are shown as dots and diamonds, respectively. The solid, dashed, and dashed-dotted lines show 1, 0.1, and 0.01 times the galactic dust-to-gas ratio, respectively.}
    \label{fig:NHvstau}
\end{figure}

\subsection{Torus dust-to-gas ratio}

The dust-to-gas ratio is found to be constant in our own Galaxy with a ratio between gas column density and optical extinction of $\rm{A_{V}/N_{H}\sim2\times10^{-21}cm^{-2}}$ \citep{Draine03}. However, this might not be the case for other galaxies \citep{Draine09}. Furthermore, the AGN might provide a source of production and/or destruction of dust \citep{Sturm05,Srinivasan17} that might yield to significantly different dust-to-gas ratio near the AGN.

Figure\,\ref{fig:NHvstau} shows the column density versus the optical depth using the preferred combination of models at X-ray and mid-IR wavelengths. We include sources with a good spectral fitting ($\rm{\chi^2/d.o.f. < 1.2}$). The solid, dashed, and dashed-dotted lines show 1, 0.1, and 0.01 times the Galactic dust-to-gas ratio\footnote{$\rm{N_{H} = 1.9 \times 10^{21} * 1.086* \tau_{\nu}}$ \citep{Bohlin78}.}, respectively.

The dust-to-gas ratio in AGN is in the range between $\rm{\sim}$0.01 and slightly above 1 times the Galactic dust-to-gas ratio. Moreover, Sy2 seems to have larger dust-to-gas ratio ($\rm{>0.1}$) than Sy1. \citet{Maiolino01} showed consistent results with a ratio in the range $\rm{0.3-0.01}$ using optical dust measurements and X-ray LOS column densities. \citet{Burtscher16} explored this relation using the absorption along the LOS measured by the BAT AGN spectroscopic survey \citep[BASS][]{Ricci17} and the optical extinction\footnote{Optical extinctions are derived from near-IR color temperatures (using spectral fitting) against the offset from the near- to mid-IR luminosity \citep{Burtscher15}.}. They found a ratio consistent with the Galactic value or below in AGN, finding a minimum ratio of $\rm{\sim}$0.02.

This range for the dust-to-gas ratio is larger than that found in nearby galaxies \citep[][]{Draine09}. There are several plausible explanations. One possibility discussed in the literature \citep{Maiolino01,Burtscher16} is that absorbing column density along the LOS is variable. However, we report here measurements of the density for the reflection component, which is constant over the years. Furthermore, many of our sources show a lack of $\rm{N_H}$ variability (see Table\,\ref{tab:observations}). Gas at the dust-free inner region or at both the dust-free and dust torus regions are also possibilities to explain a smooth distribution at X-rays and a clumpy distribution at mid-IR wavelengths. This scenarios are further discussed below.

\section{Discussion} \label{sec:Disc}

Placing constraints on the geometry of the torus parameters from SED spectral fitting is possible from near-IR, mid-IR, and sub-millimeter observations \citep{Gallimore16, Garcia-Burillo16, RamosAlmeida17, Lopez-Rodriguez18}. However, near-IR and sub-millimeter observations might be contaminated by stellar \citep{Ramos-Almeida14}, diffuse dust emission \citep{Mason06}, and synchrotron emission \citep{Pasetto19}, respectively.

Joining mid-IR and X-ray spectra of AGN might yield to a better understanding of the obscuring structure \citep[e.g.][]{Liu14,Tanimoto19, Esparza-Arredondo19}. Some parameters (e.g. the outer radius of the torus) can be recovered using only mid-IR observations but some others (e.g. viewing angle) are difficult to restrict \citep[see][]{Ramos-Almeida14}. At X-rays the half opening angle of the torus is difficult to restrict since it might be linked to the absorption along the LOS \citep[][]{Furui16,Balokovic18}. If the origin (in terms of geometrical distribution) of both emissions is the same, the combination of X-ray and mid-IR might help to solve these issues. However, before trying to attempt to a simultaneous fit, the results from independent fits need to be discussed to understand if the same distribution could explain both mid-IR and X-ray continuum emission. 
In a pilot study, \citet{Esparza-Arredondo19} show that this might be the case for the Sy2 IC\,5063. However, they tested the smooth gas torus model at X-ray while both smooth and clumpy torus models were tested at mid-IR. The X-ray clumpy gas torus model presented here was not available at the moment of that publication. In the current work, we included this source again and found that the best combination of models is clumpy at both wavelengths. This result shows the importance of testing several combinations of models. 

At mid-IR wavelengths, several works have performed a thorough comparison of the smooth and clumpy models, and have discussed the possible issues of each of them \citep{Feltre12,Tanimoto19}. However, only a few works have focused on the comparison of these two distributions of gas at X-ray wavelengths \citep[e.g.][]{Liu14,Furui16}. These works use the same code to create these models and they freeze some parameters in order to obtain better spectral fitting for a few objects. Here we analyze for the first time a sample of AGN. Based on their good statistics ($\chi^2/d.o.f. < 1.2$), we can compare the best models chosen to fit the mid-IR and X-ray spectra in 25 out of the 36 AGN (see Table\,\ref{tab:AkaikeCriterio}). We focus on these 25 AGN and we refer the reader to Appendix\,\ref{sec:app2} for a discussion of the 11 objects with poor spectral fittings. In the following sections we discuss the smooth or clumpy distribution of dust and gas and the complexity of the torus inferred from our results.

\subsection{Distribution of gas and dust}

Following the Akaike criterion, the absorption variability information, and the possible X-ray and mid-IR model combinations for each source, we found that six sources are consistent with the smooth models at both wavelengths. We also found 14 objects consistent with the clumpy models at both wavelengths. Meanwhile, a combination of the smooth and clumpy models is applicable for 17 of the sources (see Cols. 4 and 5 in Table \ref{tab:midIRvsXrays_param}). Note that in some cases, both models are suitable in a given wavelength.

The comparison of clumpy models at both wavelengths have been explored, recently, using pre-existing mid-IR results \citep{Buchner19, Ogawa20}. However, the smooth versus clumpy comparison has never been explored and it is found as a plausible scenario in this work. 
These results could suggest a scenario where the gas and dust are not necessarily distributed in the same way for some objects. Indeed, \citet{Liu14} pointed out that this kind of mismatch might occur when comparing both wavelengths. Absorption variability among AGN has been claimed as a corroboration of the clumpy nature of the absorber at X-rays \citep[e.g.][]{Markowitz14}. However, \citet{Laha20} conducted an X-ray spectral variability study of 20 Compton-thin AGN, finding $\rm{N_H}$ variations only in seven objects \citep[see also][]{Hernandez-Garcia15,Hernandez-Garcia16}. They found constant $\rm{N_H}$ over a decade for some objects. This is consistent with our results that some of the sources fit to a smooth gas torus model at X-rays. 

There are several possibilities to explain this mismatch seen for some objects between the distributions of the gas at X-ray and the dust at mid-IR wavelengths: 
(1) the gas is a smooth distribution associated with the inter-cloud dust medium; or (2) at least part of the gas comes from a smooth distribution associated with the dust-free inner region of the AGN. At mid-IR, \citet{Stalevski16} developed an AGN dust model assuming a two-phase distribution of dust in a torus-like geometry \citep[see also][]{Siebenmorgen15}; dust is distributed smoothly and a clumpy distribution is embedded within it. Therefore, in favor of the first option, it is possible that at X-ray wavelengths a two-phases models are capable to explain better this gas distribution. Unfortunately, these kind of models are not currently available.

Gas at the dust-free inner region or at both the dust-free and dust torus regions are also possibilities to explain a smooth distribution at X-rays and a clumpy distribution at mid-IR wavelengths. These scenarios might nicely explain the large dust-to-gas ratio found in a large number of objects \citep[see Section\,\ref{sec:XrayvsmidIR} and also][]{Maiolino01}. Indeed, most of the X-ray reflection models do not impose inner radii in their geometry \citep[e.g.][]{Murphy09} as naturally done for dust models due to the sublimation radius of the dust.

The stability of this component conformed by gas and dust needs to be explored through dynamical models. These models consider the torus as an axisymmetric obscuring structure that mimics a hydrostatic toroidal distribution that contains gas processed by the accretion disk and expelled outward in a disk wind \citep[see][and references therein]{Elitzur06}. In this way, the BLR and torus are originated at the inner and outer parts of the outflow, respectively. Other models have radiation-driven obscuring structures that replace the classical torus \citep[e.g.][]{Wada12}. Recently, \citet{Sarangi19} showed, through a magneto-hydrodynamic wind model, that this wind naturally produces dust with a distribution that resembles a toroidal shape with sizes matching those obtained from VLA imaging \citep{Carilli19}. Under this model, both gas and dust are smoothly distributed, with a well defined distribution of dust temperatures preventing dust from destruction \citep{Sturm05,Srinivasan17}, an issue claimed in early works \citep{Krolik88}. Thus, the smooth distribution preferred at X-rays is stable from the dynamical point of view.

\subsection{Complexity of the torus}

The comparison between the X-ray reflector and the dusty torus inferred from mid-IR spectral fitting is quite complex, when including all the available information. To illustrate this, we firstly focus on the 13 objects (six Sy1 and seven Sy2) with good spectral fits at both wavelengths and available information on the variable or non-variable LOS absorption. The rest of the sample will be discussed below. Table\,\ref{tab:midIRvsXrays_param} shows the inferred distribution of the gas and dust (i.e. clumpy or smooth), the covering factor at mid-IR, the half-opening angle of the torus at X-rays and mid-IR, the inclination angle, the number of clouds in the equatorial plane (when the clumpy model is preferred at mid-IR), the optical depth of the dust, and the hydrogen column density of the  reflection component at X-rays. We have split this sample of 13 objects into groups sharing similar properties on the distribution of gas and dust, absorption variability, mid-IR covering factor, and half opening angle of the torus:

\begin{figure}[!t]
    \centering
    \includegraphics[width=0.5\columnwidth]{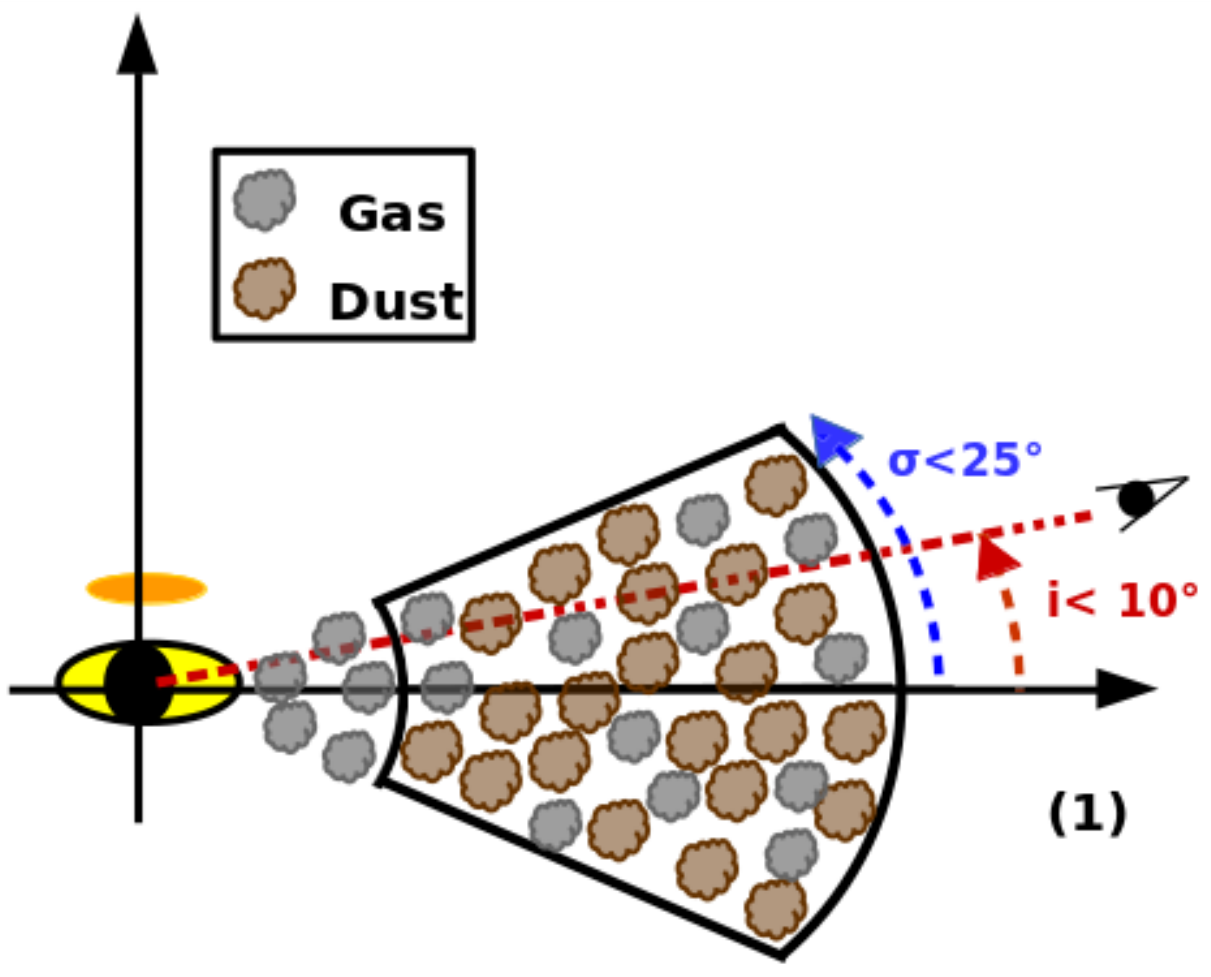}
    \includegraphics[width=0.5\columnwidth]{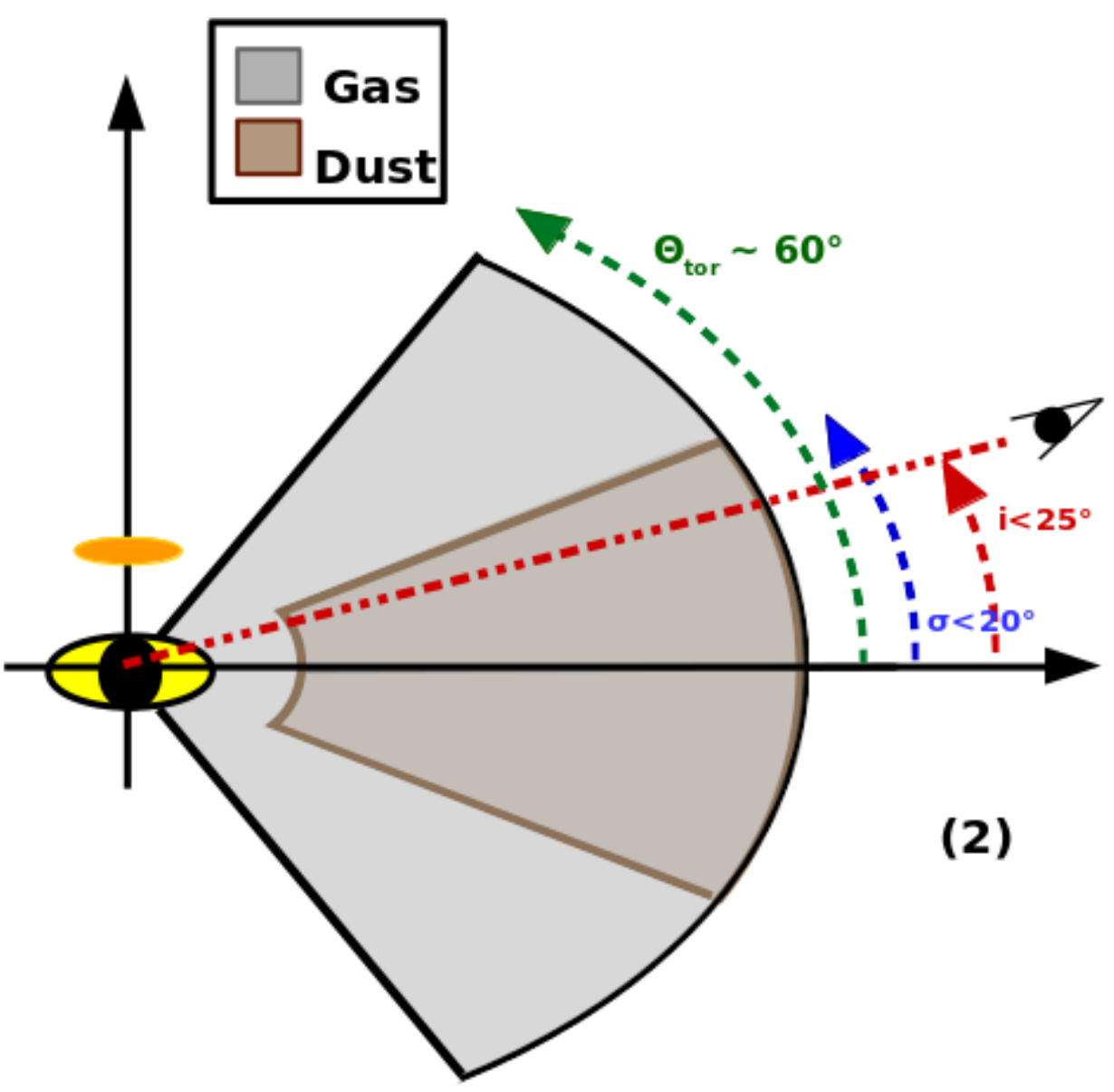}\\
    \includegraphics[width=0.5\columnwidth]{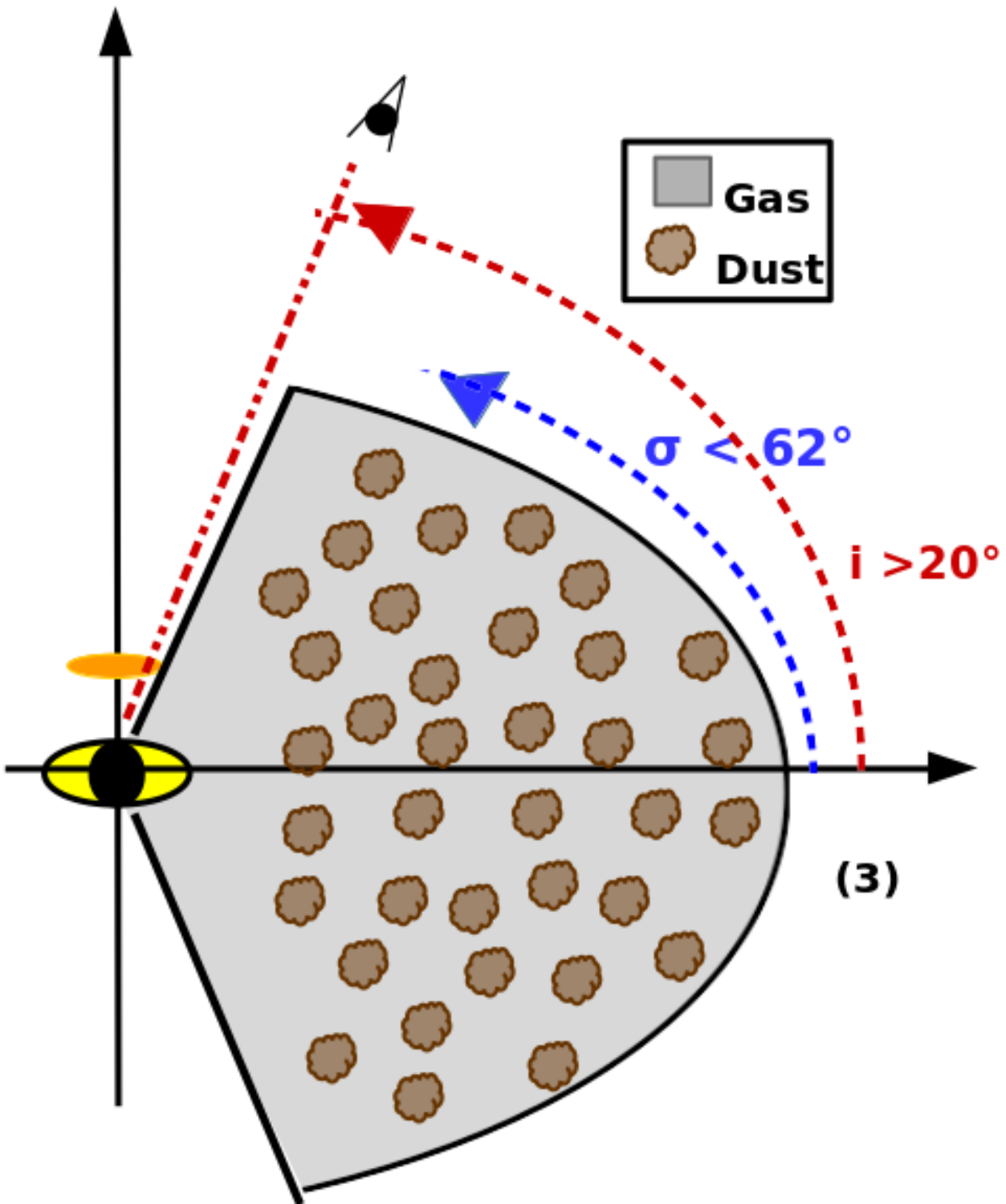}
    \includegraphics[width=0.5\columnwidth]{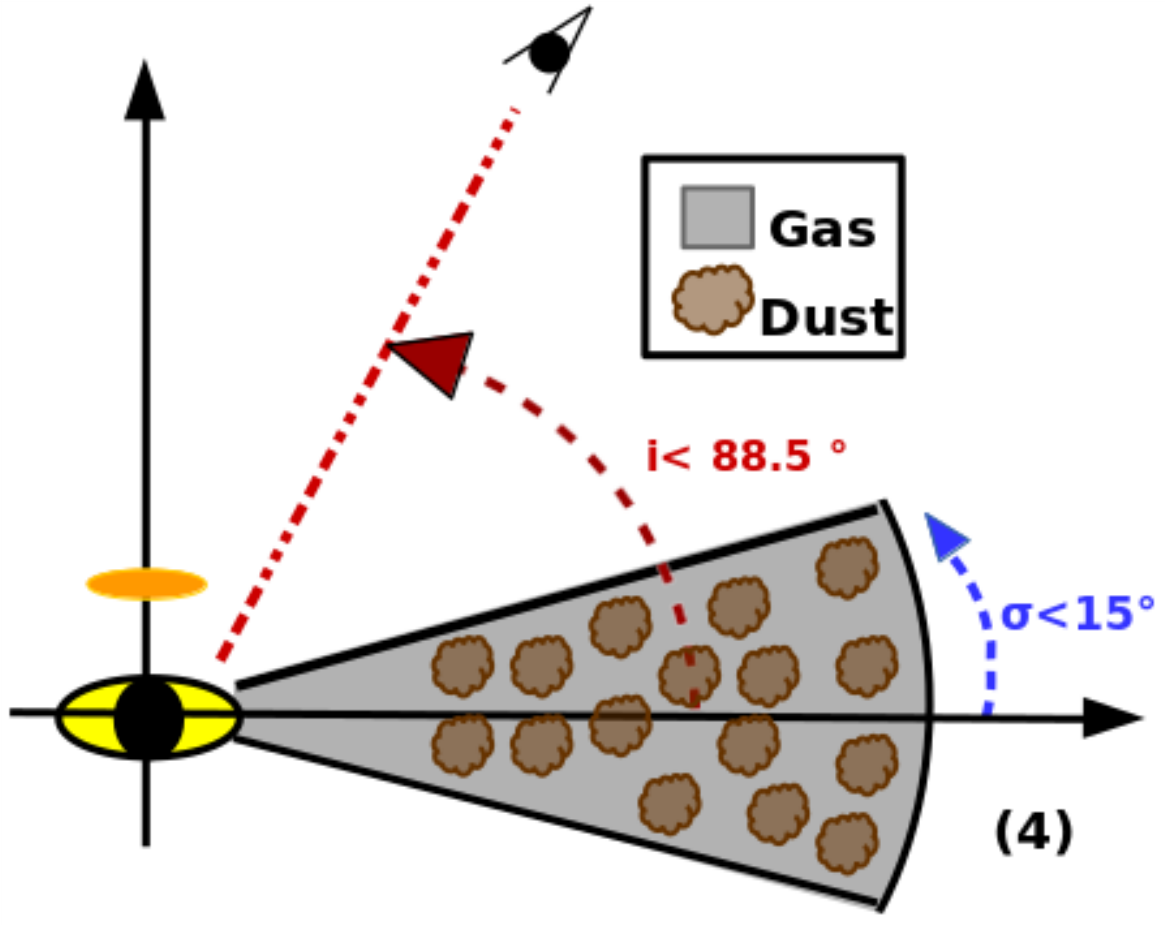}\\
\caption{Illustrations of different gas/dusty tori for some Sy1 in our sample according to the properties gathered in Table\,\ref{tab:midIRvsXrays_param}. The orange and yellow circles show the corona and accretion disk, respectively. The red dashed line shows the LOS. See text for more details.}
    \label{fig:cartoon_Sy1}
\end{figure}

\begin{figure}[!t]
    \centering
    \includegraphics[width=0.5\columnwidth]{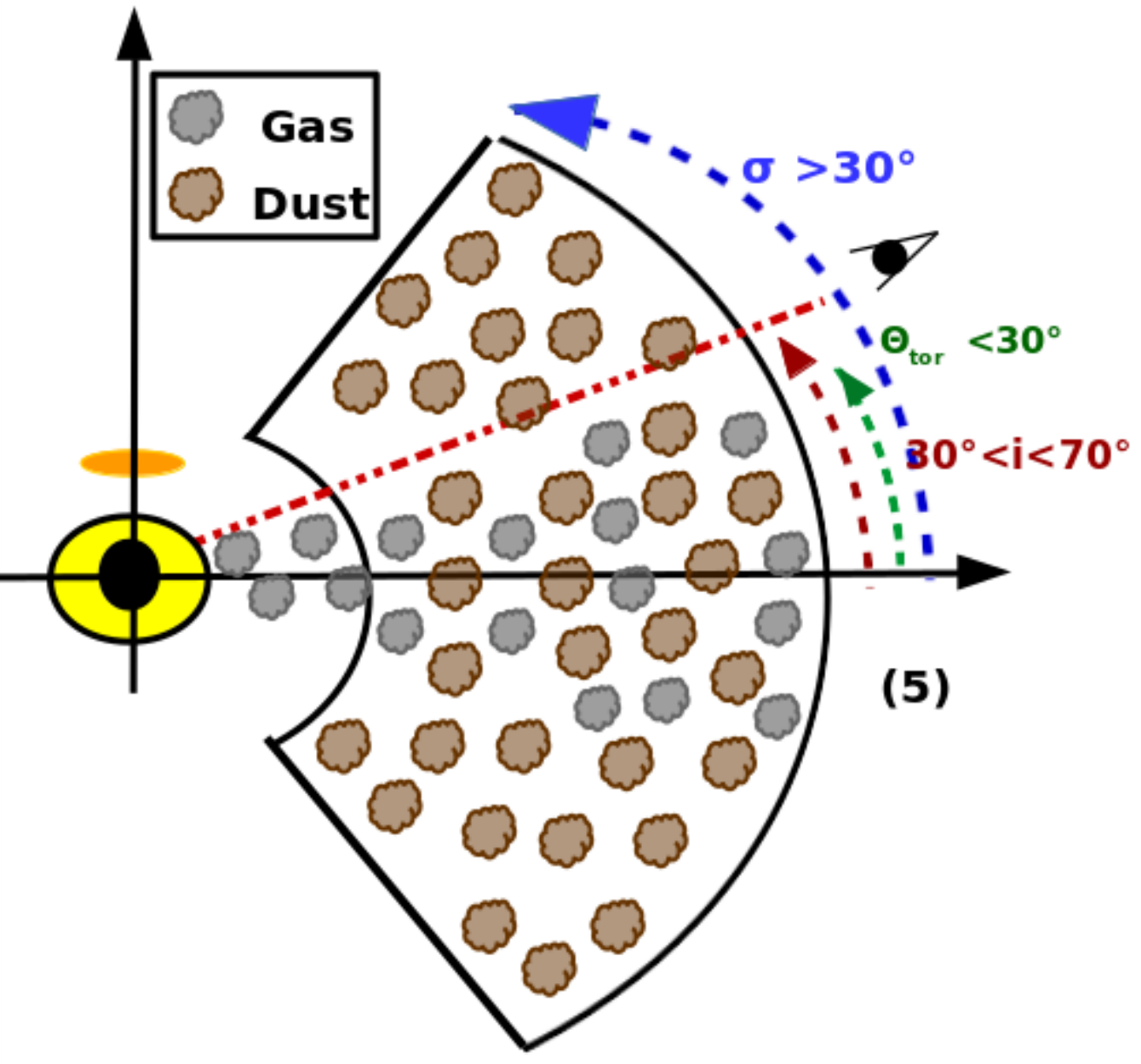}
     \includegraphics[width=0.5\columnwidth]{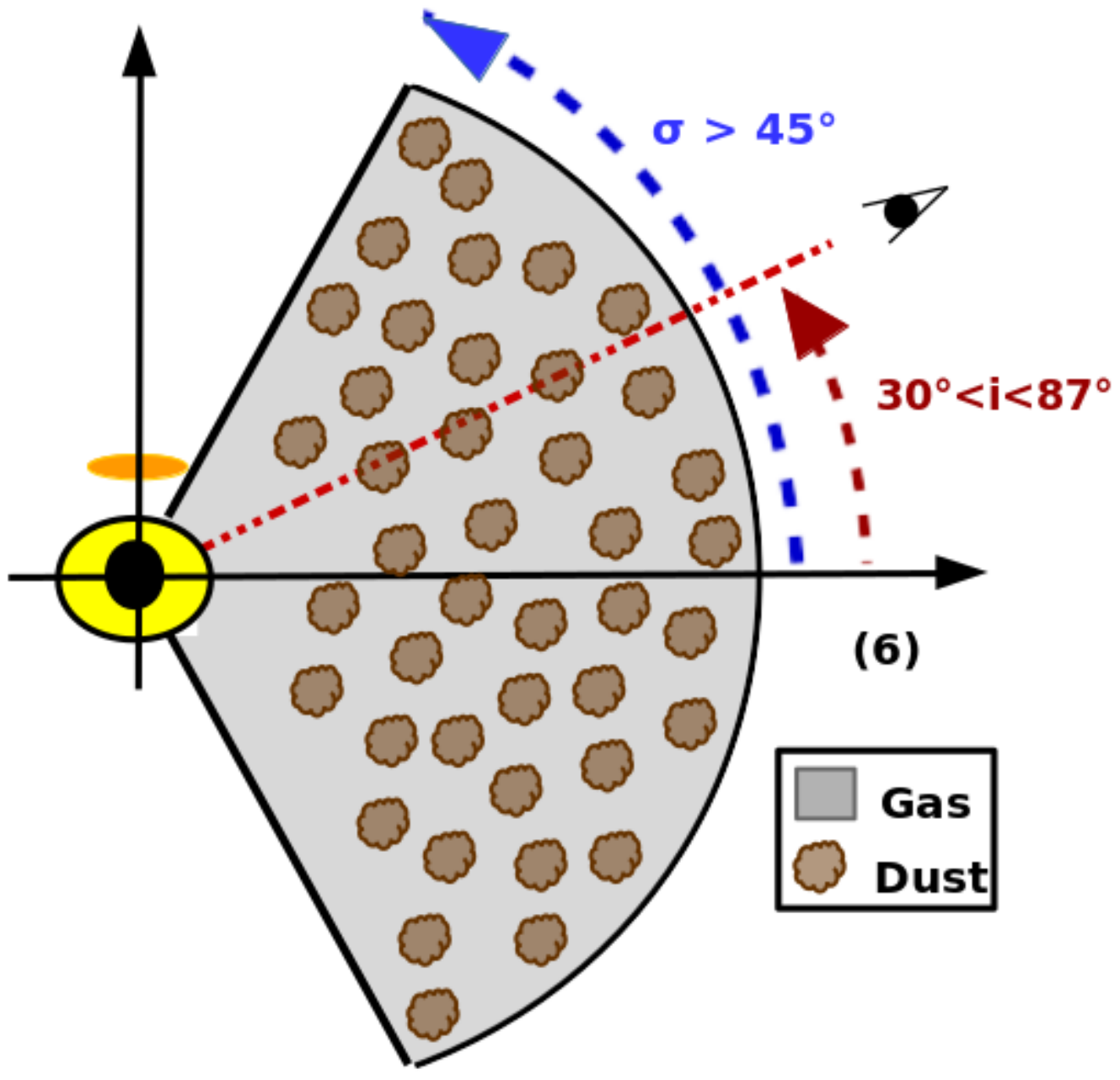}
\caption{Illustrations of different gas/dust tori for some Sy2 in our sample according to the properties gathered in Table\,\ref{tab:midIRvsXrays_param}. The orange and yellow circles show the corona and accretion disk, respectively. See text for more details.}
    \label{fig:cartoon_Sy2}
\end{figure}

\begin{itemize}
    \item Panel 1 of Figure\,\ref{fig:cartoon_Sy1} (e.g. Mrk\,1393): Sy1 source with absorption variability, intermediate mid-IR covering factor, and small half-opening angle. A clumpy distribution is preferred for both gas and dust. Interestingly, this object shows a large number of clouds at the equator and low optical depth. This object can be explained with a geometrically-thin disk of (high number of) clumps conformed by dust and gas, where the inner (dust-free) region is filled with these clumps. Note that the low inclination angle toward the structure (measured from the equator) might intercept some of the clouds close to the central engine producing the absorption variations.
    \vspace{0.15cm}
    \item Panel 2 of Figure\,\ref{fig:cartoon_Sy1} (e.g. PG\,0804+761 and PG\,1211+143): Sy1 sources without absorption variations, small covering factors, small half opening angles of the dust, and low-to-intermediate half opening angles of the gas distribution. A smooth distribution of both dust and gas is preferred for these objects. All the above signatures could be explained with a geometrically-thin and smooth disk of dust and a geometrically-thick and smooth torus of gas, coexisting within the same radii 
    Interestingly, inclination angles might intercept this thin disk along the LOS, despite its optical classification as Sy1.
    \vspace{0.15cm}
    \item Panel 3 of Figure\,\ref{fig:cartoon_Sy1} (e.g. Mrk\,1383 and Mrk\,590): Sy1 sources without absorption variability, large covering factors, intermediate-to-high half opening angles for both gas and dust. The gas is distributed smoothly while the dust is distributed into clumps. These two objects could be explained with a single geometrically-thick torus where dust is located in clumps while the gas is smoothly distributed 
    \vspace{0.15cm}
    \item Panel 4 of Figure\,\ref{fig:cartoon_Sy1} (e.g. NGC\,7213): Sy1 source without absorption variations, low covering factor, and small half opening angles of the dusty structure. A combination of a smooth distribution of gas and a clumpy distribution of dust is preferred for this object. A geometrically-thin disk of gas and dust, where gas is smoothly distributed while clouds hold most of the dust, is able to explain the observed signatures. The smooth distribution prevents from observing absorption variability of the source. Furthermore, the inclination angle is consistent with the unobstructed view of the source. 
    \vspace{0.15cm}
    \item Panel 5 of Figure\,\ref{fig:cartoon_Sy2} (e.g. NGC\,4507, NGC\,1052, NGC\,4939, and IC\,5063): Sy2 sources with absorption variability at X-rays, high covering factors, and half opening angles at both X-rays and mid-IR consistent with intermediate or high values. A clumpy distribution of both gas and dust is preferred for this group of objects. All the properties described above can be explained with a geometrically-thick torus where dust and gas is distributed in clumps. The gas is probably located at both the outer dusty and inner dust-free regions, the latter producing the observed absorption variability. Indeed the inclination angles constrained for these sources are consistent with these clumps being intercepted by the LOS.
    \vspace{0.15cm}
    \item Panel 6 of Figure\,\ref{fig:cartoon_Sy2} (e.g. ESO\,138-G1, PKS\,2356-61, and NGC\,788): Sy2 sources without absorption variability, large covering factors\footnote{Large covering factor is found in PKS\,2356-61 if it is fitted to clumpy torus at mid-IR, although the smooth model is equally good from the statistical point of view.}, and intermediate or high half-opening angles of both gas and dusty structures. The gas is smoothly distributed while the dust is located within clumps. A geometrically-thick torus of dust and gas is able to explain the observed characteristics of these three objects, where the dust is located in clumps while the gas is smoothly distributed. Interestingly these objects are seen at relatively large inclination angles but, in spite of that, the large value of the half opening angle of the torus prevents a direct view of the accretion disk. 
\end{itemize}

Ten out of the twelve remaining sources without information on the absorption variability might be classified into these groups according to the available information. We included a comment on their plausible classification in Column\,13 of Table\,\ref{tab:midIRvsXrays_param}. For example, among Sy1 sources, Mrk\,1392 have similar properties to source of panel 1 for Sy1 (Fig.\,\ref{fig:cartoon_Sy1}) such as an intermediate value for the covering factor, small half opening angle, and a large number of clouds along the equator ($\rm{N_{0}}$). RBS\,0770, and other two sources have similar properties to the object of panel 3 of Sy1 (Fig.\,\ref{fig:cartoon_Sy1}), showing large covering factors and intermediate-to-high half opening angles. Within the Sy2s class, the properties of UM\,146 and other three sources are similar to those observed in the source of panel 5 of Sy2 (Fig.\,\ref{fig:cartoon_Sy2}), showing large values of the covering factors and intermediate-to-high half opening angles. The properties of ESO\,141-G055 and J05081967+1721483 are different between them and from those observed in other sources, therefore, we are not able to classify them in any of these groups. This suggests that the complexity of these sources might extend even beyond the classes described above. The results presented here are consistent with the current paradigm of AGN in which the accretion disk is surrounded by a plethora of infalling, outflowing, and rotating material, configuring the obscurer we see with the X-ray and mid-IR observations \citep[see][for a review]{RamosAlmeida17}. 

The inclination angles are consistent with the obstructed or unobstructed view of the accretion disk, following the unification theory. However, we did not find a correspondence between different Sy1 and Sy2 groups. For example, the tori of panel 1 of Sy1 (Fig.\,\ref{fig:cartoon_Sy1}) and panel 5 of Sy2 (Fig.\,\ref{fig:cartoon_Sy2}) share the same clumpy distribution of dust and gas. However, they differ in their covering factors and half opening angles which are larger in Sy2 compared with Sy1. This is also consistent with previous works that have found that Sy2 tori tend to be thicker, and therefore, intrinsically different from that of Sy1 \citep[e.g.][]{Ramos-Almeida11,Garcia-Bernete19}. Moreover, the optical depths are higher and the number of clouds at the equator of the system is lower in Sy2 compared with Sy1. Thus, the difference between these two groups are beyond the inclination angle toward the system. Even closer resemplanse is seen for the tori of panel 3 of Sy1 (Fig.\,\ref{fig:cartoon_Sy1}) and panel 6 of figure \ref{fig:cartoon_Sy2} (Sy2).They show large covering factors and intermediate-to-high half opening angles. Still, the number of clouds is larger and the optical depth is lower for Sy2 compared with Sy1.


Interestingly, most of these scenarios are also consistent with the existence of gas in a region free of dust. This result is consistent with the recent finding for the Circinus Galaxy by \cite{Uematsu21}. Therefore, the reflection component at X-rays might be associated with the same structure producing the dust emission but including gas within the dust sublimation zone. Although we cannot rule out that X-ray and mid-IR wavelengths are tracing different components for some of the objects, this explanation is particularly appealing for objects where the half-opening widths are consistent for both gas and dust, suggesting a common origin for both dust mid-IR emission and X-ray reflection component.


\longtab{
\begin{landscape}
\begin{table*}
\caption{Comparison between mid-IR and X-ray parameters}
    \label{tab:midIRvsXrays_param}
    \centering
    \begin{scriptsize}
    \begin{tabular}{lllllllllllll}
    \hline \hline
    Objname & type & $\rm{N_H}$  & \multicolumn{2}{c}{model}  & Cf & \multicolumn{2}{c}{$\sigma$}  &  i & $\rm{N_0}$ & $\tau_{\nu}$ & $log(NH_{tor})$ & comment\\[3pt]
            &      & variable    & X-ray & mid-IR & mid-IR & X-ray & mid-IR & \multicolumn{3}{c}{mid-IR} & X-ray &  \\[3pt]
    (1) & (2) & (3) & (4) & (5) & (6) &  (7) & (8) & (9) & (10) & (11) & (12) & (13) \\[3pt] 
    \hline \hline
 &  &  &  &  &  &  Seyfert 1 &  &  &  &  &  &  \\
\hline \hline
Mrk1393   &   Sy1   &   Y   &   Clumpy*   &    Clumpy   &    0.61$_{0.47}^{0.79}$    &     -    &    $<$25.3   &    10.7$_{4.6}^{11.7}$    &    $>$14.1   &     $<$10.6   &    23.9$_{23.8}^{24.1}$    &    \\ [3pt] 
\hline
PG0804+761   &   Sy1   &   N   &   Smooth   &    Smooth   &    0.07$_{0.02}^{0.11}$   &   36.9$_{9.1}^{65.6}$    &    $<$21.1   &    $<$0.1   &    -   &      18.4$_{18.0}^{24.6}$    &    24.5$_{24.3}^{24.7}$    &    \\ [3pt] 
PG1211+143   &   Sy1   &   N   &   Smooth   &    Smooth   &    0.12$_{0.11}^{0.14}$   &   64.2$_{50.2}^{72.7}$    &    $<$20.3   &    25.3$_{24.0}^{35.3}$    &    -   &      $>$153.9   &    24.2$_{24.1}^{24.4}$    &    \\[3pt] 
\hline
Mrk1383   &   Sy1   &   N   &   Smooth   &    Clumpy   &    0.82$_{0.54}^{0.93}$    &    65.1$_{72.2}^{55.1}$    &    $<$62.9   &    $>$16.8   &    3.1$_{2.8}^{3.4}$    &     75.7$_{63.8}^{86.3}$    &    24.3$_{24.2}^{24.5}$    &    \\[3pt]
Mrk590   &   Sy1   &   N   &   Smooth*   &    Clumpy   &    0.81$_{0.51}^{0.94}$    &     -    &    $>$31.7   &    $>$55.7   &    4.2$_{3.5}^{8.2}$    &     135.0$_{66.3}^{149.2}$    &    23.7$_{23.7}^{23.8}$    &    \\[3pt]
\hline
NGC7213   &   Sy1   &   N   &   Smooth   &    Clumpy   &    0.37$_{0.35}^{0.39}$    &     -    &    $<$15.1   &    $<$88.5   &    7.0$_{5.6}^{7.3}$    &     55.8$_{50.6}^{59.2}$    &    22.8$_{22.4}^{23.2}$    &    \\[3pt] 
\hline \hline
 &  &  &  &  &  &  Seyfert 2 &  &  &  &  &  &  \\
\hline \hline
NGC4507   &   Sy2   &   Y   &   Clumpy*   &    Clumpy   &    0.81$_{0.51}^{0.91}$    &    9.9$_{0.6}^{14.2}$    &    $>$47.6   &    $>$35.3   &    3.1$_{3.0}^{3.5}$    &     70.8$_{64.8}^{76.6}$    &    23.8$_{23.8}^{23.8}$    &    \\[3pt] 
NGC1052   &   Sy2   &   Y   &   Clumpy*   &    Clumpy   &    0.87$_{0.68}^{0.95}$    &    27.8$_{0.7}^{60.4}$    &    $>$46.6   &    $>$51.0   &    3.5$_{3.3}^{4.6}$    &     199.9$_{178.6}^{218.0}$    &    23.4$_{23.3}^{23.4}$    &    \\ [3pt]
NGC4939   &   Sy2   &   Y   &   Clumpy*   &    Clumpy   &    0.95$_{0.93}^{0.96}$    &    $>$21.4   &    $>$54.2   &    $<$71.7   &    $>$8.2   &     $>$216.6   &    23.7$_{23.6}^{23.7}$    &    \\ [3pt]
IC5063   &   Sy2   &   Y   &   Clumpy   &    Clumpy   &    0.86$_{0.77}^{0.92}$    &    $>$16.5   &    38.2$_{29.9}^{49.7}$    &    $<$60.0   &    $>$8.8   &     95.3$_{79.0}^{112.3}$    &    23.5$_{23.5}^{23.5}$    &    \\[3pt]
\hline
ESO138-G1   &   Sy2   &   N   &   Smooth*   &    Clumpy   &    0.95$_{0.94}^{0.96}$    &    79.4$_{78.2}^{80.6}$    &    $>$66.8   &    $<$87.7   &    5.5$_{5.2}^{6.1}$    &     19.0$_{14.6}^{20.1}$    &    24.5$_{24.5}^{24.6}$    &    \\ [3pt]
PKS2356-61   &   Sy2   &   N   &   Smooth   &    S/C   &    0.14$_{0.12}^{0.16}$ /0.8$_{0.56}^{0.93}$    &     -    &    $<$21.7/35.0$_{23.4}^{53.1}$    &    24.8$_{21.3}^{28.2}$ /40.2$_{23.4}^{55.9}$    &    5.4$_{5.2}^{10.4}$    &     $>$204.4/40.0$_{28.3}^{51.0}$    &    23.1$_{23.0}^{23.2}$    &    \\ [3pt]
NGC788   &   Sy2   &   N   &   Smooth   &    Clumpy   &    0.92$_{0.81}^{0.96}$    &     -    &    $>$46.8   &    $>$28.9   &    10.4$_{9.6}^{12.1}$    &     43.3$_{40.6}^{46.5}$    &    23.8$_{23.8}^{23.9}$    &    \\ [3pt]
    \hline \hline
 &  &  &  &  &  &  Unclassified &  &  &  &  &  &  \\
 \hline
Mrk1392   &   Sy1   &   -   &   Clumpy   &    Clumpy   &    0.52$_{0.43}^{0.65}$    &    3.0$_{0.2}^{7.6}$    &    $<$17.9   &    6.1$_{2.8}^{12.9}$    &    $>$13.6   &     12.7$_{10.8}^{14.3}$    &    22.4$_{22.4}^{22.4}$    & Panel 1 of Figure \ref{fig:cartoon_Sy1}   \\ [3pt] 
RBS0770   &   Sy1   &   -   &   S/C   &    S/C   &    0.11$_{0.1}^{0.13}$ /0.8$_{0.52}^{0.93}$    &     - /5.4$_{2.2}^{7.7}$    &    $<$20.4/$<$57.4   &    0.0$_{0.0}^{1.9}$ /$>$16.3   &    5.0$_{3.8}^{6.0}$    &     166.4$_{155.2}^{181.7}$ /28.0$_{20.9}^{37.5}$    &    25.0$_{24.7}^{25.2}$ /22.1$_{22.1}^{22.2}$    &   Panel 3 of Figure \ref{fig:cartoon_Sy1} \\[3pt] 
RBS1125   &   Sy1   &   -   &   S/C   &    Clumpy   &    0.77$_{0.5}^{0.94}$    &     - /$>$0.1   &    $<$44.5   &    $>$4.9   &    6.2$_{5.7}^{7.3}$    &     20.0$_{17.1}^{22.4}$    &    24.3$_{24.1}^{24.4}$ /24.0$_{23.8}^{24.3}$    &  Panel 3 of Figure \ref{fig:cartoon_Sy1}  \\[3pt] 
MCG+01-57-016   &   Sy1   &   -   &   S/C   &    Clumpy   &    0.87$_{0.67}^{0.95}$    &    $>$80.5/0.3$_{0.1}^{0.4}$    &    42.6$_{29.3}^{67.7}$    &    $<$8.5   &    $>$6.7   &     198.9$_{167.8}^{259.6}$    &    24.3$_{23.9}^{24.9}$ /22.6$_{22.5}^{22.6}$    & Panel 3 of Figure \ref{fig:cartoon_Sy1}   \\[3pt]
ESO141-G055   &   Sy1   &   -   &   Smooth   &    S/C   &    0.08$_{0.07}^{0.14}$ /0.62$_{0.39}^{0.79}$    &    59.5$_{46.0}^{65.6}$    &    $<$21.6/$<$36.9   &    44.7$_{40.4}^{47.0}$ /$>$2.7   &    3.1$_{2.7}^{3.4}$    &     $>$215.9/50.4$_{44.1}^{58.6}$    &    24.4$_{24.3}^{24.6}$    & unclassified   \\[3pt]
UM146   &   Sy2   &   -   &   S/C   &    Clumpy   &    0.92$_{0.87}^{0.94}$    &     - /1.8$_{0.5}^{20.3}$    &    $>$55.4   &    $>$37.9   &    4.8$_{4.4}^{5.1}$    &     24.7$_{22.6}^{28.4}$    &    23.3$_{23.0}^{23.6}$ /23.4$_{23.0}^{23.5}$    &  Panel 5 of Figure \ref{fig:cartoon_Sy2}   \\ [3pt]
NGC1358   &   Sy2   &   -   &   S/C   &    S/C   &    0.42$_{0.16}^{0.85}$ /0.84$_{0.56}^{0.95}$    &    $>$26.1/$<$9.6   &    $<$58.6/ -    &    $<$42.1/$<$55.6   &    7.7$_{5.6}^{10.3}$    &     $>$53.0/$<$16.0   &    24.2$_{24.2}^{24.3}$ /24.2$_{24.2}^{24.3}$    & Panel 5 of Figure \ref{fig:cartoon_Sy2}   \\[3pt]
IC4518W   &   Sy2   &   -   &   S/C   &    Clumpy   &    0.92$_{0.74}^{0.96}$    &     - /2.7$_{0.3}^{9.7}$    &    $>$48.1   &    $>$51.5   &    8.3$_{6.5}^{12.1}$    &     27.2$_{22.6}^{46.9}$    &    23.4$_{23.3}^{23.4}$ /23.3$_{23.2}^{23.4}$    & Panel 5 of Figure \ref{fig:cartoon_Sy2}  \\ [3pt]
Mrk78   &   Sy2   &   -   &   S/C   &    Clumpy   &    0.94$_{0.82}^{0.96}$    &     - / -    &    $>$55.7   &    $>$21.8   &    $>$13.9   &     53.6$_{47.7}^{56.6}$    &    23.8$_{23.7}^{23.8}$ /23.8$_{23.7}^{23.8}$    & Panel 5 of Figure \ref{fig:cartoon_Sy2}   \\ [3pt]
J05081967+1721483   &   Sy2   &   -   &   Clumpy   &    Clumpy   &    0.83$_{0.57}^{0.93}$    &    $>$3.3   &    53.3$_{41.0}^{61.5}$    &    $<$75.4   &    4.3$_{3.1}^{6.7}$    &     $>$139.7   &    23.5$_{23.4}^{23.6}$    &  Unclassified  \\ [3pt]
ESO103-G35   &   Sy2   &   -   &   Smooth   &    Clumpy   &    0.85$_{0.59}^{0.95}$    &     -    &    $<$46.6   &    $>$22.7   &    8.8$_{8.7}^{11.8}$    &     46.8$_{41.9}^{50.7}$    &    24.3$_{24.3}^{24.4}$    &  Panel 6 of Figure \ref{fig:cartoon_Sy2}  \\  [3pt]
J10594361+6504063   &   Sy2   &   -   &   Smooth   &    S/C   &    0.93$_{0.77}^{0.95}$ /0.96$_{0.96}^{0.97}$    &     -    &    $<$22.1/$>$68.9   &    74.6$_{71.1}^{78.1}$ /54.2$_{62.1}^{47.5}$    &    $>$14.7   &     $>$196.7/20.1$_{19.7}^{21.2}$    &    22.7$_{22.5}^{22.9}$    & Panel 6 of Figure \ref{fig:cartoon_Sy2}  \\ [3pt]
\hline \hline
\end{tabular}
\end{scriptsize}
\tablefoot{S/C is included in Col.4 and 5 when neither of the two models is preferred. Objects marked as Clumpy* in Col.\,4 are variable at X-rays (denoting a clumpy medium) but preferred at smooth model statistically. In Col.\,6 we show the covering factors estimated to mid-IR. In Cols.\,8 and 9 we show the half opening angles using the preferred model at X-ray and mid-IR wavelengths, respectively. Both half opening angle and viewing angles are measured from the equator of the system to its pole. The optical depth is also converted to the 9.7$\rm{\mu m}$ optical depth as $\rm{\tau_{9.7\mu m}= 0.042\,\tau_{V}}$. } 
\end{table*}
\end{landscape}
}




\section{Conclusions}
\label{sec:Conclusion}
We studied the $\rm{3-70\,keV}$ \emph{NuSTAR} and the $\rm{5-30\mu m}$ \emph{Spitzer}/IRS spectra of a sample of 36 AGN. We restricted the sample to objects dominated by the AGN dust continuum at mid-IR and at least 10\% of reflection component contributing to the X-ray spectrum. We excluded from this analysis objects where disk reflection could contribute to the X-ray spectrum. We tested in this paper the comparison between smooth and clumpy torus models at both mid-IR and X-ray frequencies. We used at X-rays the smooth torus model by \citet{Balokovic18} and the clumpy torus model by \citet{Buchner19}. At mid-IR we used analog models by \citet{Fritz06} and \citet{Nenkova08a,Nenkova08b}.

We found that up to $\sim 78\%$ of the sample is well reproduced with the smooth model at X-ray wavelengths. At mid-IR wavelengths, $\sim 69\%$ is well fitted to clumpy models. Interestingly, there is a very good agreement between the clumpiness/smoothness of the X-ray reflector found in this paper and the existence/lack of variability in the column density along the line-of-sight reported in the literature.

We were also able to compare the resulting models at both wavelengths for 25 AGN with good spectral fits. We find that at least 50\% of the sample could be explained by a clumpy distribution of the dust and a smooth or clumpy distribution of the gas. Moreover, we found that dust-to-gas ratio is in a range between [0.01-1] times the Galactic ratio. This is consistent with the existence of gas located within the dust-free inner region. Considering all the information compiled in this work, we find at least six different scenarios to explain the torus properties of some sources. Inside of these scenarios three gas and dust distributions are possible: clumpy-clumpy, smooth-smooth, and smooth-clumpy. These results show the complexity of the obscuring material in the proximity of AGN.

\begin{acknowledgements}
     The authors thank the anonymous referee for careful reading and constructive suggestion that improved the paper. This work made use of data from the \emph{NuSTAR} mission, a project led by CalTech, managed by JPL, and funded by NASA. We thank the \emph{NuSTAR} Operations, Software and Calibration teams for support with the execution and analysis of these observations. This research has made use of the \emph{NuSTAR} Data Analysis Software (NuSTARDAS) jointly developed by the ASI Science Data Center (ASDC, Italy) and CalTech. This work is based in part on observations made with the \emph{Spitzer} Space Telescope, which is operated by the Jet Propulsion Laboratory, California Institute of Technology under a contract with NASA. DE-A and NO-C acknowledge support from a CONACYT scholarship. This research is mainly funded by the UNAM PAPIIT projects IN105720 and IA113719 (PI OG-M and PI DD). JM acknowledges financial support by the Spanish Ministry of Economy and Competitiveness (MEC) under grant no. AYA2016-76682-C3 and from the State Agency for Research of the Spanish MCIU through the ``Center of Excellence Severo Ochoa" award to the Instituto de Astrof\'isica de Andaluc\'ia (SEV-2017-0709). IG-B acknowledges support from STFC through grant ST/S000488/1
\end{acknowledgements}

\longtab{
\begin{landscape}
\begin{longtable}{lcccccccccccl}
\caption{\label{Tab:Bestfit_XrayValues} X-ray best-fit results per object and model.}\\
\hline\hline
    Objname & Baseline model & $\chi^2/d.o.f. $ &  &  &  & &  Parameters &  &  &  &  & \\
    &  &  & $\theta_{inc}$ & $NH_{LOS}$ &  \textit{Cf} & $\Gamma$ & $log(N_{H_{tor}})$ & $\theta_{tor}$  &  log($f_{intr}$) & log($f_{refl}$) & $log \left( \frac{f_{refl}}{f_{refl} + f_{intr}} \right)$ \\
    &    &    & $^{\circ}$ & $cm^{-2}$ &   &   & $cm^{-2}$ & $^{\circ}$ & [3-70] KeV  &[3-70] KeV &     \\
    (1) & (2) & (3) & (4) & (5) & (6) &  (7) & (8) & (9) & (10) & (11) & (12) \\
\hline
\endfirsthead
\caption{continued.}\\
\hline\hline
    Objname & Baseline model & $\chi^2/d.o.f. $ &  &  &  & &  Parameters &  &  &  &  & \\
    &  &  & $\theta_{inc}$ & $NH_{LOS}$ &  \textit{Cf} & $\Gamma$ & $log(N_{H_{tor}})$ & $\theta_{tor}$  &  log($f_{intr}$) & log($f_{refl}$) & $log \left( \frac{f_{refl}}{f_{refl} + f_{intr}} \right)$ \\
    &    &    & $^{\circ}$ & $cm^{-2}$ &   &   & $cm^{-2}$ & $^{\circ}$ & [3-70] KeV  & [3-70] KeV &     \\
    (1) & (2) & (3) & (4) & (5) & (6) &  (7) & (8) & (9) & (10) & (11) & (12) \\[3pt]
\hline
\endhead
\hline
\endfoot
Mrk1018 & Smooth$\bullet$ & 1.21 & 19.0 & 24.06* & $<$0.97 & 2.01$_{1.91}^{2.12}$  & 24.06$_{23.79}^{24.43}$  & 25.63*  & -11.52$_{-11.54}^{-11.5}$  & -12.18$_{-12.33}^{-12.07}$  & -0.75$\pm$0.25\\[3pt]
         & Clumpy$\bullet$ & 1.24 & 90.0 & 23.7* & 0.43$_{0.2}^{0.61}$  & 2.16$_{1.98}^{2.34}$  & 23.7$_{22.38}^{23.92}$  & 28.0*  & -11.5$_{-11.52}^{-11.49}$  & -12.47$_{-12.9}^{-12.25}$  & -1.01$\pm$0.57\\[3pt]
Mrk590 & Smooth & 0.98 & 19.0 & 23.74* & $<$0.42 & 1.95$_{1.9}^{2.0}$  & 23.74$_{23.66}^{23.8}$  & 0.0*  & -11.18$_{-11.19}^{-11.17}$  & -11.68$_{-11.73}^{-11.64}$  & -0.62$\pm$0.08\\[3pt] 
         & Clumpy  & 0.98 & 45.0 & 23.77* & 0.56$_{0.52}^{0.59}$  & 2.24$_{2.13}^{2.33}$  & 23.77$_{23.69}^{23.83}$  & 6.89$_{2.27}^{16.55}$  & -11.2$_{-11.22}^{-11.19}$  & -11.68$_{-11.74}^{-11.63}$  & -0.6$\pm$0.1\\[3pt]
PG0804+761 & Smooth & 0.89 & 19.0 & 24.47* & $<$0.38 & 1.93$_{1.87}^{2.0}$  & 24.47$_{24.27}^{24.66}$  & 53.1$_{24.43}^{80.92}$  & -10.8$_{-10.82}^{-10.79}$  & -11.48$_{-11.58}^{-11.4}$  & -0.76$\pm$0.17\\[3pt] 
         & Clumpy  & 0.93 & 1.0 & 24.44* & 0.27$_{0.15}^{0.37}$  & 1.89$_{1.82}^{1.97}$  & 24.44$_{24.13}^{24.65}$  & 17.6*  & -10.75$_{-10.76}^{-10.74}$  & -11.85$_{-12.14}^{-11.68}$  & -1.13$\pm$0.45\\[3pt]
RBS0770 & Smooth & 0.97 & 45.0 & 25.02* & 0.19$_{0.15}^{0.28}$  & 1.91$_{1.89}^{1.94}$  & 25.02$_{24.74}^{25.22}$  & 84.3*  & -10.33$_{-10.33}^{-10.32}$  & -11.11$_{-11.17}^{-11.06}$  & -0.85$\pm$0.11\\[3pt] 
         & Clumpy  & 0.97 & 1.0 & 22.14* & $>$0.0 & 1.94$_{1.91}^{1.97}$  & 22.14$_{22.12}^{22.17}$  & 5.42$_{2.24}^{7.68}$  & -10.37$_{-10.37}^{-10.36}$  & -10.98$_{-11.02}^{-10.94}$  & -0.70$\pm$0.07\\[3pt] 
I11119+3257 & Smooth & 1.07 & 87.0 & 22.88* & 1.0*  & 2.42$_{2.33}^{2.55}$  & 22.88$_{22.75}^{23.02}$  & 84.3*  & -11.9$_{-11.94}^{-11.86}$  & -12.61$_{-12.85}^{-12.46}$  & -0.79$\pm$0.36\\[3pt] 
         & Clumpy  & 1.03 & 45.0 & 23.09* & $>$1.0 & 2.84$_{2.72}^{2.95}$  & 23.09$_{23.03}^{23.14}$  & 5.99$_{4.74}^{7.63}$  & -11.97$_{-11.99}^{-11.95}$  & -12.41$_{-12.49}^{-12.33}$  & -0.57$\pm$0.13\\[3pt]
PG1211+143 & Smooth & 0.93 & 19.0 & 24.19* & $<$0.3 & 2.16$_{2.1}^{2.21}$  & 24.19$_{24.08}^{24.38}$  & 25.8$_{17.31}^{39.82}$  & -11.29$_{-11.3}^{-11.28}$  & -11.6$_{-11.63}^{-11.57}$  & -0.48$\pm$0.05\\[3pt] 
         & Clumpy  & 0.98 & 90.0 & 23.73* & 0.47$_{0.4}^{0.56}$  & 2.36$_{2.29}^{2.45}$  & 23.73$_{23.62}^{23.82}$  & 28.0$_{11.59}^{56.07}$  & -11.24$_{-11.25}^{-11.23}$  & -11.87$_{-11.94}^{-11.82}$  & -0.73$\pm$0.11\\[3pt]
RBS1125 & Smooth & 1.04 & 45.0 & 24.26* & 0.43$_{0.38}^{0.54}$  & 2.16$_{2.11}^{2.2}$  & 24.26$_{24.11}^{24.42}$  & 84.3*  & -11.09$_{-11.11}^{-11.08}$  & -11.94$_{-12.11}^{-11.83}$  & -0.91$\pm$0.28\\[3pt] 
         & Clumpy  & 1.08 & 1.0 & 23.99* & 0.41$_{0.21}^{0.55}$  & 2.23$_{2.07}^{2.38}$  & 23.99$_{23.81}^{24.26}$  & $>$0.07 & -11.11$_{-11.12}^{-11.09}$  & -11.94$_{-12.12}^{-11.81}$  & -0.89$\pm$0.30\\[3pt]
Mrk231 & Smooth & 1.09 & 87.0 & 23.3* & 0.6$_{0.51}^{0.82}$  & 1.57$_{1.47}^{1.68}$  & 23.3$_{23.23}^{23.49}$  & 84.3*  & -11.36$_{-11.38}^{-11.34}$  & -12.22$_{-12.43}^{-12.08}$  & -0.92$\pm$0.34\\[3pt] 
         & Clumpy  & 1.11 & 90.0 & 23.37* & 0.71$_{0.64}^{0.8}$  & 1.78$_{1.66}^{1.89}$  & 23.37$_{23.17}^{23.51}$  & $>$2.18 & -11.4$_{-11.41}^{-11.38}$  & -12.1$_{-12.25}^{-11.99}$  & -0.78$\pm$0.24\\[3pt]
Mrk1383 & Smooth & 0.91 & 19.0 & 24.34* & $<$0.4 & 2.1$_{2.05}^{2.14}$  & 24.34$_{24.24}^{24.46}$  & 24.86$_{17.81}^{34.91}$  & -10.82$_{-10.82}^{-10.81}$  & -11.26$_{-11.3}^{-11.23}$  & -0.58$\pm$0.06\\[3pt] 
         & Clumpy  & 0.95 & 45.0 & 24.2* & 0.36$_{0.28}^{0.43}$  & 2.14$_{2.09}^{2.2}$  & 24.2$_{24.14}^{24.2}$  & 7.01$_{0.01}^{23.28}$  & -10.79$_{-10.79}^{-10.78}$  & -11.4$_{-11.46}^{-11.35}$  & -0.71$\pm$0.10\\[3pt]
Mrk1392 & Smooth & 1.02 & 87.0 & 25.29* & 0.0*  & 1.92$_{1.85}^{2.0}$  & $>$25.1 & 84.3*  & -10.98$_{-11.0}^{-10.97}$  & -11.33$_{-11.39}^{-11.28}$  & -0.51$\pm$0.10\\[3pt] 
         & Clumpy  & 0.97 & 1.0 & 22.4* & $>$0.0 & 1.99$_{1.92}^{2.07}$  & 22.4$_{22.39}^{22.43}$  & 3.03$_{0.16}^{7.6}$  & -11.04$_{-11.05}^{-11.02}$  & -11.29$_{-11.34}^{-11.25}$  & -0.45$\pm$0.07\\[3pt] 
Mrk1393 & Smooth & 1.02 & 87.0 & 23.98* & 0.2$_{0.01}^{0.4}$  & 2.08$_{2.0}^{2.17}$  & 23.98$_{23.89}^{24.12}$  & 53.1*  & -10.78$_{-10.78}^{-10.77}$  & -11.69$_{-11.8}^{-11.61}$  & -0.97$\pm$0.20\\[3pt] 
         & Clumpy  & 1.03 & 90.0 & 23.94* & 0.36$_{0.24}^{0.46}$  & 2.12$_{2.02}^{2.21}$  & 23.94$_{23.85}^{24.06}$  & 27.92*  & -10.74$_{-10.75}^{-10.73}$  & $<$-12.08 & $<$-1.77\\[3pt]
PG1535+547 & Smooth$\bullet$ & 1.29 & 19.0 & 23.85* & 0.78$_{0.71}^{0.86}$  & 1.97$_{1.79}^{2.18}$  & 23.85$_{23.77}^{23.95}$  & 9.87*  & -11.75$_{-11.78}^{-11.72}$  & -12.28$_{-12.43}^{-12.18}$  & -0.64$\pm$0.22\\[3pt] 
         & Clumpy$\bullet$  & 1.27 & 90.0 & 23.77* & 0.82$_{0.76}^{0.86}$  & 2.14$_{1.99}^{2.31}$  & 23.77$_{23.64}^{23.88}$  & $>$0.09 & -11.82$_{-11.85}^{-11.79}$  & -12.12$_{-12.23}^{-12.04}$  & -0.478$\pm$0.15\\[3pt]
ESO141-G055 & Smooth & 1.03 & 19.0 & 24.39* & 0.0*  & 1.92$_{1.9}^{1.93}$  & 24.39$_{24.29}^{24.62}$  & 30.45$_{24.37}^{44.01}$  & -10.29$_{-10.29}^{-10.29}$  & -10.9$_{-10.92}^{-10.89}$  & -0.70$\pm$0.03\\[3pt] 
         & Clumpy  & 1.06 & 90.0 & 23.52* & 0.11$_{0.07}^{0.13}$  & 1.97$_{1.96}^{1.99}$  & 23.52$_{23.23}^{23.66}$  & 28.0$_{17.71}^{46.64}$  & -10.29$_{-10.29}^{-10.29}$  & -10.95$_{-10.97}^{-10.94}$  & -0.75$\pm$0.03\\[3pt]
NGC7213 & Smooth & 0.99 & 45.0 & 22.81* & 0.18$_{0.10}^{0.46}$  & 1.89$_{1.87}^{1.93}$  & 22.81$_{22.35}^{23.20}$  & 0.0*  & -10.46$_{-10.47}^{-10.45}$  & -11.70$_{-11.82}^{-11.59}$  & -1.26$\pm$0.24 \\[3pt] 
         & Clumpy  & 1.01 & 90.0 & 23.13* & 0.24$_{0.19}^{0.36}$  & 1.96$_{1.92}^{2.0}$  & 23.13$_{22.68}^{23.31}$  & $>1.05$  & -10.46$_{-10.47}^{-10.45}$  & -11.84$_{-12.28}^{-11.65}$  & -1.40$\pm$0.56\\[3pt]
MCG+01-57-016 & Smooth & 1.07 & 19.0 & 24.34* & $<$0.22 & 1.71$_{1.64}^{1.77}$  & 24.34$_{23.92}^{24.86}$  & $>$9.53 & -10.94$_{-10.95}^{-10.93}$  & -11.83$_{-12.05}^{-11.69}$  & -0.94$\pm$0.35\\[3pt] 
         & Clumpy  & 1.06 & 1.0 & 22.55* & 0.51$_{0.19}^{0.83}$  & 1.82$_{1.75}^{1.92}$  & 22.55$_{22.48}^{22.6}$  & 0.29$_{0.12}^{0.36}$  & -11.01$_{-11.02}^{-11.0}$  & -11.56$_{-11.66}^{-11.48}$  & -0.66$\pm$0.16\\[3pt]
\hline
UM146 & Smooth & 0.86 & 87.0 & 23.34* & $>$0.48 & 1.76$_{1.55}^{1.91}$  & 23.34$_{23.01}^{23.56}$  & 36.92*  & -11.47$_{-11.51}^{-11.44}$  & -11.92$_{-12.03}^{-11.82}$  & -0.58$\pm$0.18\\[3pt] 
         & Clumpy  & 0.88 & 1.0 & 23.45* & 0.77$_{0.7}^{0.98}$  & 2.2$_{1.98}^{2.33}$  & 23.45$_{23.03}^{23.53}$  & 1.82$_{0.49}^{20.27}$  & -11.6$_{-11.64}^{-11.57}$  & -11.76$_{-11.84}^{-11.68}$  & -0.38$\pm$0.12\\[3pt] 
NGC788 & Smooth & 0.9 & 87.0 & 23.84* & 0.9$_{0.89}^{0.91}$  & $<$1.53 & 23.84$_{23.82}^{23.87}$  & 84.3*  & -10.69$_{-10.73}^{-10.66}$  & -10.95$_{-11.04}^{-10.88}$  & -0.45$\pm$0.13\\[3pt] 
         & Clumpy  & 1.0 & 45.0 & 23.69* & 1.0*  & 1.56$_{1.46}^{1.63}$  & 23.69$_{23.61}^{23.76}$  & 10.44$_{0.17}^{19.92}$  & -10.8$_{-10.83}^{-10.77}$  & -10.83$_{-10.89}^{-10.78}$  & -0.32$\pm$0.07\\[3pt] 
NGC1052 & Smooth & 1.02 & 45.0 & 23.31* & 0.81$_{0.79}^{0.84}$  & 1.7$_{1.66}^{1.74}$  & 23.31$_{23.25}^{23.36}$  & 0.0*  & -10.6$_{-10.6}^{-10.59}$  & -11.53$_{-11.61}^{-11.47}$  & -0.99$\pm$0.14\\[3pt] 
         & Clumpy  & 1.02 & 1.0 & 23.37* & 0.85$_{0.84}^{0.87}$  & 1.88$_{1.84}^{1.92}$  & 23.37$_{23.32}^{23.41}$  & 27.77$_{0.71}^{60.43}$  & -10.64$_{-10.65}^{-10.64}$  & -11.32$_{-11.37}^{-11.28}$  & -0.76$\pm$0.08\\[3pt]
NGC1358 & Smooth & 0.93 & 45.0 & 24.24* & 1.0*  & 1.59$_{1.51}^{1.7}$  & 24.24$_{24.21}^{24.28}$  & $>$63.9 & -10.94$_{-10.95}^{-10.92}$  & -11.52$_{-11.55}^{-11.49}$  & -0.68$\pm$0.06\\[3pt] 
         & Clumpy  & 0.94 & 1.0 & 24.21* & 0.99$_{0.98}^{1.0}$  & 1.68$_{1.59}^{1.77}$  & 24.21$_{24.17}^{24.26}$  & $<$9.6 & -11.06$_{-11.09}^{-11.03}$  & -11.26$_{-11.3}^{-11.23}$  & -0.41$\pm$0.06\\[3pt]
J05081967+1721483 & Smooth & 1.18 & 87.0 & 23.61* & 0.35$_{0.19}^{0.45}$  & 1.9$_{1.8}^{1.98}$  & 23.61$_{23.53}^{23.7}$  & 25.76*  & -10.58$_{-10.59}^{-10.57}$  & -11.34$_{-11.43}^{-11.28}$  & -0.83$\pm$0.15\\[3pt] 
         & Clumpy  & 1.15 & 90.0 & 23.51* & 0.53$_{0.45}^{0.58}$  & 2.12$_{2.03}^{2.19}$  & 23.51$_{23.37}^{23.62}$  & $>$3.32 & -10.64$_{-10.65}^{-10.63}$  & -11.13$_{-11.19}^{-11.09}$  & -0.61$\pm$0.08\\[3pt]
Mrk3 & Smooth & 0.99 & 87.0 & 23.87* & 0.96$_{0.96}^{0.96}$  & 1.62$_{1.61}^{1.63}$  & 23.87$_{23.86}^{23.87}$  & 72.43$_{70.58}^{74.51}$  & -9.79$_{-9.8}^{-9.79}$  & -10.57$_{-10.63}^{-10.53}$  & -0.85$\pm$0.10\\[3pt] 
         & Clumpy$\bullet$  & 1.24 & 45.0 & 23.8* & 1.0*  & 1.63$_{1.61}^{1.66}$  & 23.8$_{23.78}^{23.81}$  & 4.83$_{2.1}^{7.98}$  & -9.84$_{-9.84}^{-9.83}$  & -10.37$_{-10.4}^{-10.34}$  & -0.64$\pm$0.06\\[3pt] 
ESO428-G014 & Smooth$\bullet$ & 1.42 & 19.0 & 24.38* & $<$0.99 & 2.23$_{2.03}^{2.45}$  & 24.38$_{24.3}^{24.44}$  & 13.16$_{9.94}^{17.6}$  & -12.82$_{-13.02}^{-12.68}$  & -11.2$_{-11.21}^{-11.18}$  & -0.01$\pm$0.01\\[3pt] 
         & Clumpy$\bullet$ & 1.48 & 1.0 & 20.75* & 0.0*  & 1.95$_{1.88}^{2.0}$  & 20.75$_{20.67}^{20.82}$  & $>$68.37 & -12.43$_{-12.52}^{-12.35}$  & -11.19$_{-11.21}^{-11.18}$  & -0.02$\pm$0.01\\[3pt] 
Mrk78 & Smooth & 1.13 & 45.0 & 23.78* & 0.91$_{0.89}^{0.93}$  & $<$1.55 & 23.78$_{23.72}^{23.83}$  & 45.55*  & -10.88$_{-10.91}^{-10.85}$  & -11.74$_{-11.99}^{-11.58}$  & -0.92$\pm$0.39\\[3pt] 
         & Clumpy  & 1.19 & 90.0 & 23.77* & 0.92$_{0.9}^{0.94}$  & 1.5$_{1.37}^{1.62}$  & 23.77$_{23.69}^{23.83}$  & 1.0*  & -10.88$_{-10.9}^{-10.85}$  & -11.88$_{-12.65}^{-11.61}$  & -1.04$\pm$0.76\\[3pt]
Mrk1210 & Smooth & 1.13 & 45.0 & 23.45* & 0.88$_{0.86}^{0.91}$  & 1.53$_{1.46}^{1.56}$  & 23.45$_{23.37}^{23.49}$  & 0.0*  & -10.29$_{-10.3}^{-10.28}$  & -11.18$_{-11.29}^{-11.09}$  & -0.94$\pm$0.20\\[3pt] 
         & Clumpy  & 1.09 & 90.0 & 23.46* & 0.92$_{0.9}^{0.94}$  & 1.76$_{1.71}^{1.8}$  & 23.46$_{23.41}^{23.51}$  & $>$1.76 & -10.39$_{-10.4}^{-10.38}$  & -10.81$_{-10.86}^{-10.77}$  & -0.56$\pm$0.07\\[3pt] 
J10594361+6504063 & Smooth & 0.78 & 45.0 & 22.72* & $>$0.18 & 1.57$_{1.5}^{1.66}$  & 22.72$_{22.49}^{22.92}$  & 0.0*  & -11.2$_{-11.24}^{-11.17}$  & -11.98$_{-12.21}^{-11.83}$  & -0.84$\pm$0.35\\[3pt] 
         & Clumpy  & 0.84 & 45.0 & 23.15* & $>$0.46 & 1.74$_{1.6}^{1.89}$  & 23.15$_{22.91}^{23.41}$  & 1.0*  & -11.19$_{-11.21}^{-11.17}$  & $<$-12.03 & $<$-1.18\\[3pt] 
NGC4388 & Smooth & 1.01 & 19.0 & 23.56* & 0.87$_{0.86}^{0.89}$  & 1.55$_{1.51}^{1.59}$  & 23.56$_{23.52}^{23.59}$  & 0.0*  & -10.35$_{-10.36}^{-10.34}$  & -10.76$_{-10.8}^{-10.73}$  & -0.55$\pm$0.06\\[3pt] 
         & Clumpy  & 1.14 & 45.0 & 23.44* & 1.0*  & 1.8$_{1.78}^{1.87}$  & 23.44$_{23.42}^{23.48}$  & 6.9$_{3.11}^{7.99}$  & -10.5$_{-10.52}^{-10.49}$  & -10.55$_{-10.57}^{-10.53}$  & -0.32$\pm$0.03\\[3pt]
NGC4507 & Smooth & 0.95 & 19.0 & 23.81* & 0.95$_{0.95}^{0.96}$  & 1.59$_{1.58}^{1.59}$  & 23.81$_{23.81}^{23.82}$  & 0.0*  & -9.95$_{-9.95}^{-9.94}$  & -10.55$_{-10.58}^{-10.52}$  & -0.7$\pm$0.06\\[3pt] 
         & Clumpy  & 1.09 & 45.0 & 23.78* & 1.0*  & 1.69$_{1.66}^{1.72}$  & 23.78$_{23.76}^{23.8}$  & 9.91$_{0.58}^{14.15}$  & -10.02$_{-10.03}^{-10.01}$  & -10.36$_{-10.38}^{-10.34}$  & -0.51$\pm$0.04\\[3pt] 
NGC4939 & Smooth & 0.7 & 87.0 & 23.78* & 0.95$_{0.94}^{0.96}$  & 1.61$_{1.58}^{1.65}$  & 23.78$_{23.76}^{23.8}$  & $<$51.0 & -10.79$_{-10.82}^{-10.77}$  & -11.64$_{-11.87}^{-11.49}$  & -0.9$\pm$0.36\\[3pt] 
         & Clumpy  & 0.75 & 90.0 & 23.67* & 1.0*  & 1.64$_{1.56}^{1.77}$  & 23.67$_{23.63}^{23.72}$  & $>$21.4 & -10.85$_{-10.88}^{-10.83}$  & -11.34$_{-11.46}^{-11.24}$  & -0.61$\pm$0.19\\[3pt] 
ESO-097-G013 & Smooth & 1.03 & 45.0 & 25.01* & 0.98$_{0.98}^{0.99}$  & 2.55$_{2.55}^{2.56}$  & 25.15$_{25.14}^{25.16}$  & 38.46$_{38.43}^{38.48}$  & -11.07$_{-11.08}^{-11.06}$  & -9.67$_{-9.67}^{-9.66}$  & -0.02$\pm$0.01 \\[3pt] 
         & Clumpy$\bullet$ & 1.22 & 1.0 & 24.82* & 1.0$_{0.99}^{1.0}$  & 2.54$_{2.51}^{2.56}$  & 24.82$_{24.81}^{24.83}$  & 27.76$_{13.53}^{33.88}$  & -9.83$_{-9.84}^{-9.83}$  & -10.1$_{-10.1}^{-10.1}$  & -0.43$\pm$ 0.01\\[3pt]        
IC4518W & Smooth & 1.04 & 87.0 & 23.39* & 0.76$_{0.69}^{0.94}$  & $<$1.61 & 23.39$_{23.34}^{23.45}$  & 84.3*  & -10.6$_{-10.63}^{-10.57}$  & -11.36$_{-11.61}^{-11.2}$  & -0.83$\pm$0.38\\[3pt] 
         & Clumpy  & 1.02 & 1.0 & 23.3* & 0.87$_{0.81}^{0.92}$  & 1.73$_{1.63}^{1.81}$  & 23.3$_{23.2}^{23.4}$  & 2.72$_{0.34}^{9.68}$  & -10.72$_{-10.74}^{-10.7}$  & -11.0$_{-11.08}^{-10.93}$  & -0.46$\pm$0.12\\[3pt] 
ESO138-G1 & Smooth$\circ$ & 0.99 & 19.0 & 23.87$_{23.85}^{23.89}$ & 0.96$_{0.96}^{0.97}$  & 2.51$_{2.48}^{2.54}$  & 24.51$_{24.46}^{24.57}$  & 10.57$_{9.43}^{11.75}$  & -11.16$_{-11.18}^{-11.13}$  & -10.97$_{-10.99}^{-10.94}$  & -0.22$\pm$0.03\\[3pt] 
         & Clumpy  & 0.99 & 90.0 & 23.76* & 0.88$_{0.86}^{0.9}$  & 2.23$_{2.17}^{2.29}$  & 23.76$_{23.68}^{23.82}$  & 28.07$_{15.66}^{80.37}$  & -11.25$_{-11.28}^{-11.23}$  & -10.9$_{-10.92}^{-10.88}$  & -0.16$\pm$0.02\\[3pt] 
NGC6300 & Smooth$\circ$ & 0.85 & 87.0 & 23.3$_{23.28}^{23.31}$ & 0.95$_{0.94}^{0.96}$  & 1.95$_{1.93}^{1.98}$  & 24.35$_{24.32}^{24.39}$  & $<$60.57 & -10.21$_{-10.21}^{-10.2}$  & -10.53$_{-10.55}^{-10.51}$  & -0.49$\pm$0.03\\[3pt] 
         & Clumpy  & 0.86 & 90.0 & 23.18* & $>$0.93 & 1.78$_{1.76}^{1.81}$  & 23.18$_{23.16}^{23.2}$  & 28.0$_{17.26}^{63.76}$  & -10.19$_{-10.19}^{-10.18}$  & -10.56$_{-10.59}^{-10.54}$  & -0.53$\pm$0.04\\[3pt] 
ESO103-G35 & Smooth$\circ$ & 1.04 & 87.0 & 23.46$_{23.45}^{23.47}$ & 0.93$_{0.93}^{0.94}$  & 2.07$_{2.05}^{2.09}$  & 24.34$_{24.32}^{24.36}$  & 0.0*  & -10.14$_{-10.15}^{-10.14}$  & -10.46$_{-10.48}^{-10.45}$  & -0.49$\pm$0.02\\[3pt] 
         & Clumpy  & 1.06 & 90.0 & 23.36* & 0.93$_{0.92}^{0.94}$  & 1.86$_{1.84}^{1.88}$  & 23.36$_{23.34}^{23.38}$  & 28.09$_{24.86}^{37.01}$  & -10.12$_{-10.12}^{-10.11}$  & -10.49$_{-10.5}^{-10.47}$  & -0.53$\pm$0.02\\[3pt]
MCG+07-41-03 & Smooth & 1.12 & 87.0 & 22.92* & 0.61$_{0.52}^{0.74}$  & 1.69$_{1.68}^{1.71}$  & 22.92$_{22.80}^{23.02}$  & 36.92$_{33.05}^{40.32}$ & -9.83$_{-9.83}^{-9.82}$  & -10.68$_{-10.71}^{-10.66}$  & -0.91$\pm$0.05 \\[3pt] 
         & Clumpy  & 1.18 & 1.0 & 23.0* & 0.6$_{0.57}^{0.7}$  & 1.74$_{1.71}^{1.77}$  & 23.05$_{23.00}^{23.14}$  & $> 7.40$  & -9.78$_{-9.80}^{-9.77}$  & -11.32$_{-11.93}^{-11.09}$  & -1.54$\pm$0.70\\[3pt] 
IC5063 & Smooth & 0.97 & 19.0 & 23.49* & 0.88$_{0.86}^{0.9}$  & 1.62$_{1.56}^{1.67}$  & 23.5$_{23.42}^{23.53}$  & 0.0*  & -10.29$_{-10.3}^{-10.28}$  & -11.04$_{-11.11}^{-10.98}$  & -0.82$\pm$0.12\\[3pt] 
         & Clumpy  & 0.93 & 45.0 & 23.5* & 0.92$_{0.91}^{0.94}$  & 1.88$_{1.82}^{1.92}$  & 23.5$_{23.46}^{23.54}$  & $>$16.53 & -10.4$_{-10.4}^{-10.39}$  & -10.74$_{-10.78}^{-10.71}$  & -0.51$\pm$0.05\\[3pt] 
PKS2356-61 & Smooth & 0.92 & 45.0 & 23.09* & $>$0.73 & 1.6$_{1.54}^{1.67}$  & 23.09$_{22.98}^{23.17}$  & 72.4*  & -10.95$_{-10.97}^{-10.92}$  & -11.76$_{-11.9}^{-11.65}$  & -0.87$\pm$0.24\\[3pt] 
         & Clumpy  & 0.99 & 90.0 & 23.31* & $>$0.78 & 1.81$_{1.67}^{1.92}$  & 23.31$_{23.23}^{23.4}$  & $>$0.11 & -10.95$_{-10.96}^{-10.94}$  & -11.93$_{-12.24}^{-11.75}$  & -1.02$\pm$0.46\\[3pt]
\end{longtable}
\tablefoot{Column(1): Source name. Column (2): X-ray baseline model used. The fit obtained using the baseline model with $N_{H_{tor}}$ independent to the $N_{H_{los}}$ and bat fits ($\chi^2/d.o.f. > 1.2$) are marked with white and black dots next to the baseline model name, respectively. Column(3): Reduced $\chi^2$ ($\chi^2/d.o.f.$). Columns(4-9): Final parameter values per model (see Table \ref{tab:ModelParam}). The asterisk next to the value indicate that the parameter is not constrained. Columns(10-11): Intrinsic continuum and reflection component fluxes. Column(12): Reflection fraction relative to the total flux.The confidence range of error calculated here is 1 sigma.}
\end{landscape}
}

\longtab{
\begin{landscape}
\begin{longtable}{lcccccccccc}
\caption{\label{tab:midIRParam}  Mid-IR best-fit results per object and model.}\\
\hline\hline
Objname & Baseline model &  A/S/I &  $\chi^2/d.o.f.$ & E$_{B-V}$ &   &   & Parameters &   &    &   \\[3pt]
  &   & \% &    &   &    &   &   &    &    &   \\[3pt]
  & Smooth &   &    &   & $i$  & $\sigma$ & $\gamma$ & $\beta$  & $Y$  & $\tau_{9.7}$ \\[3pt]
  & Clumpy &   &     &   & i  & $N_{0}$ & $\sigma$  & $Y$  & $q$  &  $\tau_{\nu}$  \\[3pt]
  (1) & (2) & (3) & (4) & (5) & (6) &  (7) & (8) & (9) & (10) & (11) \\[3pt]
\hline
\endfirsthead
\caption{continued.}\\
\hline\hline
Objname & Baseline model &  A/S/I &  $\chi^2/d.o.f.$ & E$_{B-V}$ &   &   & Parameters &   &    &   \\[3pt]
  &   & \% &    &   &    &   &   &    &    &   \\[3pt]
  & Smooth &   &    &   & $i$  & $\sigma$ & $\gamma$ & $\beta$  & $Y$  & $\tau_{9.7}$ \\[3pt]
  & Clumpy &   &     &   & i  & $N_{0}$ & $\sigma$  & $Y$  & $q$  & $\tau_{\nu}$  \\[3pt]
    (1) & (2) & (3) & (4) & (5) & (6) &  (7) & (8) & (9) & (10) & (11) \\[3pt]
\hline
\endhead
\hline
\endfoot
  Mrk1018                & Smooth$\circ$ &  87.6/ 12.4/  0.0 &   0.69 & $<$0.5 & $<$0.0 & $<$21.4 & 5.7$_{5.1}^{5.8}$ & $>$-0.0 & $<$10.1 & 5.6$_{4.8}^{6.4}$   \\[3pt]
                       & Clumpy$\circ$ &  88.2/ 11.8/  0.0 &   0.46 & $<$0.5 & $>$77.3 & 3.3$_{2.5}^{4.0}$ & $<$63.4 & 19.7$_{12.1}^{25.1}$ & $>$2.2 & 41.3$_{31.7}^{48.2}$   \\[3pt]
Mrk590                 & Smooth$\circ\bullet$       & 100.0/  0.0/  0.0 &   1.31 & $<$0.5 & 14.9$_{12.8}^{17.2}$ & $<$20.2 & $>$6.0 & $>$-0.0 & 21.1$_{20.7}^{21.3}$ & $>$9.8   \\[3pt]
                       & Clumpy$\circ$ &  96.5/  3.5/  0.0 &   0.23 & $<$0.5 & $>$34.3 & 4.2$_{3.5}^{8.2}$ & $>$31.7 & 10.1$_{9.6}^{13.6}$ & 1.5$_{1.3}^{1.6}$ & 135.0$_{66.3}^{149.2}$   \\[3pt]
PG0804+761             & Smooth$\circ$ &  96.8/  3.2/  0.0 &   0.76 & $<$0.5 & $<$0.1 & $<$21.1 & 4.0$_{3.97}^{5.03}$ & -0.75$_{-0.85}^{-0.7}$ & 11.44$_{11.35}^{11.55}$ & 0.77$_{0.76}^{1.03}$   \\[3pt]
                       & Clumpy$\circ\bullet$ &  85.6/ 10.4/  4.0 &   1.63 & $<$0.5 & $>$76.3 & $<$1.2 & $<$64.4 & $>$40.3 & 2.29$_{2.23}^{2.35}$ & 67.0$_{54.8}^{86.8}$   \\[3pt]
RBS0770                & Smooth$\circ$ &  94.7/  5.3/  0.0 &   0.44 & $<$0.5 & $<$1.9 & $<$20.4 & $>$6.0 & -0.2$_{-0.3}^{-0.1}$ & $<$10.1 & 7.0$_{6.5}^{7.6}$   \\[3pt] 
                       & Clumpy$\circ$ &  93.9/  6.1/  0.0 &   0.47 & $<$0.5 & $>$73.7 & 5.0$_{3.8}^{6.0}$ & $<$57.4 & 10.0$_{9.3}^{10.7}$ & 1.6$_{1.4}^{1.8}$ & 28.0$_{20.9}^{37.5}$   \\[3pt]
I11119+3257         & Smooth$\circ\bullet$ & 100.0/  0.0/  0.0 &   1.20 & 0.3$_{0.1}^{0.4}$ & $<$2.5 & $>$59.3 & 0.08$_{0.05}^{0.12}$ & $>$-0.0 & 115.9$_{108.4}^{123.9}$ & 2.3$_{2.2}^{2.4}$   \\[3pt]
                       & Clumpy$\bullet$  &  62.0/  6.6/ 31.4 &   2.54 & 0.6$_{0.6}^{0.7}$ & $<$0.0 & $>$14.9 & 48.1$_{47.7}^{49.1}$ & 80.0$_{78.4}^{81.2}$ & 0.2$_{0.1}^{0.5}$ & 80.0$_{77.8}^{82.6}$   \\[3pt]
PG1211+143             & Smooth$\circ$ &  94.9/  5.1/  0.0 &   1.01 & $<$0.5 & 25.3$_{24.0}^{35.3}$ & $<$20.3 & $>$6.0 & -0.5$_{-0.51}^{-0.49}$ & $<$10.2 & $>$6.5   \\[3pt]
                       & Clumpy$\circ\bullet$ &  92.9/  7.1/  0.0 &   1.25 & $<$0.5 & $>$84.0 & 3.0$_{2.9}^{3.4}$ & 31.8$_{16.1}^{41.7}$ & 20.0$_{18.9}^{21.2}$ & 1.82$_{1.76}^{1.89}$ & 39.7$_{35.3}^{42.6}$   \\[3pt]
RBS1125                & Smooth$\circ\bullet$       & 100.0/  0.0/  0.0 &   1.63 & $<$0.5 & 66.5$_{51.2}^{68.1}$ & 22.6$_{21.1}^{32.8}$ & 0.25$_{0.2}^{0.35}$ & -0.7$_{-0.8}^{-0.6}$ & $<$10.1 & 6.0$_{5.7}^{6.1}$   \\[3pt]
                       & Clumpy$\circ$ &  93.5/  6.5/  0.0 &   0.51 & $<$0.5 & $>$85.1 & 6.2$_{5.7}^{7.3}$ & $<$44.5 & 20.9$_{19.2}^{23.5}$ & 1.1$_{1.0}^{1.3}$ & 20.0$_{17.1}^{22.4}$   \\[3pt]
Mrk231                 & Smooth$\circ\bullet$       & 100.0/  0.0/  0.0 &   2.54 & 0.5$_{0.4}^{0.5}$ & $<$0.0 & 35.8$_{34.9}^{36.5}$ & 1.98$_{1.68}^{2.01}$ & $>$-0.0 & 117.8$_{113.3}^{121.6}$ & 2.3$_{2.26}^{2.36}$   \\[3pt]
                       & Clumpy$\bullet$ &  78.0/  2.2/ 19.8 &   2.46 & 0.9$_{0.8}^{1.0}$ & $<$0.0 & $>$14.9 & 48.1$_{46.2}^{48.6}$ & 48.4$_{46.8}^{50.9}$ & 0.5$_{0.37}^{0.51}$ & 78.7$_{69.1}^{81.0}$   \\[3pt]
Mrk1383                & Smooth$\circ$ &  82.2/  7.3/ 10.6 &   0.83 & $<$0.5 & $<$0.0 & $<$20.8 & $>$5.9 & $>$-0.0 & 12.4$_{12.1}^{12.5}$ & 7.1$_{6.9}^{7.6}$   \\[3pt]
                       & Clumpy$\circ$ &  85.0/  7.8/  7.3 &   0.42 & $<$0.5 & $>$73.2 & 3.1$_{2.8}^{3.4}$ & $<$62.9 & $>$55.1 & $>$2.4 & 75.7$_{63.8}^{86.3}$   \\[3pt]
Mrk1392                & Smooth$\circ\bullet$       &  97.9/  2.1/  0.0 &   1.69 & $<$0.5 & 16.0$_{14.9}^{20.0}$ & $<$20.4 & $>$6.0 & $>$-0.0 & 22.2$_{21.9}^{23.0}$ & $>$9.6   \\[3pt]
                       & Clumpy$\circ$ &  91.7/  2.4/  5.8 &   0.68 & $<$0.5 & 83.9$_{77.1}^{87.2}$ & $>$13.6 & $<$17.9 & 14.1$_{13.4}^{15.3}$ & $<$0.1 & 12.7$_{10.8}^{14.3}$   \\[3pt]
Mrk1393                & Smooth$\circ\bullet$       & 100.0/  0.0/  0.0 &   2.30 & $<$0.5 & 65.9$_{64.7}^{67.5}$ & $<$20.6 & $<$0.0 & $>$-0.0 & $<$10.1 & 6.0$_{5.8}^{6.3}$   \\[3pt]
                       & Clumpy$\circ$ &  95.4/  4.6/  0.0 &   0.62 & $<$0.5 & 79.3$_{78.3}^{85.4}$ & $>$14.1 & $<$25.3 & 20.0$_{17.3}^{20.3}$ & 0.7$_{0.4}^{0.8}$ & $<$10.6   \\[3pt]
PG1535+547            & Smooth$\circ$ &  83.8/  8.1/  8.1 &   0.19 & $<$0.5 & $<$1.7 & $<$25.4 & 3.8$_{2.8}^{4.3}$ & $<$-1.0 & $<$10.3 & 4.2$_{3.2}^{4.8}$   \\[3pt]
                       & Clumpy$\circ$ &  87.0/  7.5/  5.5 &   0.58 & 0.7$_{0.6}^{0.8}$ & $>$83.0 & 3.0$_{2.7}^{3.1}$ & 54.4$_{47.4}^{67.8}$ & 14.5$_{12.5}^{17.3}$ & $>$2.5 & 40.1$_{38.4}^{41.7}$   \\[3pt]
ESO141-G055            & Smooth$\circ$ &  88.0/  8.1/  3.8 &   0.99 & $<$0.5 & 44.7$_{40.4}^{47.0}$ & $<$21.6 & $>$5.9 & -0.6$_{-0.7}^{-0.5}$ & 11.3$_{10.9}^{11.5}$ & $>$9.1   \\[3pt]
                       & Clumpy$\circ$ &  90.1/  9.9/  0.0 &   0.97 & $<$0.5 & $>$87.3 & 3.1$_{2.7}^{3.4}$ & $<$36.9 & $>$67.0 & 2.11$_{2.06}^{2.17}$ & 50.4$_{44.1}^{58.6}$   \\[3pt]
NGC7213                & Smooth$\circ\bullet$       &  95.6/  4.4/  0.0 &   8.39 & $<$0.5 & 24.0$_{23.1}^{25.9}$ & $<$20.1 & $>$6.0 & $>$-0.0 & 16.1$_{15.8}^{16.6}$ & $>$9.9   \\[3pt]
                       & Clumpy$\circ$ &  97.0/  3.0/  0.0 &   1.08 & 0.8$_{0.7}^{0.8}$ & $<$1.5 & 7.0$_{5.6}^{7.3}$ & $<$15.1 & 10.2$_{10.0}^{10.6}$ & $<$0.0 & 55.8$_{50.6}^{59.2}$   \\[3pt]
MCG+01-57-016          & Smooth$\bullet$  &  76.0/  3.7/ 20.3 &   1.78 & $<$0.5 & 15.3$_{14.7}^{16.8}$ & $<$21.9 & $>$5.9 & $>$-0.0 & 37.3$_{36.6}^{38.0}$ & $>$9.0   \\[3pt]
                       & Clumpy     &  62.2/  4.7/ 33.1 &   1.00 & $<$0.5 & $<$81.5 & $>$6.7 & 42.6$_{29.3}^{67.7}$ & 10.6$_{9.3}^{12.0}$ & $<$1.5 & 198.9$_{167.8}^{259.6}$   \\[3pt]
UM146                  & Smooth$\circ$ & 100.0/  0.0/  0.0 &   0.92 & $<$0.5 & $<$0.0 & 55.4$_{49.5}^{57.1}$ & $>$5.4 & $>$-0.1 & 36.7$_{34.5}^{40.9}$ & 3.8$_{3.3}^{4.0}$   \\[3pt] 
                       & Clumpy$\circ$ & 100.0/  0.0/  0.0 &   0.56 & 0.8$_{0.7}^{0.9}$ & $>$52.1 & 4.8$_{4.4}^{5.1}$ & $>$55.4 & $>$95.9 & 1.26$_{1.21}^{1.34}$ & 24.7$_{22.6}^{28.4}$   \\[3pt] 
NGC788                 & Smooth$\circ\bullet$       & 100.0/  0.0/  0.0 &   2.45 & $<$0.5 & 9.9$_{5.5}^{10.1}$ & $<$20.2 & 5.64$_{5.57}^{5.7}$ & $>$-0.0 & 16.8$_{16.6}^{17.2}$ & $>$10.0 \\[3pt]
                       & Clumpy$\circ$ &  89.8/  3.7/  6.5 &   0.63 & $<$0.5 & $>$61.1 & 10.4$_{9.6}^{12.1}$ & $>$46.8 & 20.0$_{17.4}^{23.7}$ & $>$2.4 & 43.3$_{40.6}^{46.5}$   \\[3pt]
NGC1052                & Smooth$\circ\bullet$       &  95.8/  4.2/  0.0 &   2.53 & $<$0.5 & 22.4$_{21.7}^{22.9}$ & $<$20.4 & $>$5.9 & $>$-0.0 & 27.9$_{27.4}^{28.4}$ & $>$9.7   \\[3pt] 
                       & Clumpy$\circ$ &  82.1/  5.3/ 12.5 &   0.79 & $<$0.5 & $>$39.0 & 3.5$_{3.3}^{4.6}$ & $>$46.6 & 10.0$_{9.8}^{10.3}$ & 1.0$_{0.7}^{1.1}$ & 199.9$_{178.6}^{218.0}$   \\[3pt]
NGC1358                & Smooth$\circ$ & 100.0/  0.0/  0.0 &   0.22 & $<$0.5 & $<$42.1 & $<$58.6 & $<$0.2 & $<$-0.4 & 59.4$_{35.3}^{94.5}$ & $>$2.2   \\[3pt] 
                       & Clumpy$\circ$ & 100.0/  0.0/  0.0 &   0.23 & 3.4$_{2.6}^{3.9}$ & $<$34.4 & 7.7$_{5.6}^{10.3}$ & $>$15.0 & $>$80.3 & 0.9$_{0.6}^{1.1}$ & $<$16.0   \\[3pt] 
J05081967+1721483 & Smooth$\circ\bullet$   &  81.0/  0.0/ 19.0 &   1.66 & $<$0.5 & 20.0$_{19.5}^{20.4}$ & $<$21.4 & 4.17$_{4.12}^{4.21}$ & $<$-1.0 & 127.5$_{121.0}^{133.4}$ & 8.5$_{8.0}^{8.9}$   \\[3pt] 
                       & Clumpy &  65.3/  1.9/ 32.8 &   0.97 & 0.7$_{0.6}^{0.8}$ & $<$14.6 & 4.3$_{3.1}^{6.7}$ & 53.3$_{41.0}^{61.5}$ & $>$95.9 & 1.1$_{1.0}^{1.2}$ & $>$139.7   \\[3pt]
Mrk3                   & Smooth$\circ\bullet$       & 100.0/  0.0/  0.0 &   3.11 & $<$0.5 & 10.0$_{8.8}^{10.4}$ & $<$20.1 & $>$6.0 & $>$-0.0 & 29.5$_{29.2}^{29.9}$ & $>$9.9\\[3pt]
                       & Clumpy$\circ\bullet$ & 100.0/  0.0/  0.0 &   1.38 & $<$0.5 & 75.5$_{74.0}^{88.0}$ & $>$14.3 & 42.5$_{41.6}^{46.6}$ & 20.0$_{19.8}^{20.1}$ & 0.5$_{0.43}^{0.52}$ & $<$10.1 \\[3pt] 
ESO428-G014            & Smooth$\circ$ & 100.0/  0.0/  0.0 &   1.02 & 0.4$_{0.3}^{0.5}$ & 9.9$_{5.4}^{10.4}$ & $<$27.5 & 5.6$_{2.7}^{5.8}$ & $>$-0.0 & 33.4$_{32.8}^{34.0}$ & $>$9.6 \\[3pt] 
                       & Clumpy$\circ$ & 100.0/  0.0/  0.0 &   0.57 & 0.5$_{0.4}^{0.6}$ & $>$57.8 & $>$9.8 & $<$26.4 & 35.0$_{32.2}^{39.4}$ & 1.5$_{1.3}^{1.6}$ & 26.1$_{22.1}^{33.3}$   \\[3pt]
Mrk78                  & Smooth$\circ\bullet$       &  87.9/  1.4/ 10.7 &   1.69 & $<$0.5 & 0.8$_{0.2}^{4.3}$ & $<$20.7 & 5.22$_{5.17}^{5.3}$ & $>$-0.0 & 30.0$_{29.8}^{30.4}$ & $>$9.9   \\[3pt]
                       & Clumpy$\circ$ &  82.9/  3.6/ 13.5 &   0.82 & 0.5$_{0.4}^{0.5}$ & $>$68.2 & $>$13.9 & $>$55.7 & $>$48.3 & $>$2.4 & 53.6$_{47.7}^{56.6}$   \\[3pt]
Mrk1210                & Smooth$\circ\bullet$       & 100.0/  0.0/  0.0 &   2.40 & $<$0.5 & 12.5$_{12.2}^{13.1}$ & $<$20.1 & $>$6.0 & $>$-0.0 & 29.9$_{29.7}^{30.1}$ & $>$9.8   \\[3pt]
                       & Clumpy$\circ\bullet$ & 100.0/  0.0/  0.0 &   1.21 & $<$0.5 & 67.9$_{63.2}^{75.2}$ & 12.3$_{11.5}^{13.1}$ & 50.3$_{37.8}^{61.9}$ & 22.6$_{22.0}^{23.1}$ & 0.5$_{0.33}^{0.51}$ & $<$10.1   \\[3pt]
J10594361+6504063 & Smooth$\circ$ & 100.0/  0.0/  0.0 &   0.24 & 0.6$_{0.4}^{0.7}$ & 74.6$_{71.1}^{78.1}$ & $<$22.1 & $<$0.0 & $>$-0.1 & $<$10.6 & $>$8.3   \\[3pt]
                       & Clumpy$\circ$ &  83.3/  6.7/ 10.0 &   0.22 & 0.7$_{0.4}^{0.9}$ & 35.8$_{27.9}^{42.5}$ & $>$14.7 & $>$68.9 & 11.0$_{10.6}^{11.5}$ & $<$0.0 & 20.1$_{19.7}^{21.2}$   \\[3pt] 
NGC4388                & Smooth$\circ\bullet$       &  83.6/  0.0/ 16.4 &   2.35 & $<$0.5 & 52.4$_{51.3}^{53.2}$ & $<$20.0 & $<$0.0 & $>$-0.1 & 30.0$_{29.5}^{30.1}$ & 7.8$_{7.6}^{7.9}$   \\[3pt] 
                       & Clumpy$\circ\bullet$ &  98.2/  1.8/  0.0 &   1.75 & 0.9$_{0.8}^{1.0}$ & $>$80.2 & 11.4$_{11.1}^{11.7}$ & $>$67.0 & 52.6$_{51.1}^{54.5}$ & 0.5$_{0.46}^{0.53}$ & $<$10.1   \\[3pt] 
NGC4507                & Smooth$\bullet$   &  79.6/  4.1/ 16.4 &   2.51 & $<$0.5 & 12.9$_{10.9}^{14.0}$ & $<$20.8 & $>$6.0 & $>$-0.0 & 21.1$_{20.7}^{21.7}$ & 7.3$_{6.7}^{7.8}$   \\[3pt] 
                       & Clumpy$\circ$ &  83.2/  4.7/ 12.1 &   0.99 & $<$0.5 & $>$54.7 & 3.1$_{3.0}^{3.5}$ & $>$47.6 & $>$75.3 & 2.0$_{1.96}^{2.05}$ & 70.8$_{64.8}^{76.6}$   \\[3pt]
NGC4939                & Smooth$\circ\bullet$       &  98.7/  1.3/  0.0 &   2.06 & $<$0.5 & $<$0.0 & $<$20.2 & $>$6.0 & $>$-0.0 & 32.9$_{32.2}^{33.3}$ & 7.7$_{7.2}^{8.1}$   \\[3pt] 
                       & Clumpy$\circ$ &  90.9/  3.3/  5.9 &   0.85 & $<$0.5 & $<$18.3 & $>$8.2 & $>$54.2 & 10.0$_{9.6}^{10.1}$ & $<$0.2 & $>$216.6   \\[3pt]
ESO-097-G013           & Smooth$\bullet$    &  72.8/  0.0/ 27.2 &  32.01 & $<$0.5 & $<$12.7 & $<$20.1 & $<$0.0 & $>$-0.0 & $<$10.0 & 6.8$_{6.7}^{6.9}$   \\[3pt] 
                       & Clumpy$\circ\bullet$       & 100.0/  0.0/  0.0 &  26.71 & 2.2$_{2.2}^{2.2}$ & $<$13.9 & 13.0$_{12.3}^{13.5}$ & $>$69.0 & $>$99.3 & 2.1$_{2.0}^{2.2}$ & 23.6$_{22.6}^{24.6}$   \\[3pt] 
IC4518W                & Smooth$\bullet$             &  74.8/  0.0/ 25.2 &   1.37 & 1.4$_{1.2}^{1.5}$ & $<$1.2 & 48.7$_{44.2}^{51.8}$ & 4.0$_{3.8}^{4.1}$ & $>$-0.0 & 36.6$_{34.5}^{38.7}$ & 6.0$_{5.8}^{6.2}$   \\[3pt] 
                       & Clumpy &  73.6/  2.4/ 24.0 &   0.88 & 3.1$_{2.9}^{3.3}$ & $>$38.5 & 8.3$_{6.5}^{12.1}$ & $>$48.1 & $>$96.5 & 1.67$_{1.59}^{1.74}$ & 27.2$_{22.6}^{46.9}$   \\[3pt]
ESO138-G1              & Smooth$\circ\bullet$    &  90.5/  2.1/  7.3 &   1.67 & $<$0.5 & 11.5$_{5.9}^{13.1}$ & $<$20.2 & $>$6.0 & $>$-0.0 & 15.0$_{14.8}^{15.2}$ & $>$9.6   \\[3pt]
                       & Clumpy$\circ$ &  87.7/  0.0/ 12.3 &   0.67 & 0.8$_{0.6}^{0.8}$ & $<$2.3 & 5.5$_{5.2}^{6.1}$ & $>$66.8 & 11.0$_{10.9}^{11.3}$ & $<$0.0 & 19.0$_{14.6}^{20.1}$   \\[3pt] 
NGC6300                & Smooth$\bullet$     &  75.4/  0.0/ 24.6 &   3.73 & 2.1$_{2.0}^{2.2}$ & $<$0.0 & 50.6$_{48.2}^{51.9}$ & $>$5.9 & $>$-0.0 & 21.1$_{20.4}^{21.6}$ & $>$9.7   \\[3pt]
                       & Clumpy$\bullet$  &  78.8/  0.0/ 21.2 &   1.90 & 4.3$_{4.1}^{4.3}$ & $<$12.0 & 10.2$_{9.1}^{11.1}$ & $<$32.5 & $>$97.0 & 1.78$_{1.73}^{1.79}$ & $>$290.0   \\[3pt] 
ESO103-G35             & Smooth$\circ\bullet$       &  85.4/  0.0/ 14.6 &   3.57 & $<$0.5 & 60.0$_{59.8}^{60.1}$ & $<$20.0 & $<$0.0 & $>$-0.0 & $<$10.0 & 7.4$_{7.3}^{7.5}$   \\[3pt] 
                       & Clumpy$\circ$ &  97.7/  2.3/  0.0 &   0.53 & 1.2$_{1.2}^{1.3}$ & $>$67.3 & 8.8$_{8.7}^{11.8}$ & $<$46.6 & 28.1$_{24.6}^{35.7}$ & 2.2$_{2.1}^{2.3}$ & 46.8$_{41.9}^{50.7}$   \\[3pt]
MCG+07-41-03       & Smooth$\circ\bullet$       &  93.9/  0.0/  6.1 &   5.14 & 0.3$_{0.2}^{0.3}$ & $<$0.0 & 22.5$_{21.2}^{24.6}$ & $>$6.0 & $>$-0.0 & 42.7$_{41.8}^{43.2}$ & $>$9.9   \\[3pt] 
                       & Clumpy$\circ\bullet$ & 100.0/  0.0/  0.0 &   1.48 & 1.3$_{1.3}^{1.4}$ & $<$14.1 & 7.0$_{6.5}^{7.8}$ & $>$62.9 & $>$97.3 & 1.85$_{1.83}^{1.88}$ & 177.2$_{150.5}^{201.9}$   \\[3pt]
IC5063                 & Smooth$\circ\bullet$       &  89.3/  0.0/ 10.7 &   2.01 & $<$0.5 & $<$0.4 & 23.4$_{22.5}^{24.5}$ & $>$6.0 & $>$-0.0 & 25.4$_{24.7}^{26.1}$ & 8.5$_{8.0}^{8.8}$   \\[3pt]
                       & Clumpy$\circ$ &  93.8/  0.0/  6.2 &   0.61 & 0.6$_{0.5}^{0.6}$ & $<$30.0 & $>$8.8 & 38.2$_{29.9}^{49.7}$ & 16.4$_{14.1}^{18.6}$ & 0.8$_{0.7}^{1.0}$ & 95.3$_{79.0}^{112.3}$   \\[3pt]
PKS2356-61             & Smooth$\circ$ &  98.3/  1.7/  0.0 &   0.33 & 1.0$_{0.9}^{1.1}$ & 24.8$_{21.3}^{28.2}$ & $<$21.7 & $>$5.8 & $>$-0.0 & 27.9$_{26.7}^{29.4}$ & $>$8.6   \\[3pt]
                       & Clumpy$\circ$ & 100.0/  0.0/  0.0 &   0.31 & 1.2$_{1.1}^{1.3}$ & 49.8$_{34.1}^{66.6}$ & 5.4$_{5.2}^{10.4}$ & 35.0$_{23.4}^{53.1}$ & 18.4$_{15.8}^{21.1}$ & $<$0.9 & 40.0$_{28.3}^{51.0}$   \\[3pt]
\hline
\end{longtable}
\tablefoot{Column(1): Source name. Column(2): Mid-IR baseline model used. Column(3):Percentage contribution to the $5 - 30\, \mu m$ waveband per component (A: AGN; S: stellar; and I: Interstellar medium). Column(4): Reduced $\chi^2$ ($\chi^2/d.o.f.$). Column(5): Color excess for the foreground extinction. Column(6-11): Final parameter values per model (see Table \ref{tab:ModelParam}). The confidence range of error calculated here is 1 sigma. AGN-dominated spectra in more than 80$\%$ and the bad-fits ($\chi^2/d.o.f. > 1.2$) are marked with white and black dots next to the model name, respectively.}
\end{landscape}
}

\begin{appendix} 
\section{Disk reflection at X-rays}
\label{app:diskreflection}
In order to study the existence of disk reflection emission in our sample, we added the {\sc relxill} to the baseline model used in Eq.\,\ref{Eq:BaselineXray} (Section\,\ref{Sec:sample}). We use {\sc borus} as the distant reflector component from the torus. Thus, the baseline model is:

\begin{equation}
    M = zpcfabs \times cutoffpl + ({\sc borus02} + {\sc relxill})
    \label{Eq:BaselineXrayDisk}
\end{equation}

We also used an alternative scenario where the disk reflector is affected by the partial covering, which can be written as follows:

\begin{equation}
    M  = zpcfabs \times (cutoffpl + {\sc relxill}) + {\sc borus02}
    \label{Eq:BaselineXrayDisk2}
\end{equation}

We then compare statistically the best fit obtained using these two baseline models with that obtained without the disk reflection component. When the f-test probability is bellow $\rm{10^{-4}}$ we consider that the disk reflection component is required by the data.

The initial sample includes the 47 AGN with more than 10\% of reflection emission (see Section\,\ref{Sec:sample}). Using the baseline model included in equation \ref{Eq:BaselineXrayDisk}, we found that the disk reflection component is required in 14 out of these 47 AGN. When we used the baseline model included in equation\,\ref{Eq:BaselineXrayDisk2}, the disk reflection is statistically needed for 12 AGN. Ten objects require reflection irrespective of the baseline model used, another four AGN show signatures of the disk reflection if equation \ref{Eq:BaselineXrayDisk} is used, and another two if equation \ref{Eq:BaselineXrayDisk2} is used.

Among these 16 objects, five of these sources (ESO\,103-G035, ESO138-G1, MCG\,+07-41-03, ESO\,-097-G013 and NGC\,6300) show a resulting model not physically consistent with their type-2 AGN classification. In particular, the model preferred for MCG\,+07-41-03 has 95\% of {\sc borus}, 5\% of {\sc relxill}, and no intrinsic component (and only using equation\,\ref{Eq:BaselineXrayDisk}). A similar case is found for ESO\,138-G1 where disk reflection is needed only using  equation\,\ref{Eq:BaselineXrayDisk} and lack of intrinsic continuum. Meanwhile, the best fit obtained for ESO\,103-G035, ESO\,-097-G013, and NGC\,6300 show only {\sc relxill} component without any contribution of {\sc borus} component. In fact, ESO\,-097-G013 is one of the objects where the disk reflection is needed only when using equation\,\ref{Eq:BaselineXrayDisk} and NGC\,6300 requires it only when using equation\,\ref{Eq:BaselineXrayDisk2}. \citet{Arevalo14} found that there is an extended reflection component in ESO\,-097-G013 that might explain the complex behaviour without claiming the existence of disk reflection. Indeed, ESO\,-097-G013 and ESO\,103-G035 have already been reproduced with the {uxclumpy} model by \citet{Buchner19}. 

Thus we excluded from this analysis the nine Sy1s (FAIRALL\,9, Ark,120, NGC\,3783, MCG-06-30-15, IC\,4329A, NGC\,5548, Mrk\,841, NGC\,6814, and NGC\,7314) and two Sy2s (MCG-05-23-16 and IRAS\,13197-1627) with significant disk reflection. All of them show a contribution to the 3-70\,keV of the {\sc relxill} component in the range between $\rm{\sim}$15-40\% and less than $\rm{\sim}$5\% of {\sc borus} component (after including disk reflection). Reinforcing this analysis, disk reflection is being reported in the literature for eight out of these nine Sy1s: \citep{Lohfink12,Porquet17,Patrick12,Brenneman11,Walton18,Kammoun17,Done00,Reynolds14,Walton13,Ebrero11}. The exception is NGC\,5548. Note that interestingly among these nine Sy1s, three (NGC\,3783, Fairall\,9, Mrk\,841) require disk reflection using equation\,\ref{Eq:BaselineXrayDisk} but did not require it using equation\,\ref{Eq:BaselineXrayDisk2}. However, we kept them out of the sample since disk reflection has been reported in the literature for them (see above references). Furthermore, none of the other AGN in our sample have disk reflection reported in the literature.

\section{Poor spectral fits}
\label{sec:app2}

\begin{figure*}
    \centering
    \includegraphics[width=0.45\columnwidth]{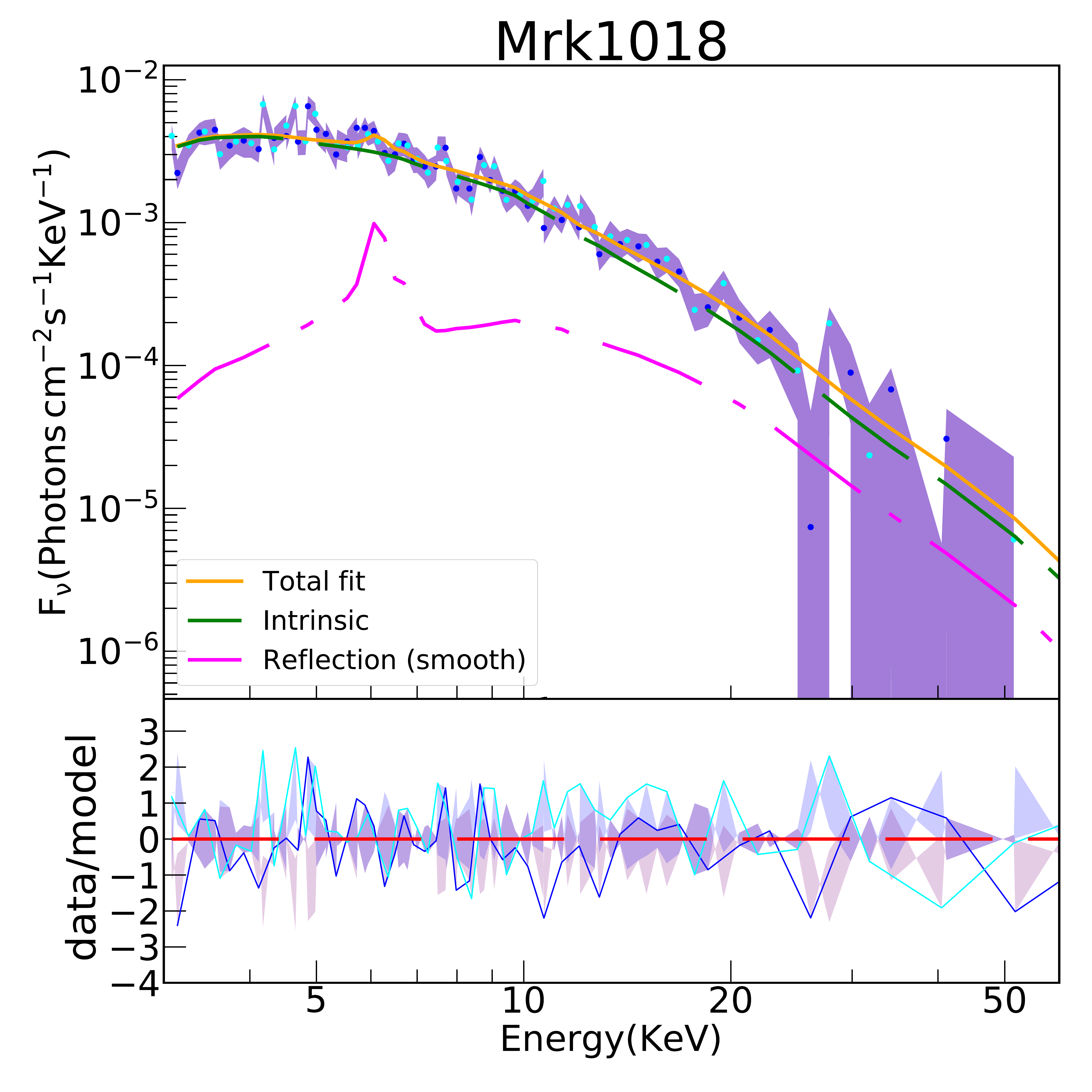}
    \includegraphics[width=0.45\columnwidth]{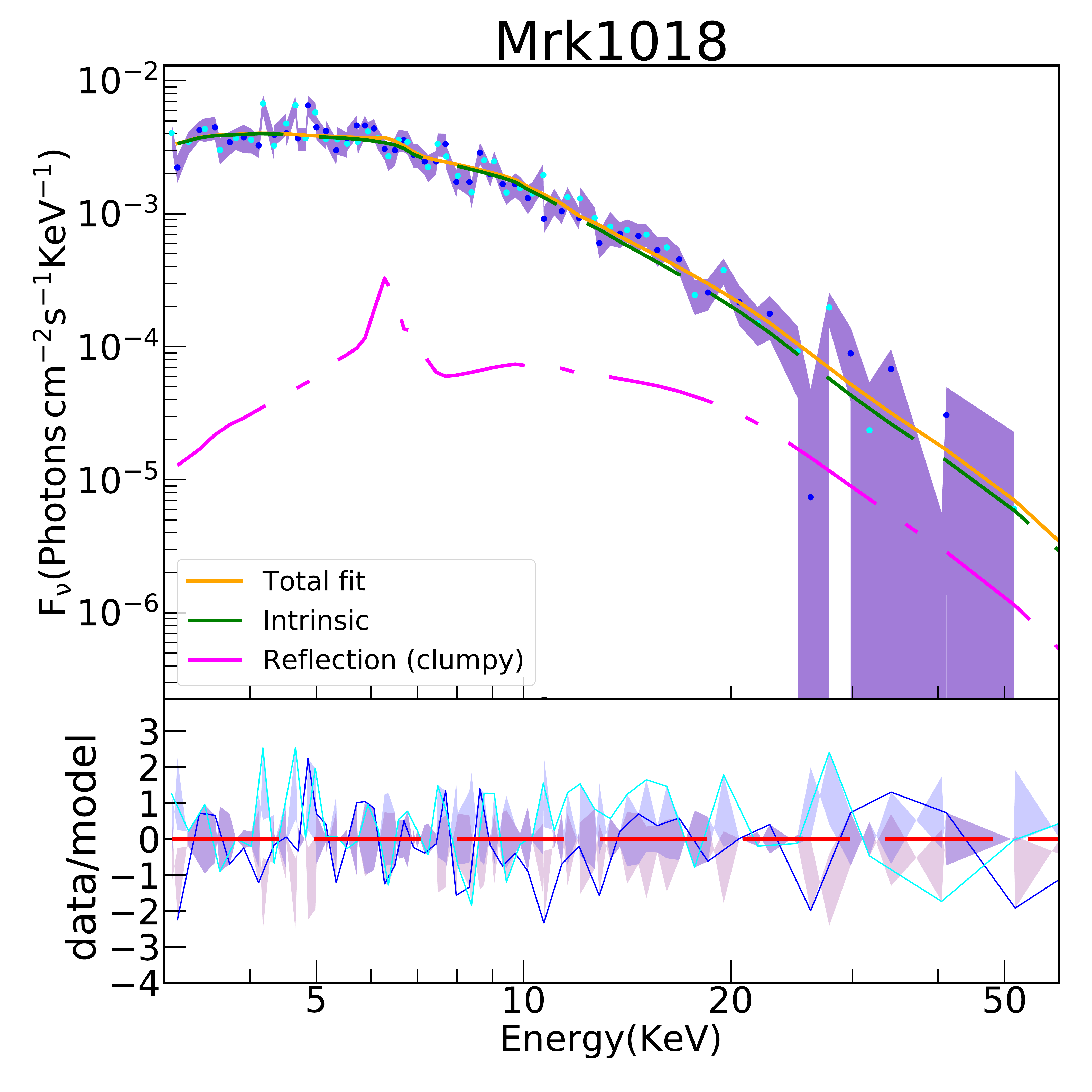}
    \includegraphics[width=0.45\columnwidth]{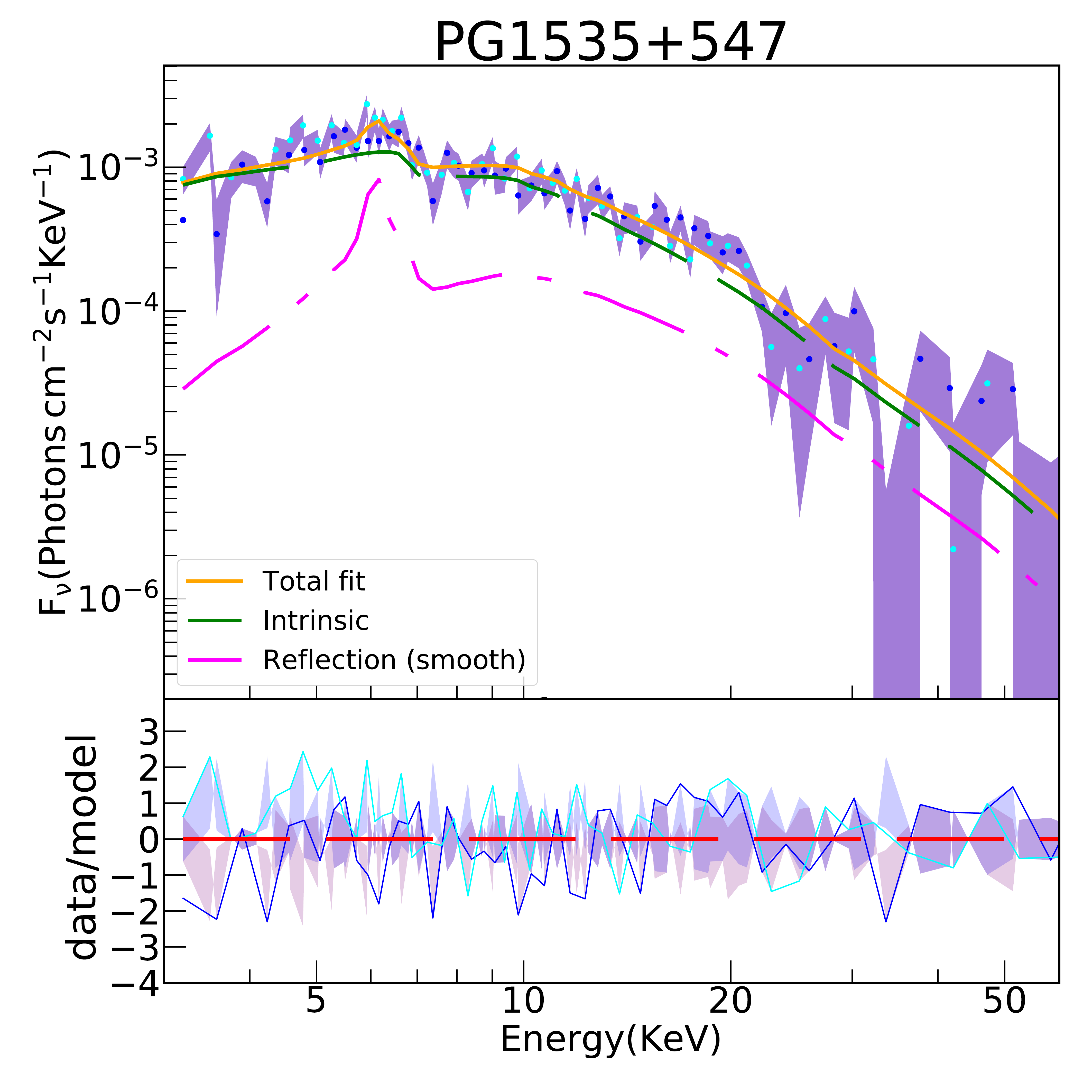}
    \includegraphics[width=0.45\columnwidth]{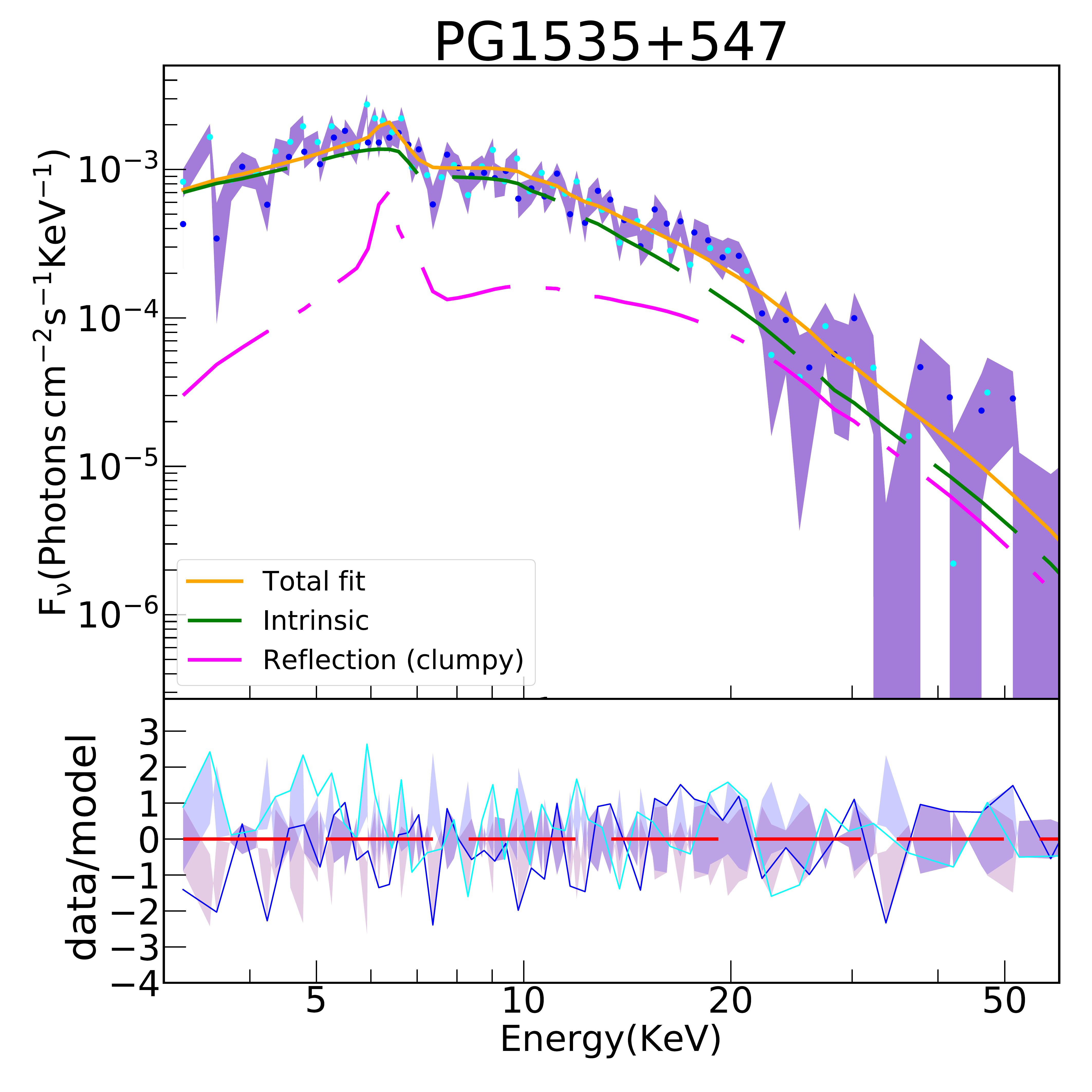}
    \includegraphics[width=0.45\columnwidth]{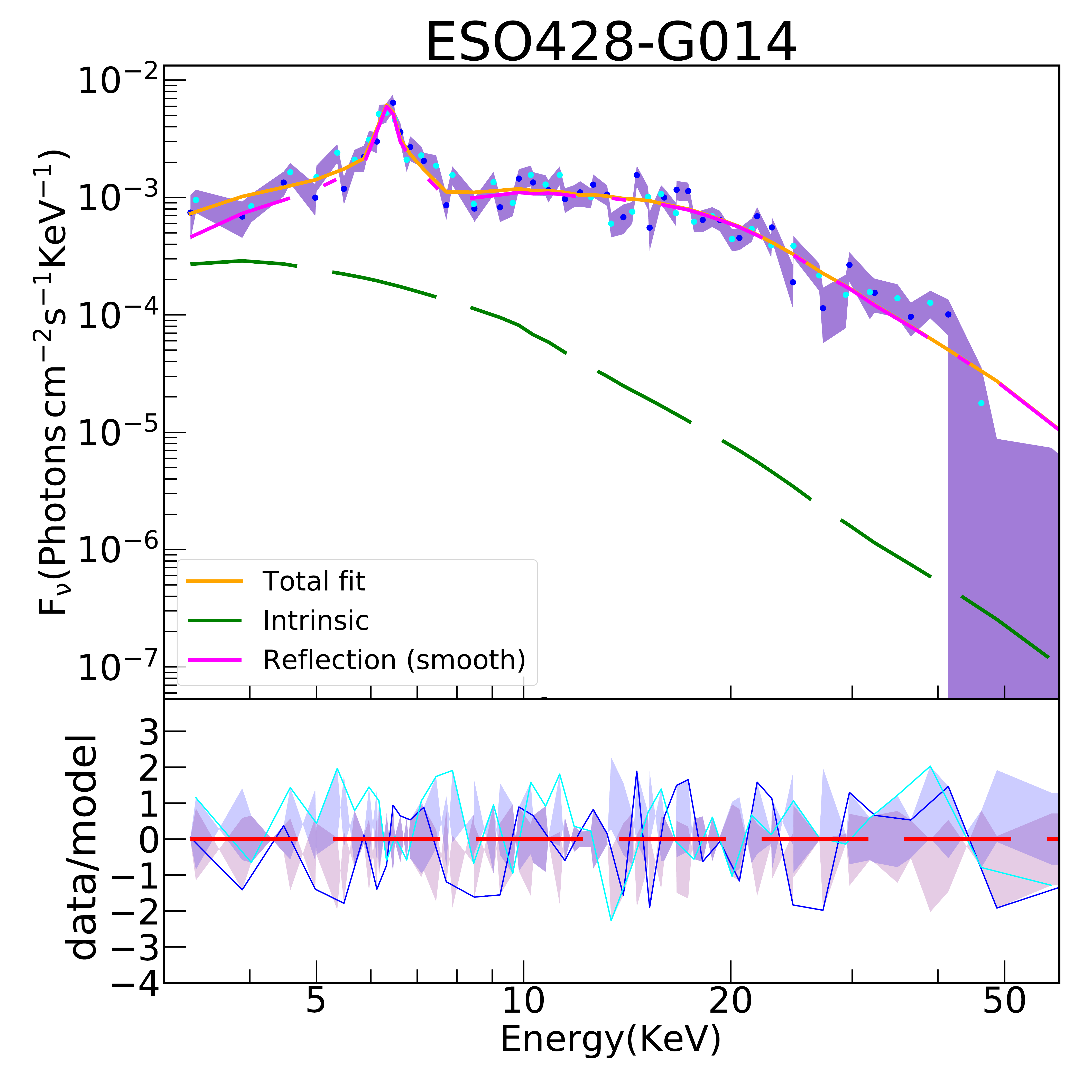}
     \includegraphics[width=0.45\columnwidth]{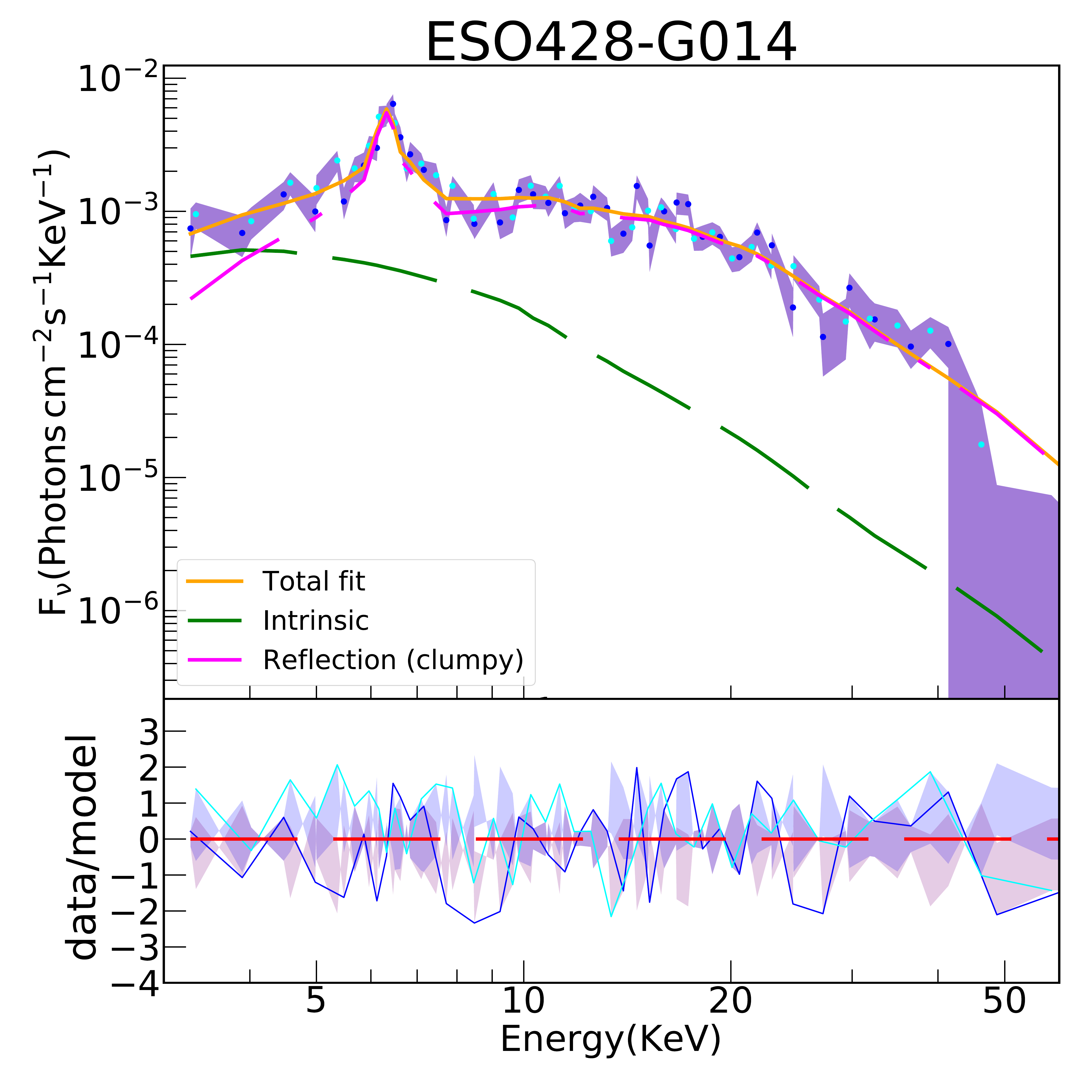}
     \includegraphics[width=0.45\columnwidth]{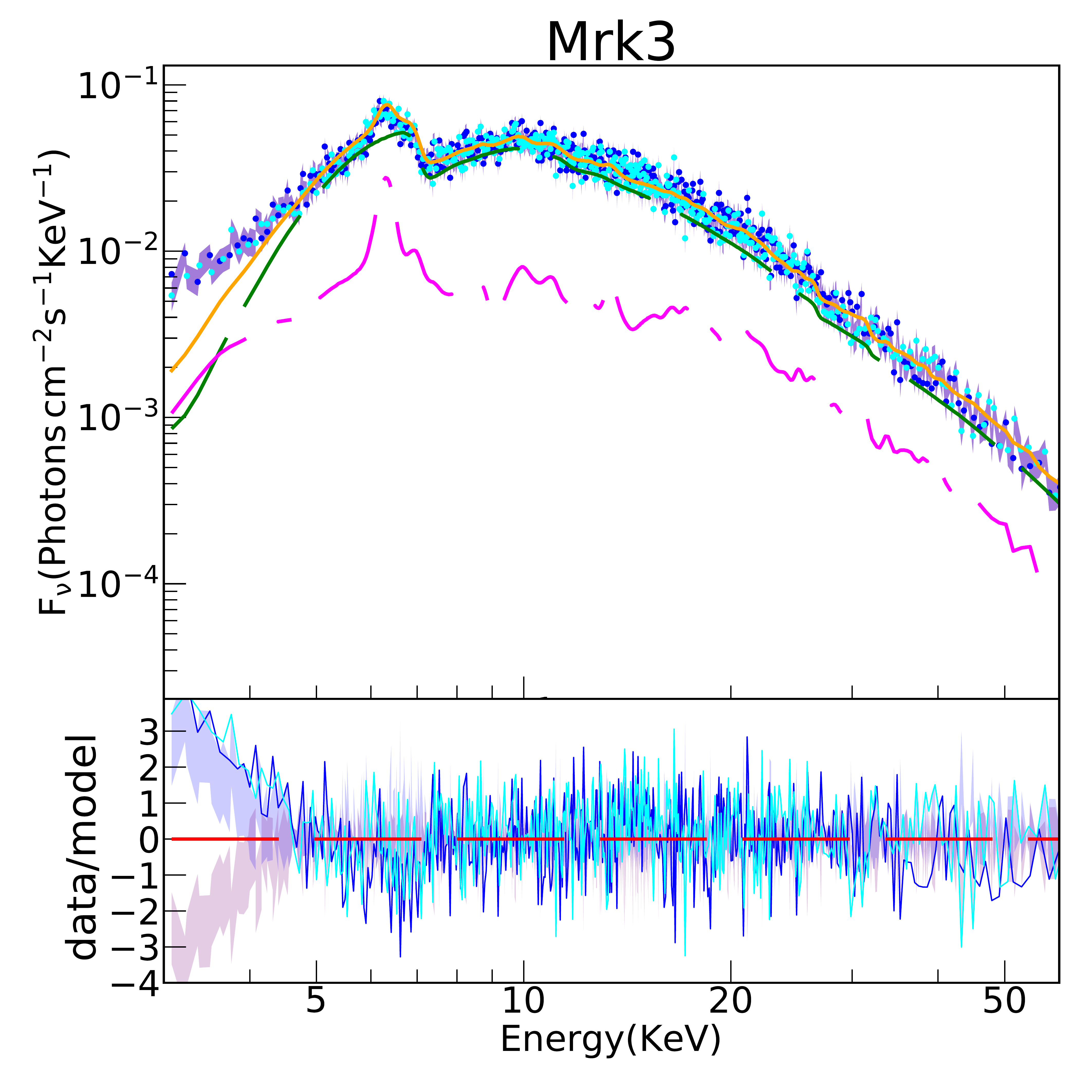}
    \caption{Unfolded spectra of Mrk\,1018, PG\,1535+547, ESO-428-G014, and MCG+07-41-03. The orange solid lines are the best-fit obtained from the smooth (top) and clumpy (bottom) baseline model at X-ray wavelengths. \emph{NuSTAR} spectra are displayed with blue and purple solid lines. The green and magenta dotted lines show the the absorbed power-law and the reprocessed components, respectively. The lower panels display the residuals between data and the best-fit model.}
    \label{fig:badfit_borus02}
\end{figure*}

\begin{figure*}
    \centering
    \includegraphics[width=0.45\columnwidth]{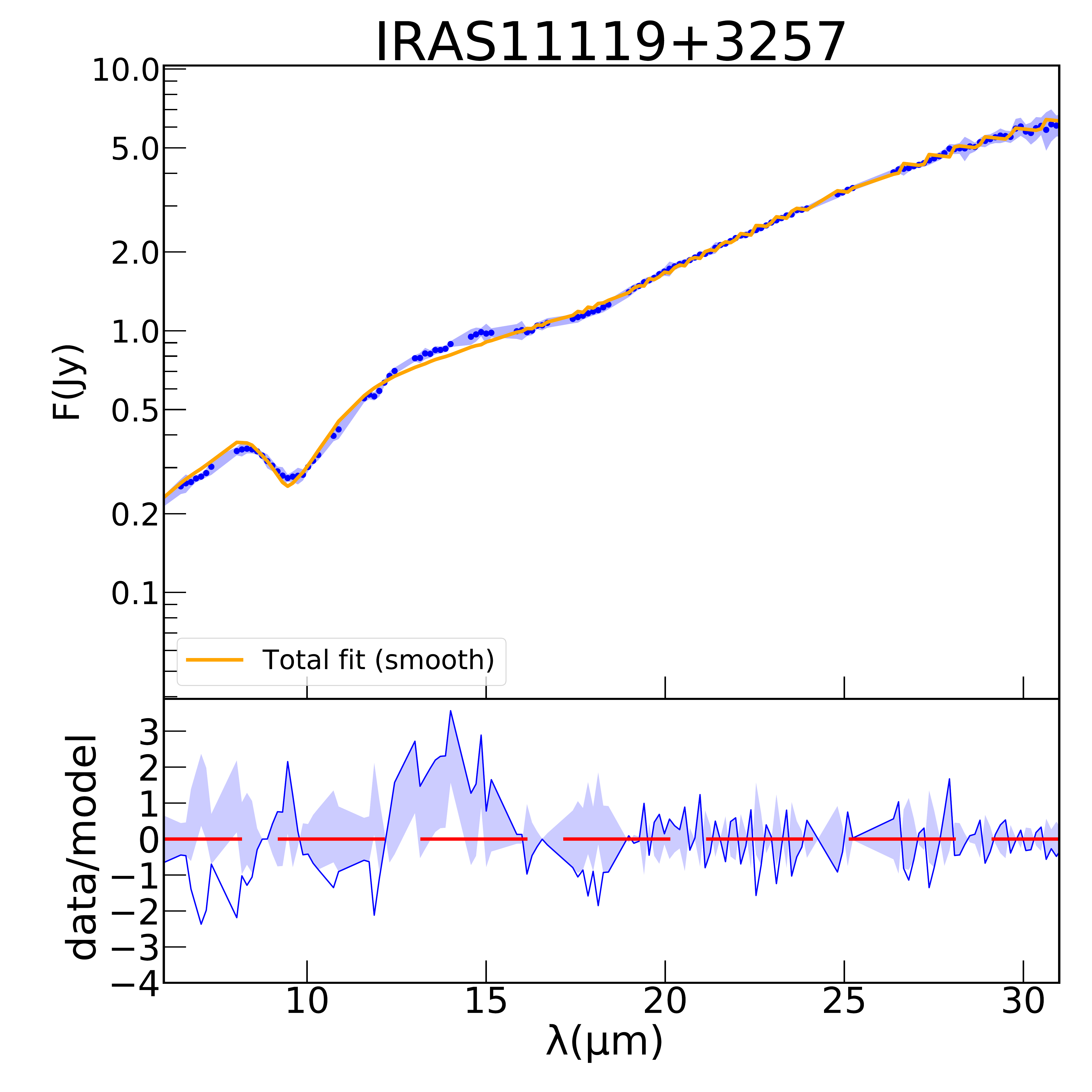}
    \includegraphics[width=0.45\columnwidth]{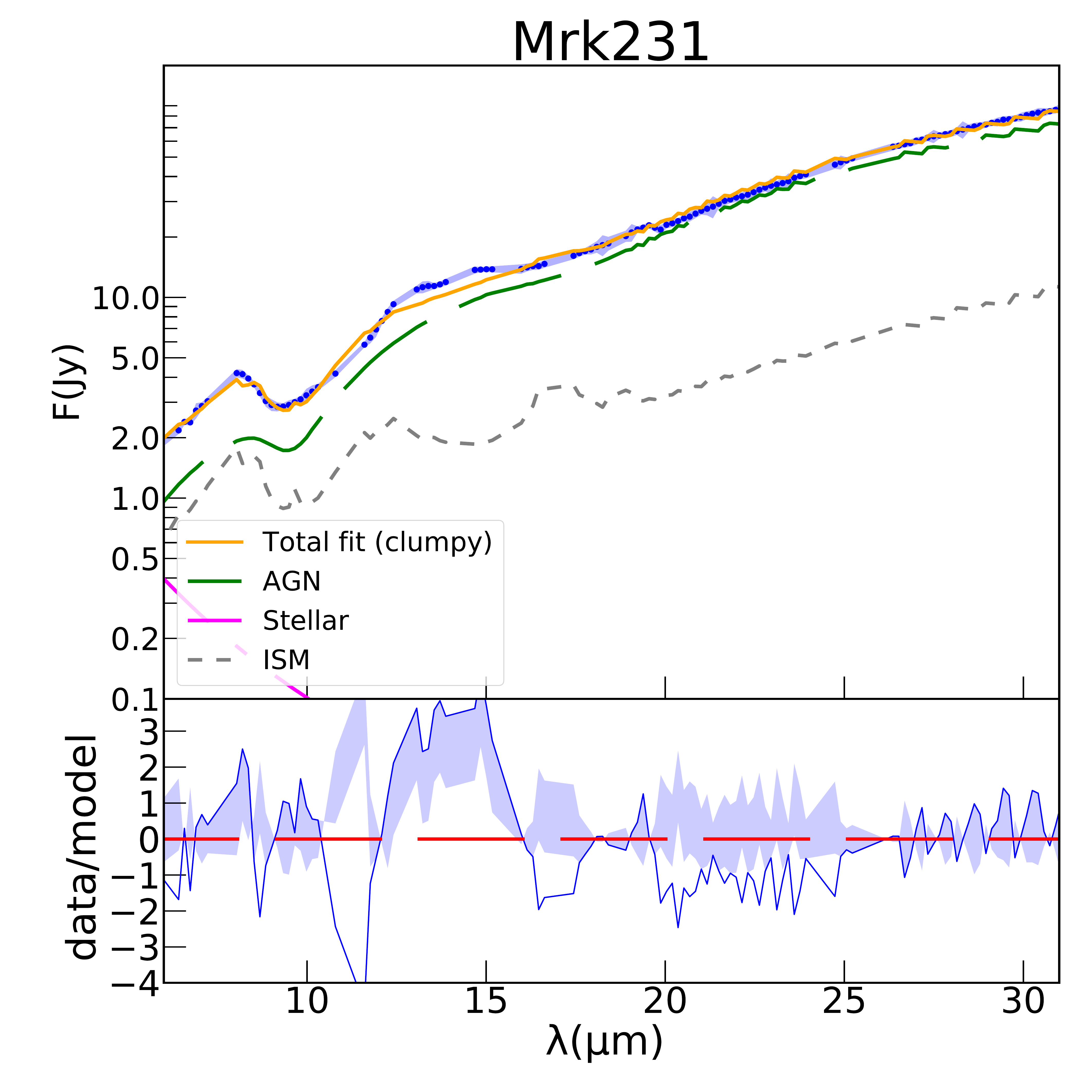}
    \includegraphics[width=0.45\columnwidth]{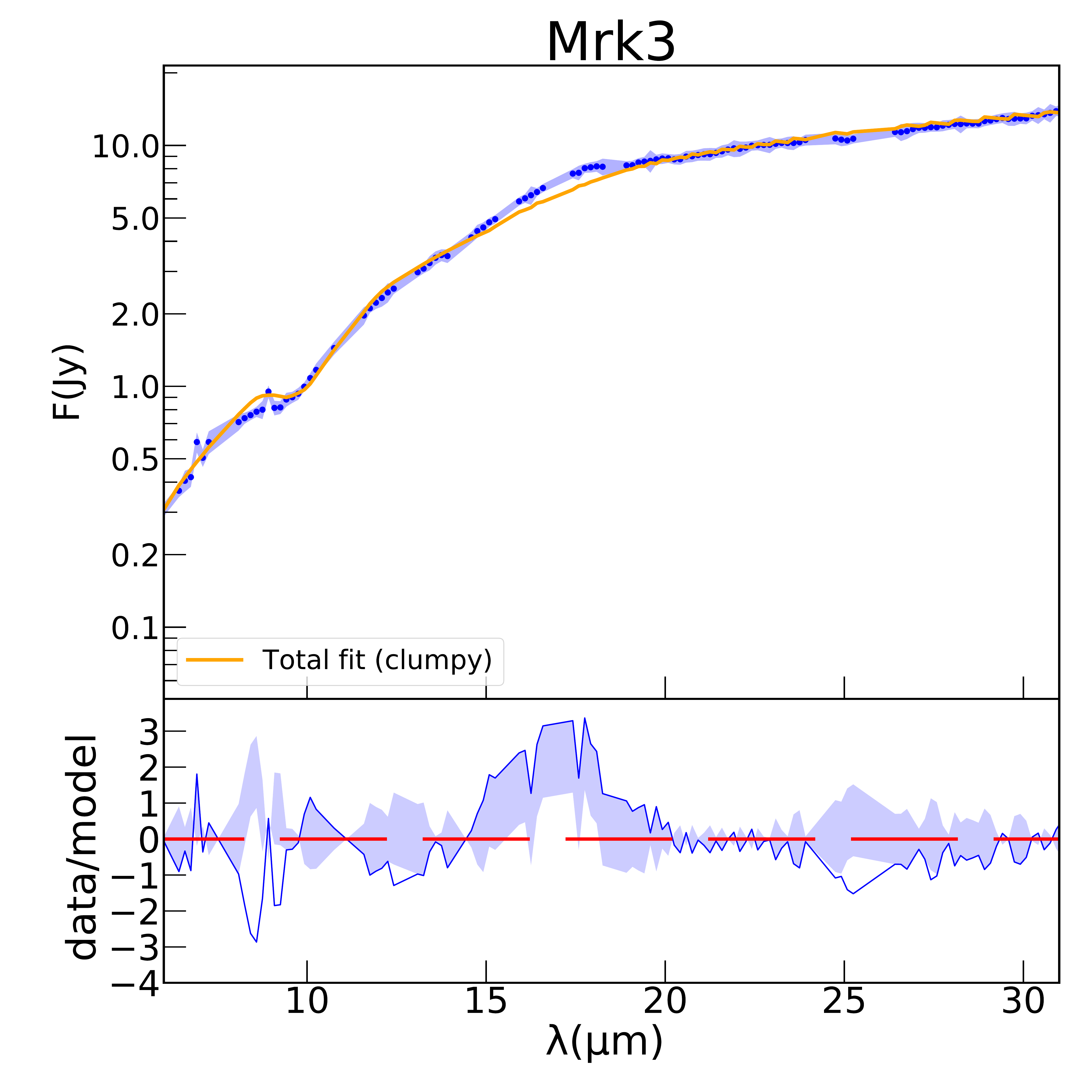}
    \includegraphics[width=0.45\columnwidth]{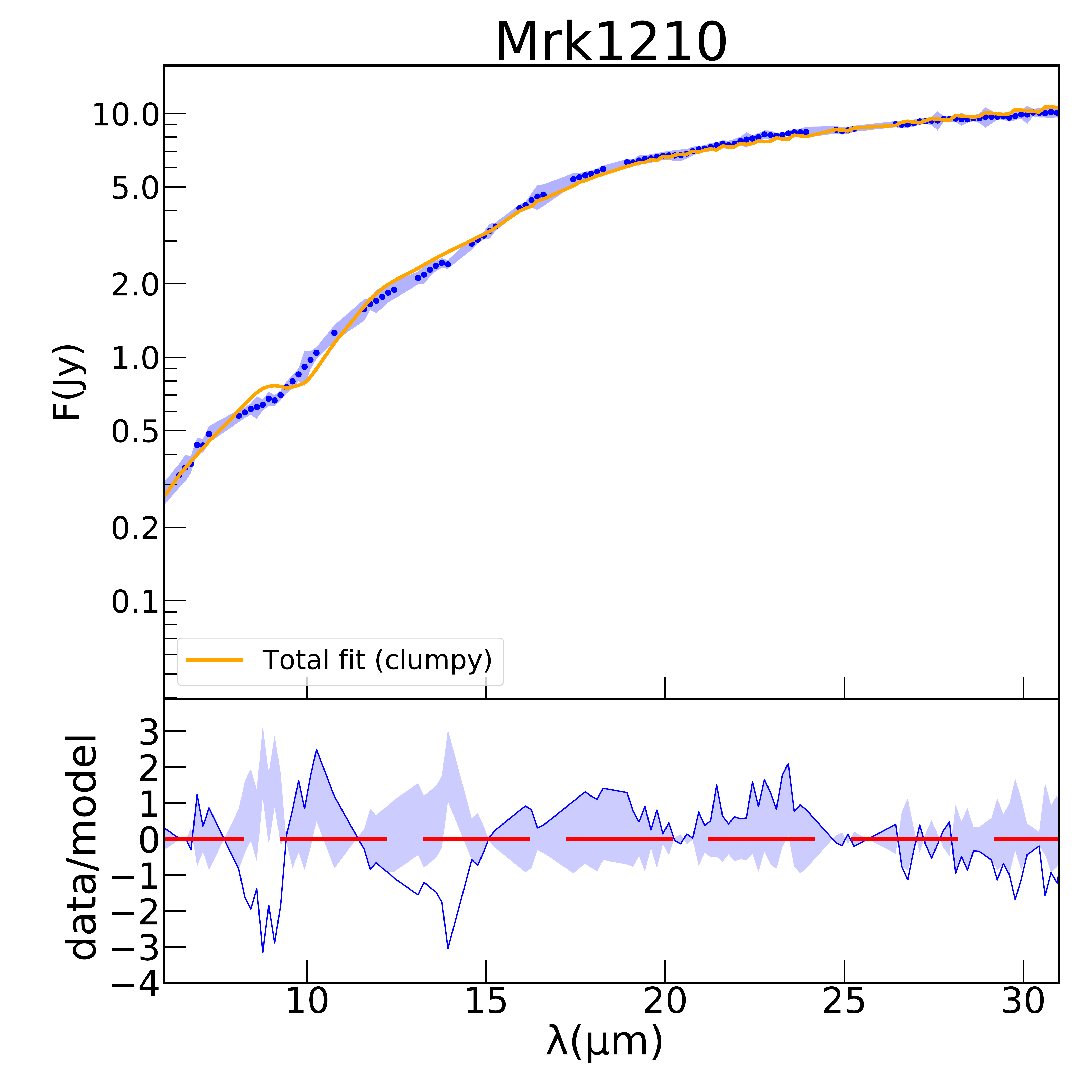}
    \includegraphics[width=0.45\columnwidth]{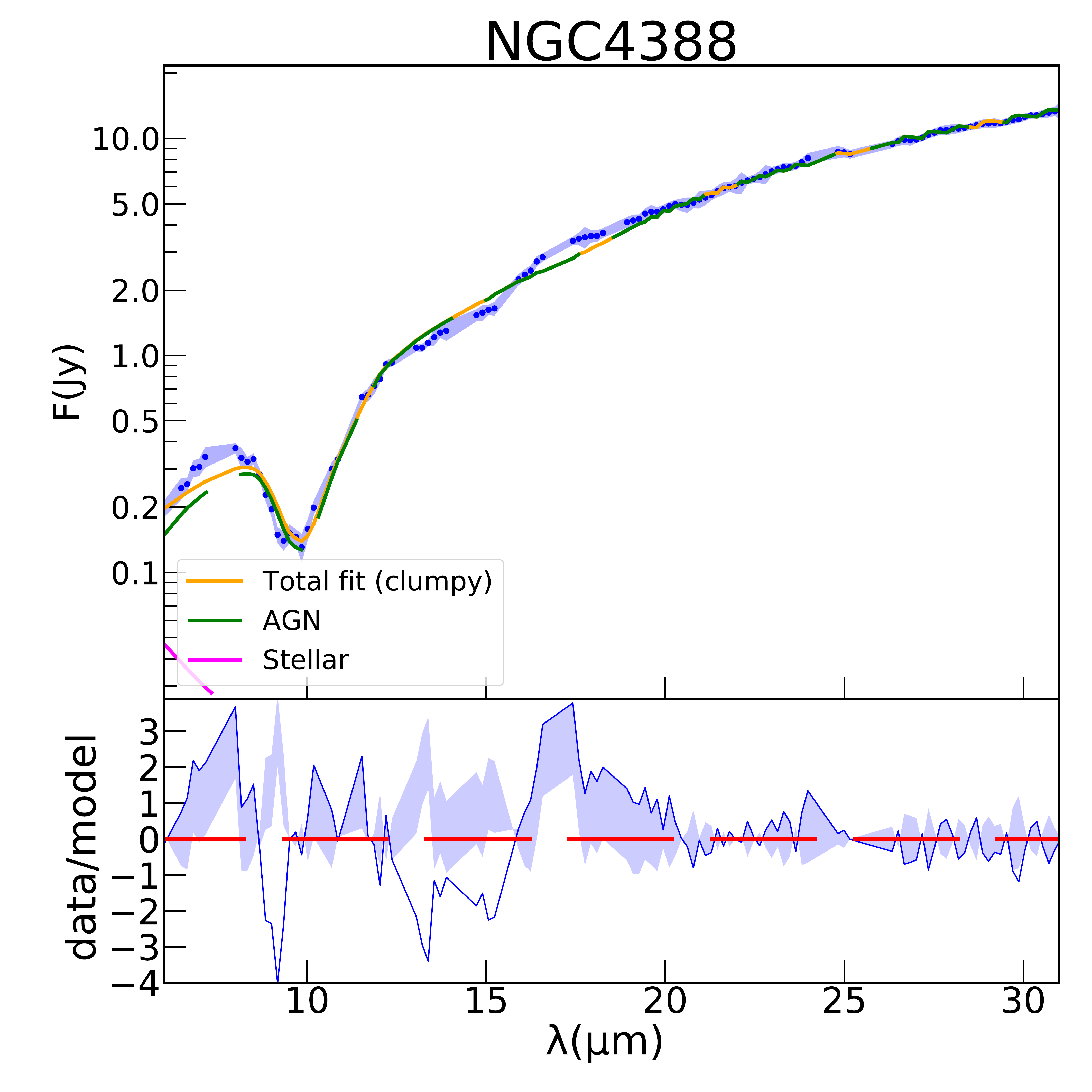}
    \includegraphics[width=0.45\columnwidth]{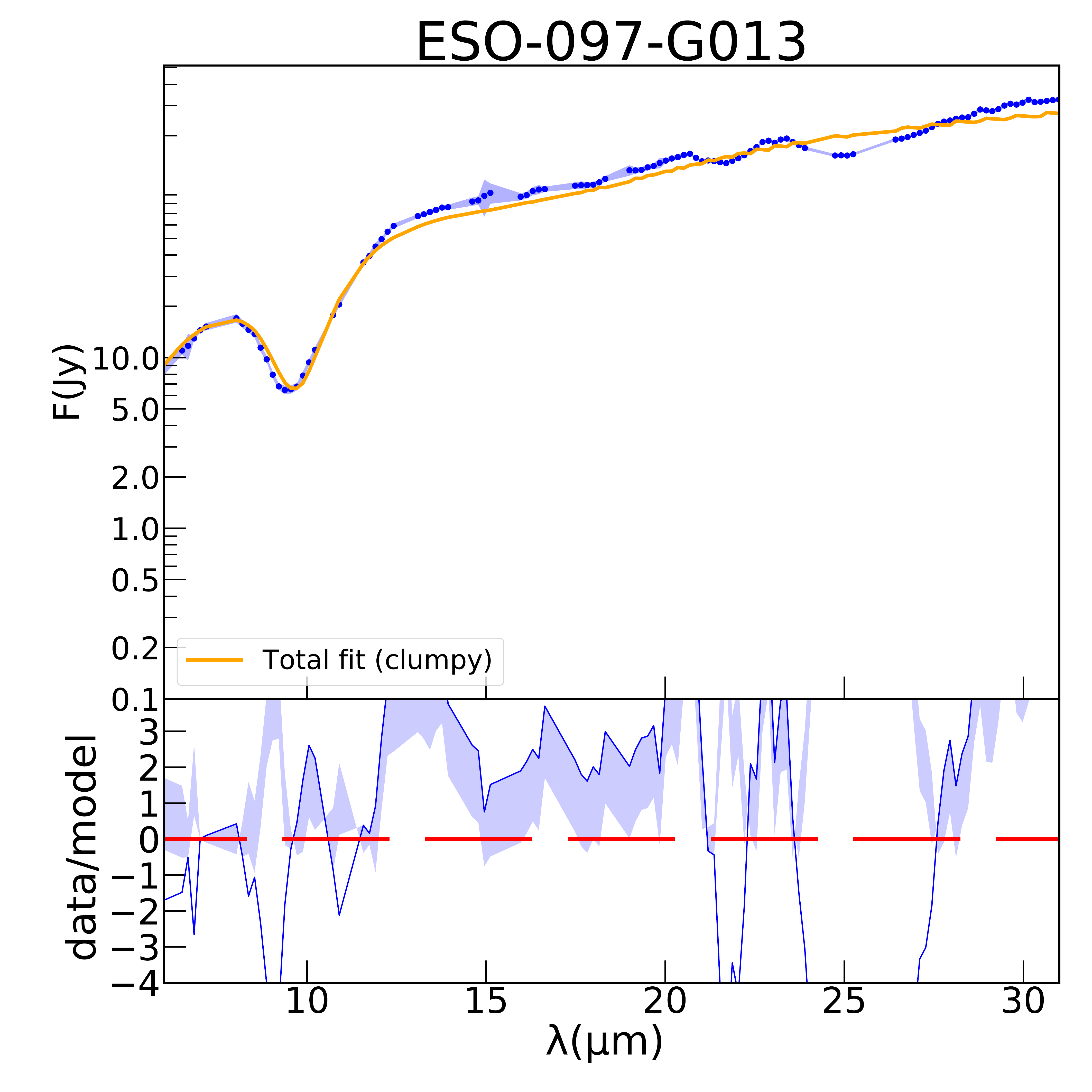}
    \includegraphics[width=0.45\columnwidth]{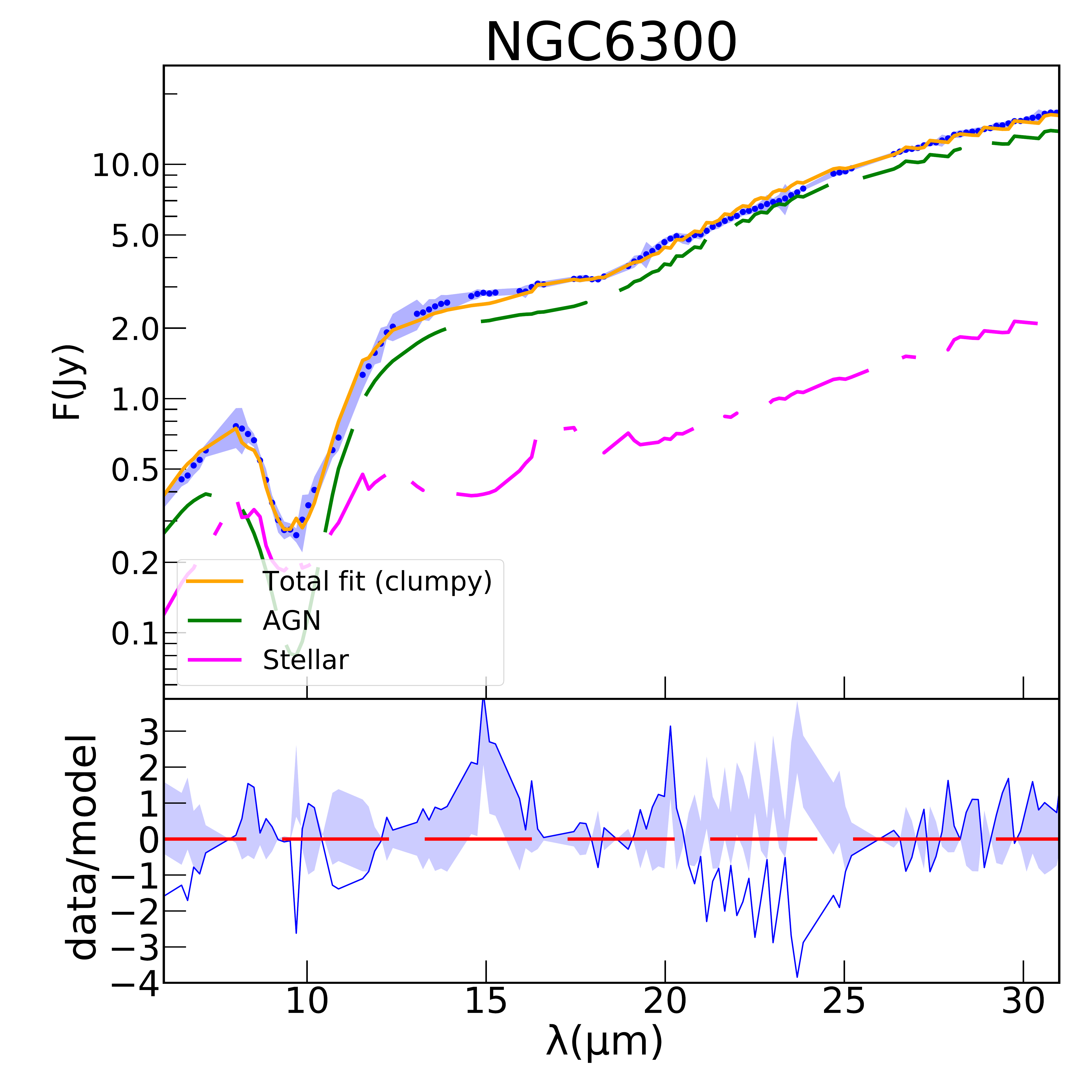}
    \includegraphics[width=0.45\columnwidth]{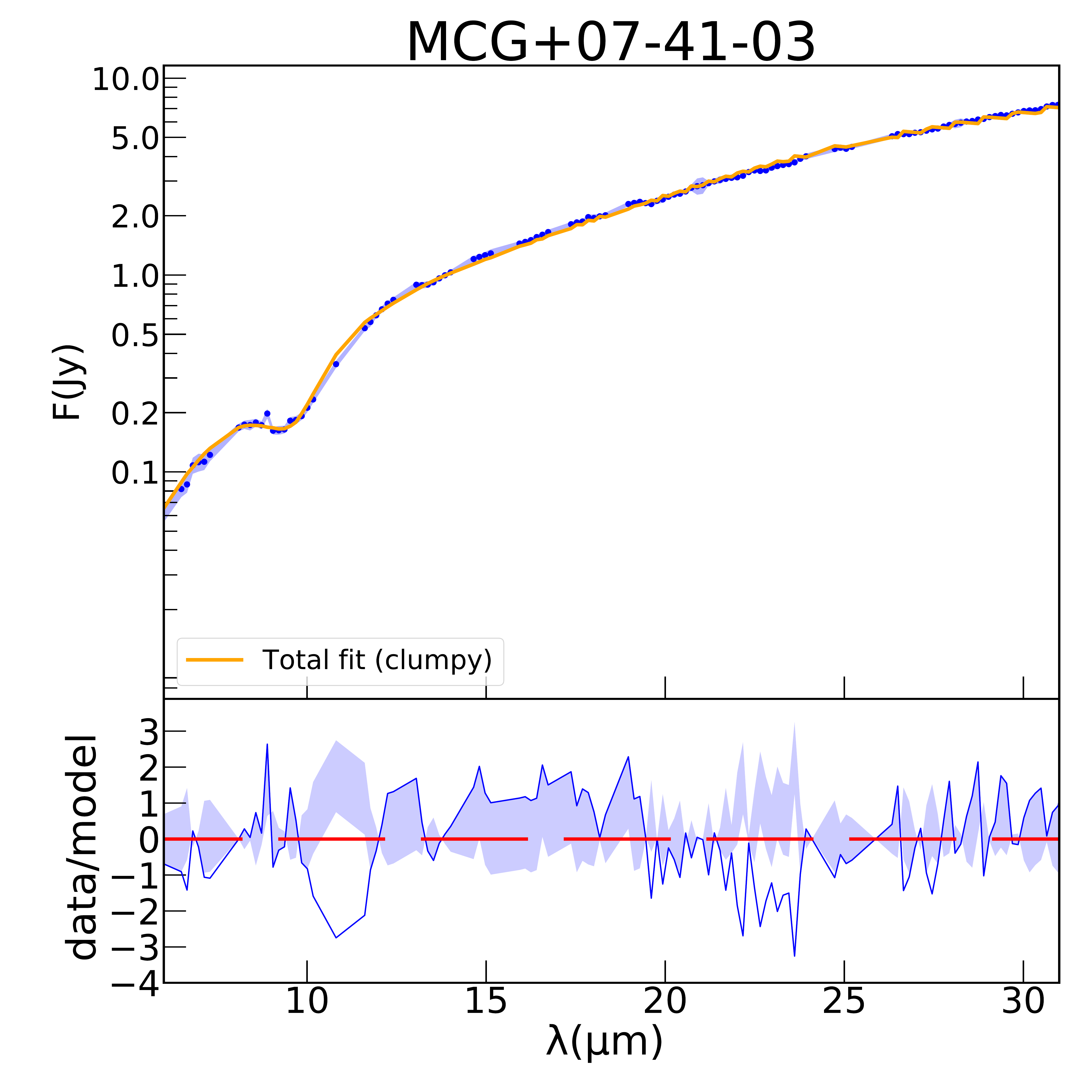}
    \caption{Unfolded spectra of I1119+3257,Mrk231,Mrk3,Mrk1210,NGC4388, ESO-097-G013, NGC6300, and MCG+07-41-03. The orange solid lines are the best-fit obtained from the smooth baseline model at mid-IR wavelengths. \emph{Spitzer} spectrum is displayed with blue points. The green,pink and gray dotted lines show the AGN, Stellar, and ISM components, respectively. The lower panels display the residuals between data and the best-fit model.}
    \label{fig:badfit_Fritz06}
\end{figure*}

We found 11 sources (4 at X-ray and 8 at mid-IR wavelengths) where the resulting fits in one or both wavelengths are bad ($\rm{\chi^2/d.o.f. > 1.2}$). 
Figures\,\ref{fig:badfit_borus02} and \ref{fig:badfit_Fritz06} show the best fit found at X-rays and mid-IR wavelengths, respectively.
The \emph{NuSTAR} spectra of Mrk\,1018 and PG\,1535+547 have less than 100 bins in the 3-70 KeV range and the FeK$\rm{\alpha}$ emission line is not clearly detected. Therefore low signal-to-noise ratio might explain the poor fit found. In the case of ESO\,428-G014, the spectra have less than 80 bins and the $\rm{N_{H}}$ value is the lowest obtained through the clumpy baseline model (see Figure\,\ref{fig:X-rayPar}, top-right panel). \citet{Fabbiano18} reported the discovery of kilo-parsec extended components in the 3-6\,keV hard continuum and FeK$\rm{\alpha}$ emission line for this source. They concluded that the hard continuum and the $\rm{FeK{\alpha}}$ emission line could be associated with circumnuclear clouds and not with a nuclear component (as the torus), which might explain why we fail to find a good fit with our baseline models. 

Regarding the statistically poor mid-IR fits, we found six sources where both baseline models are not capable to fit well the range between $\rm{10-20 \,\mu m}$. Interestingly all of them nicely fit at near-IR wavelengths, an issue that has been reported before for torus models at mid-IR \citep[e.g.][]{Netzer15}. Note that in the case of MCG\,+07-41-03, we are not capable of fitting the spectra through any of the tested X-ray and mid-IR baseline models. ESO\,097-G013 and NGC\,6300 show large residuals for the full mid-IR spectral range. In the case of NGC\,6300, we have an ISM contribution when fitting the \emph{Spitzer}/IRS spectra through the clumpy baseline model. Previously radio and near-IR observations of these two sources show evidence that most of the obscuration is associated with water masers or gas outflows, respectively \citep{Greenhill03, Gaspar19}. Therefore, the disc+wind model proposed by \citet{Hoenig17} might return better fits. However, these models are not available at X-ray wavelengths. The main discrepancies found at mid-IR wavelengths come from the 10 and 20 $\rm{\mu m}$ features. These discrepancies might be related to dust particles size and/or composition \citep{Martinez-Paredes20}. 


\section{NH vs inclination angles}
\label{sec:app3}

We investigated if the selection of viewing angles, $\theta_{inc}$, could affect the result presented in this work because it is one of the main differences expected between Sy1 and Sy2. However, the difference between the $\rm{N_{H}}$ obtained using the smooth (clumpy) baseline model with an viewing angle, $\theta_{inc}$, fixed to 19 (1) and 87 (90) is less than $\rm{log(N_{H})<}$0.5 for $95\%$ of the sample. Only Mrk\,1392 shows significant disagreement on the $\rm{N_H}$ (i.e. $\rm{log(N_{H}) > \,0.5}$) for each of the models. This source is located outside the one-to-one relation shown in Figure\,\ref{fig:X-rayPar} (top-right). Therefore, it shows a disagreement on the column densities obtained for smooth and clumpy models. If the viewing angle of the smooth and clumpy baseline models is fixed to 87 and 90, respectively, Mrk\,1392 could be located inside the relationship. However, these viewing angles are not compatible with the type of the source (Sy1). We concluded that the locus of the sources in Figure\,\ref{fig:X-rayPar} (top-right) does not depend on the viewing angle chosen in general.

\end{appendix}
\end{document}